\documentclass[twocolumn,tighten,times]{aastex63}
\hypersetup{linkcolor=magenta,citecolor=blue,filecolor=cyan}

\newcommand{\mytilde}{{\raise.17ex\hbox{$\scriptstyle\mathtt{\sim}$}}}

\usepackage{amsmath}

\received{July 23, 2021}
\accepted{January 14, 2022}
\submitjournal{The Astrophysical Journal}

\usepackage{CJKutf8}
\newcommand{\cntext}[1]{\begin{CJK*}{UTF8}{bkai}#1\end{CJK*}}

\shorttitle{The Regulated NiCu Cycles with the new $^{57}$C\lowercase{u}(\lowercase{p},$\gamma$)$^{58}$Zn reaction rate ...}
\shortauthors{Lam et al.}
\begin{document}
% \begin{CJK*}{UTF8}{bsmi}
% \begin{CJK*}{UTF8}{gbsn}

\title{The Regulated NiCu Cycles with the new $^{57}$C\lowercase{u}(p,$\gamma$)$^{58}$Zn reaction rate and \\ the Influence on Type-I X-Ray Bursts: GS~1826$-$24 Clocked Burster}

\correspondingauthor{Yi Hua Lam, Alexander Heger}
\email{lamyihua@impcas.ac.cn, alexander.heger@monash.edu}

% \author[0000-0001-6646-0745]{Yi Hua Lam}
\author[0000-0001-6646-0745]{Yi Hua Lam (\cntext{藍乙華})}
% \author[0000-0001-6646-0745]{Yi Hua Lam (蓝乙华)}
% \affiliation{CAS Key Laboratory of High Precision Nuclear Spectroscopy, Institute of Modern Physics, \\ Chinese Academy of Sciences, Lanzhou 730000, People's Republic of China}
\affiliation{Institute of Modern Physics, Chinese Academy of Sciences, Lanzhou 730000, People's Republic of China}
\affiliation{School of Nuclear Science and Technology, University of Chinese Academy of Sciences, Beijing 100049, People's Republic of China}
% \nocollaboration

% \author[0000-0001-6646-0745]{Ning Lu}
\author[0000-0002-3445-0451]{Ning Lu (\cntext{盧寧})}
% \author[0000-0002-3445-0451]{Ning Lu (卢宁)}
% \affiliation{CAS Key Laboratory of High Precision Nuclear Spectroscopy, Institute of Modern Physics, \\ Chinese Academy of Sciences, Lanzhou 730000, People's Republic of China}
\affiliation{Institute of Modern Physics, Chinese Academy of Sciences, Lanzhou 730000, People's Republic of China}
\affiliation{School of Nuclear Science and Technology, University of Chinese Academy of Sciences, Beijing 100049, People's Republic of China}
\affiliation{School of Nuclear Science and Technology, Lanzhou University, Lanzhou 730000, People's Republic of China}
% \nocollaboration
% \collaboration{1}{(AAS Journals Data Scientists collaboration)}

\author[0000-0002-3684-1325]{Alexander Heger}
\affiliation{School of Physics and Astronomy, Monash University, Victoria 3800, Australia}
\affiliation{OzGrav-Monash -- Monash Centre for Astrophysics, School of Physics and Astronomy, Monash University, VIC 3800, Australia}
\affiliation{Center of Excellence for Astrophysics in Three Dimensions (ASTRO-3D), VIC 3800, Australia}
\affiliation{The Joint Institute for Nuclear Astrophysics, Michigan State University, East Lansing, MI 48824, USA}
%\affiliation{Department of Physics and Astronomy, Michigan State University, East Lansing, MI 48824, USA}
% \nocollaboration{1}

\author[0000-0002-3580-2420]{Adam Michael Jacobs}
\affiliation{The Joint Institute for Nuclear Astrophysics, Michigan State University, East Lansing, MI 48824, USA}
\affiliation{Department of Physics and Astronomy, Michigan State University, East Lansing, MI 48824, USA}

% \collaboration{1}{(LaTeX collaboration)}

\author[0000-0001-8944-7631]{Nadezda~A.~Smirnova}
\affiliation{CENBG, CNRS/IN2P3 and University of Bordeaux, Chemin du Solarium, 33175 Gradignan cedex, France}%

\author{Teresa~Kurtukian~Nieto}
\affiliation{CENBG, CNRS/IN2P3 and University of Bordeaux, Chemin du Solarium, 33175 Gradignan cedex, France}%

\author[0000-0003-4023-4488]{Zac Johnston}
\affiliation{The Joint Institute for Nuclear Astrophysics, Michigan State University, East Lansing, MI 48824, USA}
\affiliation{Department of Physics and Astronomy, Michigan State University, East Lansing, MI 48824, USA}

% % \author{Jian Jun He}
% \author{Jian Jun He (何建軍)}
% \affiliation{Key Laboratory of Optical Astronomy, National Astronomical Observatories,\\Chinese Academy of Sciences, Beijing 100012, People's Republic of China}

% \author{Shigeru Kubono}
\author{Shigeru Kubono (\cntext{久保野\;茂})}
% \author[0000-0003-0783-5978]{Shigeru Kubono (久保野茂)}
\affiliation{RIKEN Nishina Center, 2-1 Hirosawa, Wako, Saitama 351-0198, Japan}
\affiliation{Center for Nuclear Study, University of Tokyo, 2-1 Hirosawa, Wako, Saitama 351-0198, Japan}
%

% \nocollaboration{2}



\begin{abstract}
%\limsum[1-2]
%\blindtext
During the X-ray bursts of GS 1826$-$24, ``clocked burster', the nuclear reaction flow that surges through the rapid-proton capture process path has to pass through the NiCu cycles before reaching the ZnGa cycles that moderate the further extent of hydrogen burning in the region above germanium and selenium isotopes. The $^{57}$Cu(p,$\gamma$)$^{58}$Zn reaction located in the NiCu cycles plays an important role in influencing the burst light curves as found by \citet{Cyburt2016}. We deduce the $^{57}$Cu(p,$\gamma$)$^{58}$Zn reaction rate based on the experimentally determined important nuclear structure information, isobaric-multiplet-mass equation, and large-scale shell model calculations. Based on the isobaric-multiplet-mass equation, we propose a possible order of $1^+_1$ and $2^+_3$ dominant resonance states and constrain the resonance energy of the $1^+_2$ state. The latter reduces the contribution of the $1^+_2$ dominant resonance state. The new reaction rate is up to a factor of four lower than the \citet{Forstner2001} rate recommended by JINA REACLIB v2.2 at the temperature regime sensitive to clocked bursts of GS 1826$-$24. Using the simulation from the one-dimensional implicit hydrodynamic code, \textsc{Kepler}, to model the thermonuclear X-ray bursts of GS 1826$-$24 clocked burster, we find that the new $^{57}$Cu(p,$\gamma$)$^{58}$Zn coupled with the latest $^{56}$Ni(p,$\gamma$)$^{57}$Cu and $^{55}$Ni(p,$\gamma$)$^{56}$Cu reaction rates redistributes the reaction flow in the NiCu cycles and strongly influences the burst ash composition, whereas the $^{59}$Cu(p,$\alpha$)$^{56}$Ni and $^{59}$Cu(p,$\gamma$)$^{60}$Zn reactions suppress the influence of the $^{57}$Cu(p,$\gamma$)$^{58}$Zn reaction and diminish the impact of nuclear reaction flow that by-passes the important $^{56}$Ni waiting point induced by the $^{55}$Ni(p,$\gamma$)$^{56}$Cu reaction on burst light curve.
\end{abstract}
% The influence of the newly deduced $^{56}$Ni(p,$\gamma$)$^{57}$Cu is also discussed.
%\textcolor{red}{Comprehensive study}
% reduces the production of $^{58}$Zn


% \end{CJK*}
% \begin{CJK*}
\section{Introduction} \label{sec:intro}
% \end{CJK*}
% \begin{CJK*}

Thermonuclear (Type I) X-ray bursts (XRBs) originate in the high density-temperature degenerate envelope of a neutron star in a close low-mass X-ray binary during thermonuclear runaways \citep{Woosley1976, Joss1977}. The envelope consists of stellar material accreted from the low-mass companion star. Every episode of XRBs encapsulates abundant information of the hydrodynamics and thermal states of the evolution of the degenerate envelope \citep{Woosley2004}, the structure of the accreting neutron star \citep{Steiner2010}, the rapid-proton capture (rp-) process path of synthesized nuclei \citep{Wormer1994,Schatz1998}, and the burst ashes that become compositional inertia for the succeeding bursts before sinking into the neutron-star crust \citep{Keek2011,Meisel2018}.
% stay in hydrodynamical equilibrium state and 
% The accreted H and He are the main accreted fuel for the nucleosyntheses occuring in the envelope. 
%that synthesize proton-rich nuclei which become the compositional inertia for the subsequent XRB or decay to burst ashes settling onto the neutron-star surface

XRBs are driven by the triple-$\alpha$ reaction \citep{Joss1978}, $\alpha$p-process \citep{Woosley1984}, rp-process \citep{Wallace1981,Wiescher1987}, and are constrained by $\beta$-decay and the proton dripline. After breaking out from the hot CNO cycle, the nuclear reaction flows enter the $sd$-shell nuclei region via $\alpha$p-processes, also this is the region of which the $\alpha$p-processes are dominant. Then, the reaction flows continue to the $pf$-shell nuclei region with first going through a few important cycles at the light \emph{pf}-shell nuclei, e.g., the CaSc cycle, and then reach the medium \emph{pf}-shell nuclei of which the NiCu and ZnGa cycles reside \citep{Wormer1994}. After breaking out from the ZnGa cycles and the GeAs cycle, which may transiently and weakly exist, and passing through Ge and Se isotopes, the reaction flows surge through the heavier proton-rich nuclei of where rp-processes actively burn the remaining hydrogen accreted from the companion star; and eventually the reaction flows stop at the SnSbTe cycles \citep{Schatz2001}. This rp-process path is indicated in the pioneering GS~1826$-$24 clocked burster model \citep{Woosley2004, Heger2007}.
%according to the XRB models of this work

\begin{figure}[t]
\begin{center}
%\vspace{5mm}
\includegraphics[width=5.7cm, angle=0]{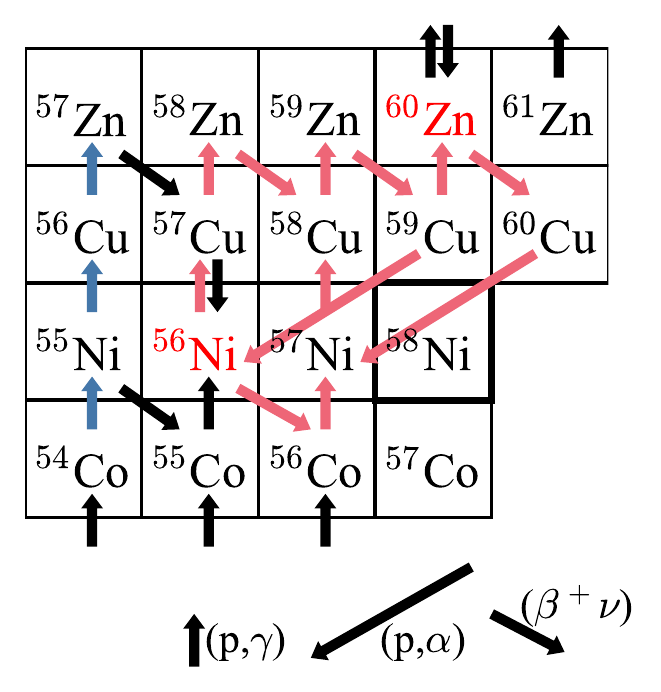}
%\vspace{-12mm}
%\caption{\label{fig:Cycles_NiCu}The rp-process path passing through the NiCu cycles. Stable nuclei are represented by thick black squares, and waiting points are shown in red texts. The NiCu cycles are displayed as red arrows. The NiCu I cycle consists of $^{56}$Ni(p,$\gamma$)$^{57}$Cu(p,$\gamma$)$^{58}$Zn($\beta^+\nu$)$^{58}$Cu(p,$\gamma$)$^{59}$Zn($\beta^+\nu$)$^{59}$Cu(p,$\alpha$) $^{56}$Ni reactions, and the NiCu II cycle is a series of $^{57}$Ni(p,$\gamma$) $^{58}$Cu(p,$\gamma$)$^{59}$Zn($\beta^+\nu$)$^{59}$Cu(p,$\gamma$)$^{60}$Zn($\beta^+\nu$)$^{60}$Cu(p,$\alpha$)$^{57}$Ni reactions \citep{Wormer1994}. The other sub-NiCu~II cycle, $^{56}$Ni($\beta^+\nu$)$^{56}$Co(p,$\gamma$)$^{57}$Ni(p,$\gamma$)$^{58}$Cu(p,$\gamma$)$^{59}$Zn($\beta^+\nu$)$^{59}$Cu (p,$\alpha$)$^{56}$Ni, can also be established. The matter flow induced by the $^{55}$Ni(p,$\gamma$)$^{56}$Cu reaction to bypass the $^{56}$Ni waiting point is illustrated in blue arrows. (All color figures in this paper are available in the online journal.)}
\singlespace
\caption{\label{fig:Cycles_NiCu}{\footnotesize The rp-process path passing through the NiCu cycles. Stable nuclei are represented by thick black squares,~and~waiting points are shown in red texts. The NiCu cycles are displayed~as red arrows. The NiCu I cycle consists of $^{56}$Ni(p,$\gamma$)$^{57}$Cu(p,$\gamma$)$^{58}$Zn ($\beta^+\nu$)$^{58}$Cu(p,$\gamma$)$^{59}$Zn($\beta^+\nu$)$^{59}$Cu(p,$\alpha$)$^{56}$Ni reactions, and the NiCu~II cycle is a series of $^{57}$Ni(p,$\gamma$)$^{58}$Cu(p,$\gamma$)$^{59}$Zn($\beta^+\nu$)$^{59}$Cu(p,$\gamma$)$^{60}$Zn($\beta^+\nu$) $^{60}$Cu(p,$\alpha$)$^{57}$Ni reactions \citep{Wormer1994}. The other sub-NiCu~II cycle, $^{56}$Ni($\beta^+\nu$)$^{56}$Co(p,$\gamma$)$^{57}$Ni(p,$\gamma$)$^{58}$Cu(p,$\gamma$)$^{59}$Zn($\beta^+\nu$)$^{59}$Cu (p,$\alpha$)$^{56}$Ni, can also be established. The matter flow induced by the $^{55}$Ni(p,$\gamma$)$^{56}$Cu reaction to bypass the $^{56}$Ni waiting point is illustrated in blue arrows. (All color figures in this paper are available in the online journal.)}}
%Pointing upward arrows indicate the (p,$\gamma$) reactions, whereas pointing downward arrows show the photodisintegration ($\gamma$,$p$) reactions. Slanting arrows from left to right depict ($\beta^+ \nu$) decays and long slanting arrows from right to left represent (p,$\alpha$) reactions.
\end{center}
\vspace{-5mm}
\end{figure}

The $^{57}$Cu(p,$\gamma$)$^{58}$Zn reaction that draws material from the $^{56}$Ni waiting point via the $^{56}$Ni(p,$\gamma$)$^{57}$Cu branch is located in the NiCu I cycle (Fig.~\ref{fig:Cycles_NiCu}). The influence of this reaction on XRB light curve and on burst ash abundances was studied by \citet{Cyburt2016}, and they concluded that the $^{57}$Cu(p,$\gamma$)$^{58}$Zn reaction is the fifth most influential (p,$\gamma$) reaction that affects the light curve of GS~1826$-$24 clocked burster \citep{Makino1988, Tanaka1989, Ubertini1999}. %whereas \citet{Schatz2017} found that 
\citet{Forstner2001} constructed the $^{57}$Cu(p,$\gamma$)$^{58}$Zn reaction rate based on shell-model calculation and predicted the properties of important resonances. Later, \citet{Langer2014} experimentally confirmed some low-lying energy levels of $^{58}$Zn, which are dominant resonances contributing to the $^{57}$Cu(p,$\gamma$)$^{58}$Zn reaction rate at temperature range $0.3\lesssim T\mathrm{(GK)}\lesssim 2.0$. With the high precision measurement of these energy levels, Langer et al. largely reduced the rate uncertainty up to 3 orders of magnitude compared to Forstner et al. reaction rate. Nevertheless, the order of $1^+_1$ and $2^+_3$ dominating resonance states was unconfirmed, and the $1^+_2$ resonance state, which is one of the dominant resonances at XRB temperature range, $0.8\lesssim T\mathrm{(GK)}\lesssim2$, was not detected in their experiment.

\begin{figure}[t]
\begin{center}
%\hspace{14mm}
%\includegraphics[width=\columnwidth, angle=0]{rp_57Cu_58Zn_grace_contri.eps}
\includegraphics[width=8.7cm, angle=0]{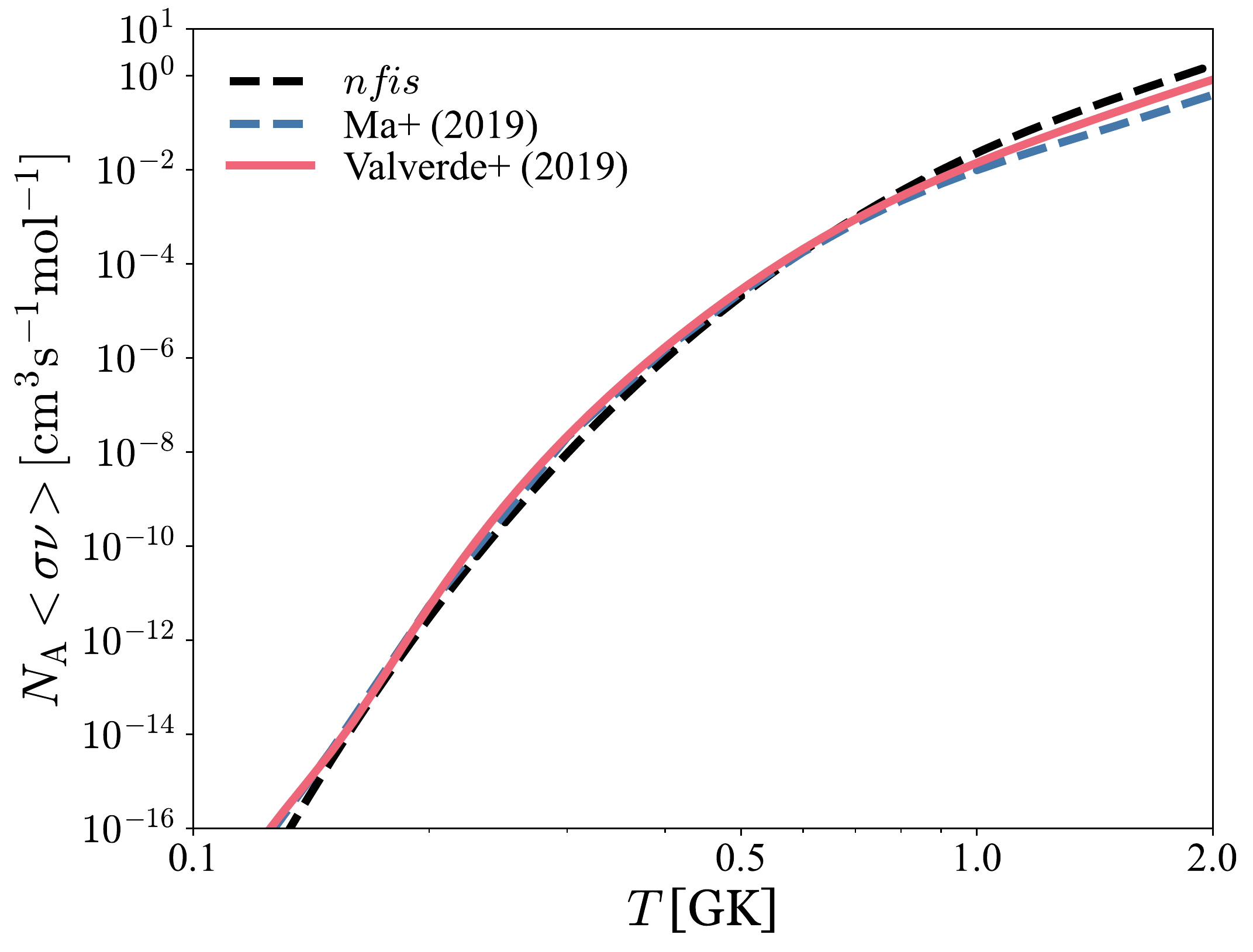}
% \includegraphics[width=9.0cm, angle=0]{rp_57Cu_58Zn_grace_contri.eps}
% \vspace{-12mm}
\singlespace
\caption{\label{fig:rp_55Ni_56Cu}{\footnotesize The $^{55}$Ni(p,$\gamma$)$^{56}$Cu thermonuclear reaction rates. \citet{Valverde2019} corrected the $^{55}$Ni(p,$\gamma$)$^{56}$Cu reaction rate (red solid line) after \citet{Ma2019} proposed a $^{55}$Ni(p,$\gamma$)$^{56}$Cu reaction rate (blue dashed line) based on the $^{56}$Cu proton separation energy, $S_\mathrm{p}$($^{56}$Cu)~$=579.8$~keV, and pointed out the incorrect penetrability scaling factor implemented by \citet{Valverde2018}. The \citet{Fisker2001} rate (black dashed line) is recommended by JINA REACLIB v2.2 \citep{Cyburt2010}.}}
%(A color version of this figure is available in the online journal.)
\end{center}
\end{figure}

The $^{55}$Ni(p,$\gamma$)$^{56}$Cu reaction rate was recently determined by \citet{Valverde2019} and \citet{Ma2019} with the highly-precisely measured $^{56}$Cu mass \citep{Valverde2018} and the precisely measured excited states of $^{56}$Cu \citep{Ong2017}. In fact, \citet{Ma2019} found that the $^{55}$Ni(p,$\gamma$)$^{56}$Cu reaction rate was up to one order of magnitude underestimated by \citet{Valverde2018} due to the incorrect penetrability scaling factor, causing a set of wrongly determined burst ash abundances of nuclei $A =55$~--~60. Figure~\ref{fig:rp_55Ni_56Cu} presents the comparison of the $^{55}$Ni(p,$\gamma$)$^{56}$Cu reaction rates deduced by \citet{Valverde2019}, \citet{Ma2019}, and \citet{Fisker2001}. The reaction rate was then corrected by \citet{Valverde2019} and used in their updated one-zone XRB model indicating that the reaction flow by-passing the important $^{56}$Ni waiting point could be established. Based on the updated zero-dimensional one-zone hydrodynamic XRB model, the extent of the impact the newly corrected $^{55}$Ni(p,$\gamma$)$^{56}$Cu reaction induces on the by-passing reaction flow, however, causes merely up to 5\% difference in the productions of nuclei $A=55$~--~65 \citep{Valverde2019}. Moreover, due to the zero-dimensional feature of one-zone XRB model, the distribution of synthesized nuclei along the mass coordinate in the accreted envelope is unknown, and importantly, the one-zone hydrodynamic XRB model does not match with any observation. %considered not significant as referred to the findings of the sensitivity study performed by \citet{Cyburt2016}. 
%the impact of the corrected $^{55}$Ni(p,$\gamma$)$^{56}$Cu reaction rate on the abundance of nuclei heavier than $^{56}$Cu is actually insignificant.
%based on the new compared to the \citet{Fisker2001} rate, which is the recommended rate in REACLIB v2.2.  

%In this work, Section~\ref{sec:rate_formalism}, we present the formalism in calculating the reaction rates and the isobaric-multiplet-mass equation \citep{Wigner1957} that we use to determine the order of $1^+_1$ and $2^+_3$ (states of $^{58}$Zn) dominating resonance states and to estimate the resonance energy of $1^+_2$ resonance state. Then, the discussion of the reconstructed and newly deduced $^{57}$Cu(p,$\gamma$)$^{58}$Zn reaction rates are shown in Section~\ref{sec:rates}.
In the present work, we re-analyze the nuclear structure information and perform simulations with the aim to constrain the reaction flows in the NiCu cycles and to analyze their impact on the clocked bursts of the GS~1824$-$26 burster. In Section~\ref{sec:rate_formalism}, we present the formalism for the reaction rate calculation and introduce the isobaric-multiplet-mass equation (IMME) that we use to cross-check the order of the $1^+_1$ and $2^+_3$ states in $^{58}$Zn, dominating resonances for the $^{57}$Cu(p,$\gamma$)$^{58}$Zn reaction, and to estimate the energy of the $1^+_2$ resonance state. The deduced $^{57}$Cu(p,$\gamma$)$^{58}$Zn reaction rate is discussed in detail in Section~\ref{sec:rates}. Using the one-dimensional multi-zone hydrodynamic \textsc{Kepler} code \citep{Weaver1978,Woosley2004,Heger2007}, we model a set of XRB episodes matched with the GS~1826$-$24 burster with the newly deduced $^{57}$Cu(p,$\gamma$)$^{58}$Zn, \citet{Valverde2019} $^{55}$Ni(p,$\gamma$)$^{56}$Cu, and \citet{Kahl2019} $^{56}$Ni(p,$\gamma$)$^{57}$Cu reaction rates. We study the influence of these rates, and also investigate the effect of the $^{56}$Ni-waiting-point bypassing matter flow induced by the $^{55}$Ni(p,$\gamma$)$^{56}$Cu reaction. The implication of the new $^{57}$Cu(p,$\gamma$)$^{58}$Zn, $^{55}$Ni(p,$\gamma$)$^{56}$Cu, and $^{56}$Ni(p,$\gamma$)$^{57}$Cu reaction rates on XRB light curve, the nucleosyntheses in and evolution of the accreted envelope of GS~1826$-$24 (clocked burster) along the mass coordinate, is presented in Section~\ref{sec:Astro}. The conclusion of this work is given in Section~\ref{sec:summary}. 
% in Section~\ref{sec:Astro}
% and $^{55}$Ni(p,$\gamma$)$^{56}$Cu 

\section{Reaction rate calculations}
\label{sec:rate_formalism}
The total thermonuclear proton-capture reaction rate is expressed as the sum of resonant- (res) and direct-capture (DC) on the ground state and thermally excited states in the target nucleus, and each capture with given initial and final states is weighted with its individual population factor~\citep{Fowler1964,Rolfs1988}, 
\begin{eqnarray}
\label{eq:total}
N_\mathrm{A}\langle \sigma v \rangle = && \sum_i(N_\mathrm{A}\langle \sigma v \rangle_\mathrm{res}^i+N_\mathrm{A}\langle \sigma v \rangle_\mathrm{DC}^i)\nonumber\\
&&\times\frac{(2J_i+1)e^{-E_i/kT}}{\sum_n(2J_n+1)e^{-E_n/kT}} \, ,
\end{eqnarray}
%P(E_\mathrm{x})=(2J+1)\times\mathrm{exp}\left(-\frac{E_\mathrm{x}}{kT} \right),
where $J$ are the angular momenta of initial states of target nucleus and $E$ are the energies of these initial states. 
%The reaction rate is expressed in units of $\mathrm{cm^3s^{-1}mol^{-1}}$.

%\subsection{\label{sec:Resonant}Resonant rates}
\paragraph{Resonant rate}
\label{sec:Resonant}
The resonant reaction rate for proton capture on a target nucleus in its initial state, $i$, $N_\mathrm{A}\langle \sigma v \rangle_\mathrm{res}^i$, is a sum over all respective compound nucleus states~$j$ above the proton separation energy~\citep{Rolfs1988,Iliadis2007}. The resonant rate can be expressed as~\citep{Fowler1967,Schatz2005}, 
%\begin{widetext}
\begin{eqnarray}
\label{eq:res}
N_\mathrm{A}\langle \sigma v \rangle_\mathrm{res}^i = && 1.54 \times 10^{11} (\mu T_9)^{-3/2} \nonumber\\
                                && \times \sum_j \omega\gamma_{ij} \mathrm{exp} \left (-\frac{11.605E^{ij}_\mathrm{res}}{T_9} \right)\, ,
\end{eqnarray}
%\end{widetext}
in units of $\mathrm{cm^3s^{-1}mol^{-1}}$, where the resonance energy in the center-of-mass system, $E^{ij}_\mathrm{res}=E^j_\mathrm{x} -S_\mathrm{p} -E_i$ (in MeV in Eq.~(\ref{eq:res})), is the energy difference between the compound nucleus $E^j_\mathrm{x}$ state and the sum of the excitation energies of the initial state $E_i$ and the respective proton threshold, $S_\mathrm{p}$. 
For the capture on the ground state, $E_i=0$.
%The resonance energy for the ground-state capture is simply expressed as $E^{ij}_\mathrm{res}$=$E_\mathrm{x}^j-S_\mathrm{p}$, where $E_i$=0 and $E^j_\mathrm{x}$ is the excited-state energy of the compound nucleus. 
$\mu$ is the reduced mass of the entrance channel in atomic mass units ($\mu=A_\mathrm{T}/(1+A_\mathrm{T})$, with $A_\mathrm{T}$ the target mass number), and $T_9$ is the temperature in Giga Kelvin (GK). The resonance energy and strength in Eq.~(\ref{eq:res}) are given in units of MeV. The resonance strength, $\omega\gamma_{ij}$, taken in MeV in Eq.~(\ref{eq:res}), reads
\begin{eqnarray}
\label{eq:resonantStrength}
\omega\gamma_{ij}=\frac{2J_j+1}{2(2J_i+1)}\frac{\Gamma_\mathrm{p}^{ij}\times\Gamma_\gamma^j}{\Gamma_\mathrm{total}^j} \, ,
\end{eqnarray}
where $J_i$ is the target spin and $J_j$, $\Gamma_\mathrm{p}^{ij}$, $\Gamma_\gamma^j$, and $\Gamma_\mathrm{total}^j$ are a spin, proton-decay width, $\gamma$-decay width, and total width of the compound nucleus state~$j$, respectively. Assuming that other decay channels are closed~\citep{AME2016} in the considered excitation energy range of the compound nuclei, the total width becomes $\Gamma_\mathrm{total}^j=\Gamma_\gamma^j+\Gamma_\mathrm{p}^{ij}$. Within the shell-model formalism which we use here, the proton width can be expressed as
\begin{eqnarray}
\label{eq:ProtonWidth_WS}
\Gamma_\mathrm{p} = \sum_{\mathtt{nlj}} C^2S(\mathtt{nlj}) \, \Gamma_\mathrm{sp}(\mathtt{nlj}) \, ,
\end{eqnarray}
%\mathsf{nlj}
where  $\Gamma_\mathrm{sp}$ is a single-particle width for the capture of a proton with respect to a given $(\mathtt{nlj})$ quantum orbital in a spherically-symmetric mean-field potential, while $C^2 S(\mathtt{nlj})$ denotes a corresponding spectroscopic factor containing information of the structure of the initial and final states. The $\Gamma_\mathrm{sp}$ can either be estimated from proton scattering cross sections in a Woods-Saxon potential with the adjusted potential depth to reproduce known proton energies~\citep{WSPOT}; or alternatively, it can also be obtained from the potential barrier penetrability calculation as~\citep{Wormer1994,Herndl1995},
\begin{eqnarray}
\label{eq:ProtonWidth_Tunnel}
\Gamma_\mathrm{sp}=\frac{3\hbar^2}{\mu R^2}P_{\mathtt{l}}(E) \, ,
\end{eqnarray}
where, $R=r_0\times(1+A_\mathrm{T})^{1/3}$ fm (with $r_0=1.25$~fm) is the nuclear channel radius; and the Coulomb barrier penetration factor $P_{\mathtt{l}}$ is
\begin{eqnarray}
P_{\mathtt{l}}(E)=\frac{kR}{F^2_{\mathtt{l}}(E)+G^2_{\mathtt{l}}(E)},
\label{eq6}
\end{eqnarray}
where $k=\sqrt{2\mu E}/\hbar$ and $E$ is the proton energy in the center-of-mass system; $F_{\mathtt{l}}$ and $G_{\mathtt{l}}$ are the regular and irregular Coulomb functions, respectively. In the present work, we follow the same procedure as was used by~\citet{Lam2016} to get the proton widths of the important $^{57}$Cu(p,$\gamma$)$^{58}$Zn resonances up to the Gamow window. The maximum difference between the $\Gamma_\mathrm{sp}$ described by the two methods above is below $40$~\% for the present work.
%Eq.~\ref{eq:ProtonWidth_WS} or ~\ref{eq:ProtonWidth_Tunnel} 

Gamma decay widths are obtained from electromagnetic reduced transition probabilities $B$($\Omega L; J_i$$\rightarrow$$ J_f$) ($\Omega $ stands for electric or magnetic), which contain the nuclear structure information of the resonance states and the final bound states. The corresponding gamma decay widths for the most contributed transitions (M1 and E2) can be expressed as~\citep{Brussaard1977} 
%$\Gamma_\gamma$=$\Gamma_{M1}+\Gamma_{E2}$~\citep{Brussaard1977} that can be expressed as%~\citep{Herndl1995},
\begin{eqnarray}
\label{eq:GammaWidth}
\Gamma_\mathrm{M1} =&& 1.16\times 10^{-2}E_\gamma^3 B(\mathrm{M1}) \, ,~\mathrm{and} \nonumber \\
\Gamma_\mathrm{E2} =&& 8.13\times 10^{-7}E_\gamma^5 B(\mathrm{E2}) \, ,
\end{eqnarray}
where $B(\mathrm{M1})$ are in $\mu_N^2$, $B(\mathrm{E2})$ are in e$^2$fm$^4$, $E_\gamma$ are in keV, while $\Gamma_\mathrm{M1}$ and $\Gamma_\mathrm{E2}$ are in units of eV. The $B(\mathrm{M1})$ values have been obtained from free $g$-factors, i.e., $g^s_\mathrm{p}=5.586$, $g^s_\mathrm{n}=-3.826$ and $g^l_\mathrm{p}=1$, $g^l_\mathrm{n}=0$; whereas the $B(\mathrm{E2})$ values have been obtained from standard effective charges, $e_\mathrm{p}=1.5e$, and $e_\mathrm{n}=0.5e$~\citep{Honma2004}. We use experimental energies, $E_\gamma $, when available. The total electromagnetic decay width is obtained from the summation of all partial decay widths for a given initial state.

\paragraph{Information of nuclear structure}
\label{sec:Nuclear_info}
The essential information needed to estimate the resonant rate contribution of $^{57}$Cu(p,$\gamma$)$^{58}$Zn consists of the resonance energies of the compound nucleus $^{57}$Cu$+$p, one-proton transfer spectroscopic factors, and proton- and gamma-decay widths. The properties of resonances sensitive to the $^{57}$Cu(p,$\gamma$)$^{58}$Zn reaction rate of XRB temperature range are provided by \citet{Langer2014}. Nevertheless, the order of the $1^+_1$ and $2^+_3$ states of $^{58}$Zn was undetermined by Langer et al. In order to reproduce Langer et al. rate, we find that the dominant resonances for the temperature range from $1$ to $2$~GK sensitive to XRB are not limited to the measured $2^+_4$ state. The $1^+_2$ and $2^+_5$ resonance states, which were not observed by Langer et al., also contribute to the total reaction rate at temperatures $0.8\lesssim T\mathrm{(GK)}\lesssim2$. 
%as estimated using shell model 

In the present study, we use the isobaric-multiplet-mass equation (IMME) to constrain the energies of experimentally unknown, but important resonance state in $^{58}$Zn, i.e., the order of the $1^+_1$ and $2^+_3$ states of $^{58}$Zn and the energy of $1^+_2$ state. A similar method was exploited earlier by \citet{Richter2011,Richter2012,Richter2013} to provide missing experimental information of the nuclear level schemes. Also, the same method was used by \citet{Schatz2017} to estimate the unknown nuclear masses important for reverse (p,$\gamma$) rates.
%and the respective resonance energies needed for constructing the $^{25}$Al(p,$\gamma$)$^{26}$Si, $^{35}$Ar(p,$\gamma$)$^{36}$K, and $^{29}$P(p,$\gamma$)$^{30}$S resonance rates. 
Assuming that the isospin-symmetry breaking forces are two-body operators of the isovector and isotensor character, the mass excesses of the members of an isobaric multiplet ($I=1$, $I_\mathrm{z}=-I,-I+1,\ldots,I$) show at most a quadratic dependence on $I_\mathrm{z}$, as expressed by the IMME~\citep{Wigner1957},
\begin{eqnarray}
\label{eq:IMME}
M_{I_\mathrm{z}}(\alpha,I)= a(\alpha,I) + b(\alpha,I) I_\mathrm{z} + c(\alpha,I) I_\mathrm{z}^2 \, ,
\end{eqnarray}
where $M_{I_\mathrm{z}}(\alpha,I)$ is the mass excess of a quantum state of isospin $(I,I_\mathrm{z})$, and $\alpha=(A, J^\pi, N_\mathrm{exc}, \dots)$ are the nuclear mass number $A$, excited state number $N_\mathrm{exc}$, and all other quantum numbers labeling the quantum state. The $a$, $b$, and $c$ coefficients reflect contributions from the isoscalar, isovector, and isotensor parts of the effective nucleon-nucleon interaction, respectively (see ~\citet{Ormand1989} or \citet{Lam2013a,Lam2013b} for details). For an isobaric-triplet states ($I=1, I_\mathrm{z}=-1, 0, +1$), we can form from Eq.~(\ref{eq:IMME}) a system of three linear equations, and therefore, 
express the IMME $c$ coefficient in terms of three mass excesses as
\begin{eqnarray}
\label{eq:isotensor}
c = \left[M_{-1}(\alpha,1) + M_{+1}(\alpha,1) - 2M_{0}(\alpha,1)\right] / 2 \, .
\end{eqnarray}
In turn, if we know the mass excesses of $I_\mathrm{z}=0$ and $I_\mathrm{z}=1$ isobaric multiplet members and a theoretical $c$ coefficient, the mass excess of a proton-rich member ($I_\mathrm{z}=-1$) can be found via a simple relation:
\begin{eqnarray}
\label{eq:protonRich}
M_{-1}(\alpha,1)= 2M_{0}(\alpha,1) - M_{+1}(\alpha,1) + 2c(\alpha,1) \, .
\end{eqnarray}
This equation defines the method which we use in the present paper.

We first obtain a set of theoretical IMME $c$ coefficients for the lowest and excited $A=58$ triplets,  including those which involve the dominant resonances. To this end, we perform large-scale shell-model calculations in the full \emph{pf} shell-model space  using the \textsc{NuShellX@MSU} shell-model code~\citep{NuShellX} with the charge-dependent Hamiltonian, which is constructed from the modern isospin-conserving Hamiltonian (GXPF1a;~\citealt{Honma2004,Honma2005}), the two-body Coulomb interaction, strong charge-symmetry-breaking and charge-independence-breaking terms~\citep{Ormand1989}, and the \emph{pf} shell-model space isovector single-particle energies~\citep{Ormand1995}. The Hamiltonian is referred to as ``cdGX1A'' and was used by \citet{Smirnova2016,Smirnova2017} to investigate the isospin mixing in $\beta$-delayed proton emission of \emph{pf}-shell nuclei. The IMME $c$ coefficients of these dominant resonances permit us to determine the order of $1^+_1$ and $2^+_3$ states of $^{58}$Zn and to estimate the resonance energy of the $1^+_2$ resonance state. Properties of all other resonances situated within the Gamow window corresponding to the XRB temperature range are computed using the \textsc{KShell} code~\citep{Shimizu2019} in a full \emph{pf} shell-model space with the GXPF1a Hamiltonian. For $A=57$ and $58$, Hamiltonian matrices of dimensions up to $1.58\times10^9$ have been diagonalized using thick-restart block Lanczos method.

The theoretical IMME $c$ coefficients are then compared with the available experimental data compiled in \citet{Lam2013b} and updated in the present work by the recently re-evaluated mass excesses of $^{58}$Zn, $^{58}$Cu, and $^{58}$Ni (\citealt{AME2016}; AME2016). For excited multiplets, the experimental information on level schemes have been taken from ~\citet{Langer2014} for $^{58}$Zn, from \citet{Rudolph1973,Rudolph1998,Rudolph2000} for $^{58}$Cu, and from \citet{Jongsma1972,Honkanen1981,Johansson2009,Rudolph2002} for $^{58}$Ni. The uncertainty of the measured $^{58}$Zn mass \citep{Seth1986} dominates the experimental IMME $c$ coefficients uncertainties and propagates to the proton separation energy of $^{58}$Zn, $S_\mathrm{p}$($^{58}$Zn)~$=2.280\pm0.050$~MeV (AME2016). In general, theoretical $c$ coefficients are seen to be in robust agreement with the respective experimental values. The comparison yields root-mean-square (rms) deviation of about $22$~keV, which we assign as theoretical uncertainty to the calculated values, see Table~\ref{tab:IMME}. 
%\textcolor{red}{I suggest to remove this sentence: We reanalysis \citet{Fujita2007} data and find that the 3540-keV state of $^{58}$Cu could be not a $1^+$ state, and this may yield a rather large IMME $c$ coefficient of 264$\pm$32~keV if it is assigned as the $3^+_1$, $I$=1 state.}
% which is lower than the experimental uncertainties of $c$ coefficient of $A$=58, $I$=1

\renewcommand{\arraystretch}{0.85}
\LTcapwidth=\textwidth
\begin{table}[tb]
\caption{Experimental isospin $I=1$ states in $^{58}$Zn, $^{58}$Cu, and $^{58}$Ni organized in isobaric multiplets and the corresponding experimental and theoretical IMME $c$ coefficients. Tentatively spin and parity assignments are proposed on the basis of the IMME theory for the states (bold texts) without firm experimental assignments.}
%Theoretical $c$ coefficients are given for comparison.
%(denoted by asterisks)
\label{tab:IMME}
\footnotesize
\begin{tabular*}{\linewidth}{@{\hspace{2mm}\extracolsep{\fill}}llcllc@{\hspace{2mm}}}
  % % \cmidrule[0.40pt](r{.75em}l{.25em}){3-8}
\hline
\hline
&&&&&\\
$J^{\pi}_i$  & \multicolumn{3}{c}{$E_\mathrm{x}$ [keV]$^{a,b}$} & \multicolumn{2}{c}{IMME $c$ [keV]} \\
             & $^{58}$Zn  & $^{58}$Cu & $^{58}$Ni &  Exp.$^b$  & Theo.$^c$ \\
&&&&&\\
\hline
$0^+_{1}$             & $0   $        & $ 203$          & $0   $  & $200$ $(25)$ & $235$ $(22)$ \\
$2^+_{1}$             & $1356$ $(3)$  & $1653$          & $1454$  & $156$ $(25)$ & $179$ $(22)$ \\
$4^+_{1}$             & $2499$ $(4)$  & $2750$          & $2459$  & $133$ $(25)$ & $154$ $(22)$ \\
$2^+_{2}$             & $2609$ $(6)$  & $2931$          & $2775$  & $166$ $(25)$ & $162$ $(22)$ \\
%$\mathbf{1^+_{1}}$  & 2861 (4)  & [2900 - 3100]  & 2902  & 337 (25)$^d$&  133 (22) \\
$\mathbf{1^+_{1}}$$^d$& $2861$ $(4)$  & $\mytilde 3100$~--~$\mytilde3200$ & $2902$  &  &  $133$ $(22)$ \\
%$\mathbf{2^+_{3}}$  & 2904 (5)  & [3230(20)]     & 3038  & 145 (32) &  192 (22) \\
%$\mathbf{2^+_{3}}$  & 2904 (5)  & [3230(20)]$^d$ & 3038  &  145 (32) &  192 (22) \\
$\mathbf{2^+_{3}}$$^d$& $2904$ $(5)$  &                & $3038$  &          &  $192$ $(22)$ \\
%$0^+_{2}$           & 2995      &                &       &          &           \\
%$4^+_{2}$           & 3263      &                &       &          &           \\
$2^+_{4}$             & $3265$ $(6)$  & $3513$           & $3264$  & $156$ $(25)$ &  $143$ $(22)$ \\
%$0^+_{3}$           & 3349      &                &       &          &           \\
%$3^+_{1}$           & 3378 (5)  & $\mytilde$3540 & 3421  & 234 (32) &  141 (22) \\
%$3^+_{1}$           & 3378 (5)  & $\mytilde$3540$^e$ & 3421  &          &  141 (22) \\
%$\mathbf{1^+_{2}}$  &           & 3820           & 3594  &          &  142 (22) \\
%$\mathbf{2^+_{5}}$  &           & 4210 (20)      & 3898  &          &  161 (22) \\
$3^+_{1}$             & $3378$ $(5)$  &                & $3421$  &          &  $141$ $(22)$ \\
$\mathbf{1^+_{2}}$    &               & $\mytilde 3900$~--~$\mytilde 4100$  & $3594$  &          &  $142$ $(22)$ \\
$\mathbf{2^+_{5}}$    &               &                & $3898$  &          &  $161$ $(22)$ \\
\hline
\end{tabular*}
\begin{minipage}{\columnwidth}
%\begin{minipage}{\textwidth}
\vskip5pt
{\sc Note}---\\
% \textbf{Note.}\\
$^a$ Only uncertainties of (or more than) 1~keV based on the evaluation of \citet{Nesaraja2010} are shown.\\
$^b$ Presently compiled from the evaluated nuclear masses (AME2016), and experimentally measured levels \citep{Jongsma1972,Honkanen1981,Rudolph1973,Rudolph1998,Rudolph2000,Rudolph2002,Johansson2009,Langer2014} according to the procedure implemented by \citet{Lam2013b}. \\
$^c$ Presently calculated with the cdGX1A Hamiltonian based on the full \emph{pf} shell-model space. The $1^+_1$, $2^+_3$, $3^+_1$, $1^+_2$, and $2^+_5$ triplets are not taken into comparison yielding the rms. \\
$^d$ An alternative order of the $1^+_1$ and $2^+_3$ states according to IMME dominance to the previous order proposed by \citet{Langer2014}.
%\vspace{10mm}
\end{minipage}
% \tablecomments{\\
% % \begin{singlespace}
% \footnotesize
% $^a$ Only uncertainties of (or more than) 1~keV based on the evaluation of \citet{Nesaraja2010} are shown.\\
% $^b$ Presently compiled from the evaluated nuclear masses (AME2016), and experimentally measured levels \citep{Jongsma1972,Honkanen1981,Rudolph1973,Rudolph1998,Rudolph2000,Rudolph2002,Johansson2009,Langer2014} according to the procedure implemented by \citet{Lam2013b}.\\
% $^c$ Presently calculated with the cdGX1A interaction based on the full \emph{pf} shell-model space. The $1^+_1$ and $2^+_3$ triplets are not taken into comparison yielding the rms.\\
% $^d$ The 2949-keV state of $^{58}$Cu and the respective $c$ coefficient are merely for comparison, see text.
% % \end{singlespace}
% }
\end{table}
\renewcommand{\arraystretch}{1.0}

According to the recent compilation of IMME $c$ coefficients of isobaric multiplets with $A=6$~--~$58$~\citep{Lam2013b}, the IMME $c$ coefficients exhibit a gradually decreasing trend as a function of $A$ with values ranging between about $400$ and $150$~keV. As is well known from the data, the $c$ coefficients of triplets show a prominent staggering effect, being split in two families: the values of $c$ coefficients inherent to isobars with $A=4n+2$ appear to be systematically higher than those for their $A=4n$ neighbors with $n$ being a positive integer. These average values decrease with increasing $A$ approximately as $A^{-1/3}$ as suggested by a uniformly charged liquid drop model. It has also been noticed that the amplitude of staggering decreases with increasing excitation energy manifesting the weakening of the pairing effects in higher excited states \citep{Lam2013a}. In the present study, we extend the compilation of \citet{Lam2013b} and tentatively propose excited isobaric multiplets in the $A=58$ triplet. Although the dependence of $c$ coefficients on excitation energy is less known, from theoretical studies in the $sd$-shell nuclei, the amplitude of staggering in isobaric triplets is expected to gradually diminish in the $pf$-shell nuclei. Recently, more precise nuclear mass measurements confirmed the persistence of these trend in the $pf$-shell nuclei \citep{Zhang2018,Surbrook2019,Fu2020}. We find that the values of $c$ coefficients provide a very stringent test for isobaric multiplets as we will see below.

%\textcolor{red}{Ormand recent $c$ coefficients.}
%which is the limit of available experimental data, we found that 

\subparagraph{The order of $1^+_1$ and $2^+_3$ states.}
\label{sec:1+1_2+3}
As was mentioned before, the order of $1^+_1$ and $2^+_3$ states stays undetermined in the work by \citet{Langer2014} with two plausible energies, $2861$~keV and $2904$~keV. The character of electromagnetic decay of those states weakly supports the assignment proposed in that work, that the lower state is a $2^+_3$ state and the higher one is the $1^+_1$, $I=1$ state. Indeed, we can get the ratio of the partial electromagnetic widths for the decay of these states to the $0^+_\mathrm{g.s.}$ and  $2^+_1$ to be more in reasonable agreement with that assignment as seen from Table~\ref{tab:Zn58em}.

%to those states that the proposed assignment does not completely exclude, the agreement between theoretical and experimental values of $c$ coefficients is worsen.
%Before we come to this analysis, let us comment on the selection of the $1^+$ states in $^{58}$Cu.
%-151,663
%-150,528
%-149,482
%-148,881
%-148,365
%-148,310
%-148,237
%-148,113
%-148,051
%-147,896
Alternatively, a certain constraint can also be imposed by the IMME. Although the $1^+_1$ and $2^+_3$, $I=1$ states of $^{58}$Cu are not assigned \citep{Nesaraja2010}, from the existing data we find that the best candidate for $2^+_3$ could be a state at $3230\pm20$~keV as measured by \citet{Rudolph1973} via the ($^3$He,t) reaction on the $^{58}$Ni target. Taking into account the $2^+_3$ ($3037.86\pm0.16$~keV) state of $^{58}$Ni \citep{Nesaraja2010}, we check for resulting values of the corresponding $c$ coefficients for the two states of question in $^{58}$Zn. Thus, we obtain $c=145\pm32$~keV assuming that the $2^+_3$ state in $^{58}$Zn is at $2904$~keV, or $c=124\pm32$~keV assuming that it is at $2861$~keV. The former value is closer to the theoretical $c$ coefficient of $192\pm22$~keV (Table~\ref{tab:IMME}). Based on this indication, we suggest here that the $2904$-keV state could be tentatively assigned as $2^+_3$.
%we check for plausible values of the corresponding $c$ coefficients assuming both states of question in $^{58}$Zn; we then obtain $c=145\pm32$~keV assuming $2904$~keV, or $c=124\pm32$~keV assuming $2861$~keV for the $2^+_3$ of $^{58}$Zn. With the theoretical uncertainties of $c$ coefficients, $22$~keV (Table~\ref{tab:IMME}), we propose the $2904$-keV state could be tentatively assigned as $2^+_3$.

For $1^+_1$, $I=1$ state in $^{58}$Cu, only an interval of energies can be proposed. Indeed, no low-lying  $1^+$, $I=1$ states have been observed by \citet{Fujita2002} and by \citet{Fujita2007}. To understand this fact, we have calculated a Gamow-Teller (GT) strength distribution from the $^{58}$Ni ground state to the $1^+$ states in $^{58}$Cu using the GXPF1A Hamiltonian. The results are summarized in Table~\ref{tab:BG}. 
First, we remark that there is a relatively good agreement with the data found in \citet{Fujita2002} and in \citet{Fujita2007}. For example, the $B$(GT) values of $1^+$, $I=0$ at low energies are comparable. In particular, we find also large intensities populated two lowest states, as well as our calculation reproduces a relatively large strength fragment at $3.4$~MeV which may be split between two states in experiment. Second, it can be noticed that the two lowest $1^+$, $I=1$ states carry a very small amount of the GT strength, similar to what \citet{Fujita2007} found also using the KB3G Hamiltonian~\citep{Poves2001}. It is therefore well probable that those states either were not observed by charge-exchange experiments, or correspond to low statistic counts at around $3.1$~--~$3.2$~MeV in Fig.~5 of \citet{Fujita2007}. With the tentative assignment of $2^+_3$ of $^{58}$Zn above, we propose an alternative assignment as compared to the work of \citet{Langer2014}, and hence the $2861$-keV state could be proposed as $1^+_1$.
% and the 2904-keV state could be tentatively assigned as $2^+_3$
% although nothing can be concluded unambiguously on the $1^+_1$ state in $^{58}$Zn from this analysis, 
% In either case, the impact on the final rate is small. ???  Yi Hua, is it so ?

%isobaric analogue states to form the ($A$=58, $2^+_3$, $I$=1) triplet are $^{58}$Zn (2904$\pm$5~keV), $^{58}$Cu (3230$\pm$20~keV), and $^{58}$Ni (3037.86$\pm$0.16~keV)\footnote{The $2^+_3$ state of $^{58}$Ni is cited from the levels evaluated by \citet{Nesaraja2010}.} as summarized in Table~\ref{tab:IMME}. The $^{58}$Cu (3230$\pm$20~keV) was measured by \citet{Rudolph1973} based on the ($^3$He,$t$) reaction on the $^{58}$Ni target. The selection of either the nearby 2931-keV or 3281-keV (possibly 0$^+$ to 4$^+$) state of $^{58}$Cu \citep{Lisetskiy2003} yields the IMME $c$ coefficients which are out of bounds of the systematic trend in $pf$-shell region. Besides, the uncertainty of theoretically predicted $c$ coefficient for the $2^+_3$ state overlaps with the uncertainty of presumed experimental $c$ coefficient, supporting the predicted assignment of the $2^+_3$ (2904$\pm$5~keV) state of $^{58}$Zn.

\renewcommand{\arraystretch}{0.85}
\begin{table}[tb]
\caption{Experimental decay intensities and theoretical partial widths for electromagnetic decay of the states at $2861$~keV and $2904$~keV in $^{58}$Zn.}
\label{tab:Zn58em}
\footnotesize
\begin{tabular*}{\linewidth}{@{\hspace{2mm}\extracolsep{\fill}}cccc@{\hspace{2mm}}}
  % % \cmidrule[0.40pt](r{.75em}l{.25em}){3-8}
\hline
\hline
% &&&\\
$J^{\pi}_i$& $E_\mathrm{x}$ [keV] & \multicolumn{2}{c}{Partial electromagnetic widths, $\Gamma_\gamma$ [meV]} \\
           &             & $J^{\pi}_i \to 0^+_\mathrm{g.s.}$ & $J^{\pi}_i \to 2^+_1$   \\
% &&&\\
\hline
$2^+_{3}$ & $2861$ & $2.18$      & $3.12$ \\
$1^+_{1}$ & $2904$ & $0.34$      & $7.5 $ \\
\hline
$1^+_{1}$ & $2861$ & $0.33$      & $7.0 $ \\
$2^+_{3}$ & $2904$ & $2.35$      & $3.38$ \\
\hline
&&&\\
&  $E_\mathrm{x}$ [keV] & \multicolumn{2}{c}{Electromagnetic decay intensities, $I_{\gamma}$ [\%]} \\
&              & $J^{\pi}_i \to 0^+_\mathrm{g.s.}$ & $J^{\pi}_i \to 2^+_1$ \\
% &&&\\
\hline
Exp. & $2861$   & $7$ $(2)$    & $8$  $(2)$  \\
Exp. & $2904$   & $3$ $(1)$    & $13$ $(2)$  \\
\hline
\end{tabular*}
\end{table}
\renewcommand{\arraystretch}{1.0}

\renewcommand{\arraystretch}{0.85}
\begin{table}[tb]
\caption{Theoretical Gamow-Teller strength, $B$(GT), populating the $1^+$ states in $^{58}$Cu deduced from the cdGX1A Hamiltonian.}
\label{tab:BG}
\footnotesize
\begin{tabular*}{\linewidth}{@{\hspace{2mm}\extracolsep{\fill}}crc@{\hspace{2mm}}}
  % % \cmidrule[0.40pt](r{.75em}l{.25em}){3-8}
\hline
\hline
% &&\\
%$J^{\pi}_i$ & $E_\mathrm{x}$ [keV] &\multicolumn{2}{c}{Gamow-Teller strengths, $B$(GT)} & Isospin, $I$\\
%  & &  \emph{Exp.} & \emph{Theo.} & \\
%$J^{\pi}_i$ & Isospin, $I$ & $E_\mathrm{x}$ [keV] & Gamow-Teller strengths, $B$(GT) \\
Isospin, $I$ & $E_\mathrm{x}$ [keV] & Gamow-Teller strengths, $B$(GT)$^a$ \\
%&&&\\
\hline
%&&&\\
%$1^+_1$   & 0     & 0.155 (1)  &  0.373 &  0 \\
%($1^+_2$) & 1051  & 0.265 (13) &  0.321 &  0 \\
%          & 2949  & 0.025 (3)  &  0.022 & (1) \\
%          & 3460  & 0.173 (11) &  0.022 &  0 \\
%          &       &            &  0.040 &  0 \\
%          & 3678  & 0.155 (10) &  0.365 &  0 \\
%          & 3717  & 0.050 (5)  &  0.003 &  0 \\
%          &       &            &  0.    &  1 \\
%          & 4720  & 0.042 (4)  &  0.002 &  1 \\
%          & 5065  & 0.040 (4)  &  0.002 &  1 \\
%          &       &            &  0.002 &  1 \\
$0$  & $0.000$  & $0.221$ \\ % 0.2212 \\
$0$  & $1.135$  & $0.190$ \\ % 0.1904 \\
$0$  & $2.181$  & $0.013$ \\ % 0.0130 \\
$0$  & $2.782$  & $0.024$ \\ % 0.0235 \\
$1$  & $3.298$  & $0.001$ \\ % 0.0013 \\
$0$  & $3.353$  & $0.002$ \\ % 0.0020 \\
$0$  & $3.426$  & $0.217$ \\ % 0.2166 \\
$0$  & $3.550$  & $0.000$ \\ % 0.0000 \\
$1$  & $3.612$  & $0.020$ \\ % 0.0195 \\
$0$  & $3.767$  & $0.015$ \\ % 0.0145 \\
$0$  & $3.860$  & $0.076$ \\ % 0.0755 \\
$0$  & $4.321$  & $0.005$ \\ % 0.0046 \\
$0$  & $4.565$  & $0.051$ \\ % 0.0514 \\
$0$  & $4.871$  & $0.151$ \\ % 0.1505 \\
$0$  & $5.035$  & $0.081$ \\ % 0.0813 \\
$0$  & $5.130$  & $0.010$ \\ % 0.0102 \\
$0$  & $5.260$  & $0.023$ \\ % 0.0232 \\
$0$  & $5.358$  & $0.028$ \\ % 0.0281 \\
$0$  & $5.491$  & $0.003$ \\ % 0.0027 \\
$0$  & $5.528$  & $0.000$ \\ % 0.0000 \\
\hline
\end{tabular*}
\begin{minipage}{\columnwidth}
%\begin{minipage}{\textwidth}
\vskip5pt
{\sc Note}---\\
% \textbf{Note.}\\
$^a$ The theoretical $B$(GT) is quenched with the standard quenching factor of $0.77$ \citep{Horoi2007}.
%\vspace{10mm}
\end{minipage}
\end{table}
\renewcommand{\arraystretch}{1.0}

\subparagraph{The energies of $1^+_2$ and $2^+_5$ states.}
\label{sec:1+2_2+5}
%Similarly, the energy of the  $1^+_2$, $I=1$ state is not known in $^{58}$Cu.
The predicted $B$(GT) intensity to this state by theory could, in principle, have been seen in the data of the charge-exchange experiment performed by \citet{Fujita2007}. Three possible candidates have been reported between $3.6$ and $4$ MeV as can be seen from Figs. 5 and 7 of that article. Taking any of them and using the theoretically predicted IMME $c$ coefficient of ($A=58$, $1^+_2$, $I=1$) triplet, $142\pm22$~keV, we estimate that the energy of the $^{58}$Zn, $1^+_2$ state cannot be below about $3664\pm22$~keV. The uncertainty is based on the comparison presented in Table~\ref{tab:IMME}. This IMME estimated $1^+_2$ state is $309\pm22$~keV higher than the one estimated by using GXPF1A Hamiltonian that was used by \citet{Langer2014} to obtain the contribution from the $1^+_2$ resonance state for the $^{57}$Cu(p,$\gamma$)$^{58}$Zn reaction rate. 

There is no best candidate $2^+_5$ isobaric analogue state in $^{58}$Cu to estimate the $2^+_5$ state of $^{58}$Zn. The GXPF1a Hamiltonian predicts the $2^+_5$ state to be at $3605$~keV excitation energy and we adopt this value as a lower limit for $^{58}$Zn, being aware that in the mirror nucleus,  $^{58}$Ni, its analogue is found at $3.898$~MeV. Applying the theoretical IMME $c$ coefficient of ($A=58$, $2^+_5$, $I=1$) triplet, $161\pm22$~keV, we can expect that the $2^+_5$, $I=1$ state in $^{58}$Cu to be in the energy interval of $3.9$~--~$4.1$~MeV.
Future high precision experiment measuring the level schemes of $^{58}$Cu and $^{58}$Zn in this energy region may provide more information of the $1^+_1$, $1^+_2$, and $2^+_5$ isobaric analogue states, and the $1^+_2$ and $2^+_5$ states of $^{58}$Zn.
% 3614~keV (cdGX1A)

\renewcommand{\arraystretch}{0.85}
\LTcapwidth=\textwidth
% \setlength{LTcapwidth}{5.2in}
% \begin{center}
\begin{table*}[tb]
% \scriptsize
\footnotesize
\caption{Properties of $^{58}$Zn for the ground-state proton capture in the present $^{57}$C\lowercase{u}(p,$\gamma$)$^{58}$Z\lowercase{n} resonant rate calculation.}
\label{tab:58Zn}
% \begin{center}
  % % \begin{ruledtabular}
\begin{tabular*}{\linewidth}{@{\hspace{2mm}\extracolsep{\fill}}cllcccclll@{\hspace{2mm}}}
  % % \cmidrule[0.40pt](r{.75em}l{.25em}){3-8}
\hline
\hline
% &&&&&&&&&\\ %\tnote{e}
$J^{\pi}_i$  & $E_\mathrm{x}$ [MeV]$^a$ & $E_\mathrm{res}$ [MeV]$^d$ & $C^2S_{7/2}$ & $C^2S_{3/2}$ & $C^2S_{5/2}$ &  $C^2S_{1/2}$ & $\Gamma_\gamma$ [eV]  & $\Gamma_\mathrm{p}$ [eV]  & $\omega\gamma$ [eV] \\
             &  &  & $(l=3)$      & $(l=1)$      & $(l=3)$      &  $(l=1)$      &  &  &  \\
% &&&&&&&&&\\
\hline
$0^+_{1}$  &$0.000$     &             &          & $1.1001$ &          &          &         ---          &                       &                       \\
$2^+_{1}$  &$1.356$$^b$ &             & $0.0351$ & $0.8381$ & $0.1459$ & $0.0913$ & $6.736\times10^{-4}$ &                       &                       \\
$4^+_{1}$  &$2.499$$^b$ & $0.219$     & $0.0123$ &          & $0.6737$ &          & $1.741\times10^{-4}$ & $1.002\times10^{-17}$ & $1.127\times10^{-17}$ \\
$2^+_{2}$  &$2.609$$^b$ & $0.329$$^e$ & $0.0027$ & $0.5776$ & $0.0063$ & $0.1144$ & $9.034\times10^{-3}$ & $1.695\times10^{-10}$ & $1.059\times10^{-10}$ \\
$1^+_{1}$  &$2.861$$^b$ & $0.581$$^e$ &          & $0.0000$ & $0.6522$ & $0.0867$ & $7.300\times10^{-3}$ & $4.662\times10^{-6 }$ & $1.747\times10^{-6 }$ \\
$2^+_{3}$  &$2.904$$^b$ & $0.624$$^e$ & $0.0020$ & $0.0131$ & $0.0103$ & $0.1649$ & $5.278\times10^{-3}$ & $3.380\times10^{-5 }$ & $2.099\times10^{-5 }$ \\
$0^+_{2}$  &$2.995$     & $0.715$     &          & $0.4393$ &          &          & $4.758\times10^{-5}$ & $1.097\times10^{-3 }$ & $5.700\times10^{-6 }$ \\
$4^+_{2}$  &$3.263$     & $0.983$     & $0.0016$ &          & $0.0048$ &          & $3.589\times10^{-4}$ & $1.074\times10^{-5 }$ & $1.173\times10^{-5 }$ \\
$2^+_{4}$  &$3.265$$^b$ & $0.985$$^e$ & $0.0005$ & $0.1174$ & $0.4953$ & $0.0002$ & $4.131\times10^{-3}$ & $4.338\times10^{-2 }$ & $2.357\times10^{-3 }$ \\
$0^+_{3}$  &$3.349$     & $1.069$     &          & $0.0419$ &          &          & $8.758\times10^{-4}$ & $4.827\times10^{-2 }$ & $1.075\times10^{-4 }$ \\
$3^+_{1}$  &$3.378$$^b$ & $1.098$     & $0.0016$ & $0.0000$ & $0.6849$ &          & $3.466\times10^{-3}$ & $4.649\times10^{-3 }$ & $1.737\times10^{-3 }$ \\
$2^+_{5}$  &$3.605$     & $1.325$     & $0.0012$ & $0.0011$ & $0.0302$ & $0.2498$ & $1.387\times10^{-2}$ & $3.677              $ & $8.637\times10^{-3 }$ \\
$1^+_{2}$  &$3.664$$^c$ & $1.384$$^e$ &          & $0.0000$ & $0.1011$ & $0.5980$ & $4.526\times10^{-2}$ & $1.527\times10^{+1 }$ & $1.692\times10^{-2 }$ \\
$3^+_{ 2}$ &$3.670$     & $1.390$     & $0.0042$ & $0.0000$ & $0.0033$ &          & $3.680\times10^{-4}$ & $1.745\times10^{-3 }$ & $2.659\times10^{- 4}$ \\
$4^+_{ 3}$ &$3.969$     & $1.689$     & $0.0169$ &          & $0.0077$ &          & $1.275\times10^{-2}$ & $6.250\times10^{-2 }$ & $1.191\times10^{- 2}$ \\
$5^+_{ 1}$ &$4.009$     & $1.729$     & $0.0012$ &          &          &          & $5.162\times10^{-4}$ & $4.589\times10^{-3 }$ & $6.380\times10^{- 4}$ \\
$2^+_{ 6}$ &$4.077$     & $1.797$     & $0.0000$ & $0.0019$ & $0.0030$ & $0.0671$ & $2.016\times10^{-3}$ & $3.024\times10^{+1 }$ & $1.260\times10^{- 3}$ \\
$3^+_{ 3}$ &$4.168$     & $1.888$     & $0.0192$ & $0.0020$ & $0.0312$ &          & $1.202\times10^{-2}$ & $2.044              $ & $1.046\times10^{- 2}$ \\
$4^+_{ 4}$ &$4.188$     & $1.908$     & $0.0079$ &          & $0.0087$ &          & $1.430\times10^{-3}$ & $1.468\times10^{-1 }$ & $1.593\times10^{- 3}$ \\
$0^+_{ 4}$ &$4.242$     & $1.962$     &          & $0.0000$ &          &          & $8.661\times10^{-3}$ & $0.000              $ & $0.000              $ \\
$4^+_{ 5}$ &$4.268$     & $1.988$     & $0.0034$ &          & $0.0584$ &          & $1.323\times10^{-2}$ & $6.240\times10^{-1 }$ & $1.458\times10^{- 2}$ \\
$2^+_{ 7}$ &$4.270$     & $1.990$     & $0.0035$ & $0.0007$ & $0.0357$ & $0.0001$ & $3.740\times10^{-3}$ & $1.477              $ & $2.332\times10^{- 3}$ \\
$0^+_{ 5}$ &$4.363$     & $2.083$     &          & $0.0198$ &          &          & $1.606\times10^{-2}$ & $4.158\times10^{+1 }$ & $2.007\times10^{- 3}$ \\
$4^+_{ 6}$ &$4.520$     & $2.240$     & $0.0051$ &          & $0.0000$ &          & $6.485\times10^{-3}$ & $3.290\times10^{-1 }$ & $7.154\times10^{- 3}$ \\
$3^+_{ 4}$ &$4.546$     & $2.266$     & $0.0003$ & $0.0000$ & $0.0016$ &          & $3.161\times10^{-2}$ & $7.852\times10^{-2 }$ & $1.972\times10^{- 2}$ \\
$5^+_{ 2}$ &$4.594$     & $2.314$     & $0.0065$ &          &          &          & $7.146\times10^{-3}$ & $5.545\times10^{-1 }$ & $9.701\times10^{- 3}$ \\
$3^+_{ 5}$ &$4.653$     & $2.373$     & $0.0003$ & $0.0000$ & $0.0000$ &          & $1.428\times10^{-2}$ & $3.279\times10^{-2 }$ & $8.705\times10^{- 3}$ \\
$2^+_{ 8}$ &$4.708$     & $2.428$     & $0.0092$ & $0.0000$ & $0.0323$ & $0.0059$ & $2.609\times10^{-2}$ & $4.483\times10^{+1 }$ & $1.630\times10^{- 2}$ \\
$4^+_{ 7}$ &$4.832$     & $2.552$     & $0.0050$ &          & $0.0050$ &          & $1.378\times10^{-2}$ & $1.715              $ & $1.538\times10^{- 2}$ \\
$5^+_{ 3}$ &$4.909$     & $2.629$     & $0.0001$ &          &          &          & $1.751\times10^{-3}$ & $2.902\times10^{-2 }$ & $2.271\times10^{- 3}$ \\
$4^+_{ 8}$ &$4.964$     & $2.684$     & $0.0000$ &          & $0.0090$ &          & $2.452\times10^{-3}$ & $1.692              $ & $2.755\times10^{- 3}$ \\
$2^+_{ 9}$ &$5.013$     & $2.733$     & $0.0002$ & $0.0027$ & $0.0002$ & $0.0022$ & $3.656\times10^{-3}$ & $9.118\times10^{+1 }$ & $2.285\times10^{- 3}$ \\
$3^+_{ 6}$ &$5.040$     & $2.760$     & $0.0060$ & $0.0000$ & $0.0006$ &          & $1.759\times10^{-2}$ & $2.944              $ & $1.530\times10^{- 2}$ \\
$4^+_{ 9}$ &$5.184$     & $2.904$     & $0.0073$ &          & $0.0037$ &          & $2.096\times10^{-2}$ & $6.858              $ & $2.351\times10^{- 2}$ \\
$5^+_{ 4}$ &$5.208$     & $2.928$     & $0.0001$ &          &          &          & $7.299\times10^{-3}$ & $7.758\times10^{-2 }$ & $9.173\times10^{- 3}$ \\
$2^+_{10}$ &$5.227$     & $2.947$     & $0.0001$ & $0.0327$ & $0.0016$ & $0.0023$ & $2.958\times10^{-2}$ & $1.230\times10^{+3 }$ & $1.849\times10^{- 2}$ \\
$3^+_{ 7}$ &$5.250$     & $2.970$     & $0.0001$ & $0.0008$ & $0.0002$ &          & $1.045\times10^{-2}$ & $2.904\times10^{+1 }$ & $9.142\times10^{- 3}$ \\
% &&&&&&&&&\\
\hline
\end{tabular*}
\begin{minipage}{\linewidth}
\footnotesize
\vskip5pt
{\sc Note}---\\
%\textbf{Note.}\\
$^a$ The energy levels of $^{58}$Zn obtained from the present full \emph{pf}-model space shell-model calculation with cdGX1A Hamiltonian, except otherwise quoted from experiment or predicted from IMME. \\
$^b$ The experimentally determined energy levels of $^{58}$Zn~\citep{Langer2014}. \\
$^c$ The theoretical energy levels of $^{58}$Zn predicted from IMME, see text. \\
$^d$ Calculated by $E_\mathrm{res}=E_\mathrm{x}-S_\mathrm{p}-E_i$ with $S_\mathrm{p}$($^{58}$Zn)~$=2.280\pm0.050$~MeV deduced from AME2016~\citep{AME2016}.\\
$^e$ Resonances dominantly contributing to the total rate within temperature region of $0.1$~--~$2$~GK.
\end{minipage}
\vspace{3mm}
% \end{center}
\end{table*}
\renewcommand{\arraystretch}{1.0}

\subparagraph{Properties of resonances.}
\label{sec:resonances}
With the information on nuclear structure described above, we deduce a set of resonance properties of $^{58}$Zn to construct the new $^{57}$Cu(p,$\gamma$)$^{58}$Zn resonant reaction rate within the typical XRB temperature range, e.g., the GS~1826$-$24 burster. We only consider the proton-capture on the $3/2^-_\mathrm{g.s.}$ ground state (g.s.) of $^{57}$Cu as the contribution from proton resonant captures on thermally excited states of $^{57}$Cu are negligible due to rather high lying excited states. Hence, it is adequate to just present the newly deduced resonance properties of $^{57}$Cu(p,$\gamma$)$^{58}$Zn reaction rate up to the $3^+_7$ state ($5.250$~MeV) in Table~\ref{tab:58Zn} within the Gamow window corresponding to the XRB temperature range.

By comparing the $\Gamma_\gamma^{2^+_3}$ produced from the full $pf$-model space used in the present work with the $\Gamma_\gamma^{2^+_3}$ generated from the four-particle-four-hole truncated scheme used in \citet{Langer2014} calculation, we notice that the present $\Gamma_\gamma^{2^+_3}$ of $^{58}$Zn (Table~\ref{tab:58Zn}) is one order of magnitude lower than the one calculated by \citet{Langer2014}. Nevertheless, the respective $\Gamma_\mathrm{p}$ is two orders of magnitude lower than $\Gamma_\gamma^{2^+_3}$, and thus, such difference in the $\Gamma_\gamma^{2^+_3}$ state does not impact the respective $\omega\gamma$.

We note that the inverse assignment of the $1^+_1$ and $2^+_3$ states compared to \citet{Langer2014} assignment, in fact, changes the contributions of the $1^+_1$ and $2^+_3$ resonance states. This is mainly because the main contributions for the $2^+_3$ and $1^+_1$ states are the $p_{1/2}$ and $f_{5/2}$ particle captures, respectively. For higher values of the orbital angular momentum $\mathtt{l}$ of the captured proton, the corresponding width becomes more sensitive to the proton energy because barrier penetrability varies faster. Once the $1^+$ state, governed by the $f$-capture, is assigned at a lower excitation energy, its contribution to the resonant rate becomes drastically reduced.

%\subsection{\label{sec:DC}Direct-capture rates}
\paragraph{Direct-capture rate}
\label{sec:DC}
Comparing the direct-capture rate deduced by \citet{Fisker2001} (or by \citealt{Forstner2001}) with the presently deduced resonant capture rate, we notice that the contribution of direct capture is exponentially lower than the contribution of the dominating resonances throughout XRB related temperature range from $0.3$ to $2$~GK. Hence, the contribution of the direct-capture rate is negligible for the $^{57}$Cu(p,$\gamma$)$^{58}$Zn reaction rate, see Fig.~\ref{fig:rp_57Cu_58Zn_contri} which only presents Fisker et al. direct-capture rate.

\newpage
\section{New $^{57}$C\lowercase{u}(\lowercase{p},$\gamma$)$^{58}$Z\lowercase{n} reaction rate}
\label{sec:rates}

\renewcommand{\arraystretch}{0.8}
% in units of \MakeLowercase{cm$^{3}$s$^{-1}$mol$^{-1}$}
\begin{table}
\footnotesize
\caption{\label{tab:rates} Thermonuclear reaction rates of $^{57}$Cu(p,$\gamma$)$^{58}$Zn.}
%\begin{ruledtabular}
% \begin{tabular}{|c|ccc|ccc|}
% \begin{tabular*}{\linewidth}{@{\hspace{2mm}\extracolsep{\fill}}|c|ccc|ccc|@{\hspace{2mm}}}
\begin{tabular*}{\linewidth}{@{\hspace{2mm}\extracolsep{\fill}}clll@{\hspace{2mm}}}
\hline
\hline
% \toprule[1.0pt]
% \midrule[0.25pt]
% &&&\\ % $N_{A}\langle\sigma v\rangle$
% &\multicolumn{3}{c}{$^{61}$Ga(p,$\gamma$)$^{62}$Ge} \\
% &&&\\
$T_{9}$ & centroid & lower limit & upper limit \\
&[cm$^{3}$s$^{-1}$mol$^{-1}$]&[cm$^{3}$s$^{-1}$mol$^{-1}$]&[cm$^{3}$s$^{-1}$mol$^{-1}$]\\
% &&&\\
\hline
% \midrule[0.25pt]
% &&&\\
%%%
%%% 1+_2 is determined from IMME formalism 
%%% The uncertainty is merely from 50 keV (AME2016 Sp) and omitted the 168 keV (GXPF1a).
%%% We only take into account of main contributing resonances for computing the uncertainty.
%%%
$0.1$  &   $1.44\times10^{-20}$ &   $1.28\times10^{-21}$ & $6.62\times10^{-20}$ \\     
$0.2$  &   $9.72\times10^{-13}$ &   $2.39\times10^{-13}$ & $1.62\times10^{-12}$ \\     
$0.3$  &   $1.23\times10^{-9 }$ &   $6.53\times10^{-10}$ & $2.78\times10^{-9 }$ \\     
$0.4$  &   $2.34\times10^{-7 }$ &   $2.19\times10^{-7 }$ & $2.93\times10^{-7 }$ \\     
$0.5$  &   $6.08\times10^{-6 }$ &   $4.23\times10^{-6 }$ & $9.30\times10^{-6 }$ \\     
$0.6$  &   $5.45\times10^{-5 }$ &   $3.08\times10^{-5 }$ & $9.90\times10^{-5 }$ \\     
$0.7$  &   $2.80\times10^{-4 }$ &   $1.40\times10^{-4 }$ & $5.60\times10^{-4 }$ \\     
$0.8$  &   $1.05\times10^{-3 }$ &   $5.08\times10^{-4 }$ & $2.16\times10^{-3 }$ \\     
$0.9$  &   $3.22\times10^{-3 }$ &   $1.61\times10^{-3 }$ & $6.43\times10^{-3 }$ \\     
$1.0$  &   $8.42\times10^{-3 }$ &   $4.50\times10^{-3 }$ & $1.60\times10^{-2 }$ \\     
$1.1$  &   $1.94\times10^{-2 }$ &   $1.11\times10^{-2 }$ & $3.46\times10^{-2 }$ \\     
$1.2$  &   $3.99\times10^{-2 }$ &   $2.43\times10^{-2 }$ & $6.72\times10^{-2 }$ \\     
$1.3$  &   $7.51\times10^{-2 }$ &   $4.81\times10^{-2 }$ & $1.20\times10^{-1 }$ \\     
$1.4$  &   $1.31\times10^{-1 }$ &   $8.73\times10^{-2 }$ & $2.00\times10^{-1 }$ \\     
$1.5$  &   $2.13\times10^{-1 }$ &   $1.48\times10^{-1 }$ & $3.14\times10^{-1 }$ \\     
$1.6$  &   $3.30\times10^{-1 }$ &   $2.35\times10^{-1 }$ & $4.70\times10^{-1 }$ \\     
$1.7$  &   $4.86\times10^{-1 }$ &   $3.56\times10^{-1 }$ & $6.76\times10^{-1 }$ \\     
$1.8$  &   $6.89\times10^{-1 }$ &   $5.15\times10^{-1 }$ & $9.36\times10^{-1 }$ \\     
$1.9$  &   $9.44\times10^{-1 }$ &   $7.19\times10^{-1 }$ & $1.26              $ \\     
$2.0$  &   $1.26              $ &   $9.71\times10^{-1 }$ & $1.65              $ \\  
% &&&\\ 
\hline
% \bottomrule[1.0pt]
\end{tabular*}
% \end{tabular}
%\end{ruledtabular}
\end{table}
\renewcommand{\arraystretch}{1.0}

Table~\ref{tab:rates} shows the presently calculated total reaction rate of $^{57}$Cu(p,$\gamma$)$^{58}$Zn as a function of temperature. The present (\emph{Present}, hereafter) thermonuclear rate is parameterized in the format proposed by~\citet{Rauscher2000} with the expression below,
\begin{eqnarray}
\label{eq:parameterization}
N_\mathrm{A}\langle\sigma v\rangle &= \sum_i \mathrm{exp}(a^i_0 + \frac{a^i_1}{T_9} + \frac{a^i_2}{T_9^{1/3}} + a^i_3 T_9^{1/3} \nonumber \\
&+ a^i_4 T_9 + a^i_5 T_9^{5/3} + a^i_6 \ln{T_9}) \, .
\end{eqnarray}
These parameters, i.e., $a_0, a_1, a_2, a_3, a_4, a_5$, and $a_6$ are listed in Table~\ref{tab:parameter}. The running index $i$ is up to $6$ for the \emph{Present} rate for the temperature region, $0.1$~--~$2$~GK. The parameterized \emph{Present} rate is evaluated according to an accuracy quantity proposed by \citet{Rauscher2000}, 
\begin{eqnarray}
\label{eq:accuracy}
\zeta = \frac{1}{n}\sum_{m=1}^n ( \frac{r_{m} - f_{m}}{f_{m}} )^2 \, , \nonumber
\end{eqnarray}
where $n$ is the number of data points, $r_m$ are the original \emph{Present} rate calculated for each respective temperature, and $f_m$ are the fitted rate at that temperature. With $n=297$, $\zeta$ is $4.45\times10^{-3}$, and the fitting error is $5.90$~\% for the temperature range from $0.01$~GK to $3$~GK. The parameterized rate is obtained with aid from the Computational Infrastructure for Nuclear Astrophysics (CINA;~\citealt{CINA}). For the rate above $3$~GK, one may refer to statistical model calculations to match with the \emph{Present} rate, which is only valid within the mentioned temperature range and fitting errors, see NACRE~\citep{Angulo1999}.

\renewcommand{\arraystretch}{0.8}
\begin{table*}
\footnotesize
\caption{\label{tab:parameter} Parameters of $^{57}$Cu(p,$\gamma$)$^{58}$Zn centroid reaction rate.}
%\begin{tabular*}{\linewidth}{@{\hspace{2mm}\extracolsep{\fill}}lllllll@{\hspace{2mm}}}
\begin{tabular*}{\linewidth}{@{\hspace{2mm}\extracolsep{\fill}}crrrrrrr@{\hspace{2mm}}}
%\begin{tabular*}{\linewidth}{llllllld{5}}
%\begin{tabular*}{\linewidth}{l@{\extracolsep{\fill}}*{5}{d{-7}}}
%\begin{tabular*}{\linewidth}{@{\hspace{2mm}\extracolsep{\fill}}crrrrrr@{\hspace{2mm}}}
\hline
\hline
% \toprule[1.0pt]
% \midrule[0.25pt]
%&&&&&\\
%$i$ & 1 & 2 & 3 & 4 & 5 & 6\\
%&&&&&\\
%\hline
%&&&&&\\
%$a_0$ & $ -27.0569 $ & $ -10.9645 $  &  $ 1.4516 $   &  $  6.1710 $   & $  7.9311 $   &  $  7.1151$ \\
%$a_1$ & $  -2.5415 $ & $  -3.8183 $  &  $-7.2530 $   &  $-11.5201 $   & $-16.2517 $   &  $-15.4875$ \\
%$a_2$ & $  -0.0005 $ & $   0.0180 $  &  $-0.0115 $   &  $ -0.0108 $   & $  0.1040 $   &  $  0.0555$ \\
%$a_3$ & $   0.0045 $ & $  -0.0401 $  &  $ 0.0455 $   &  $  0.1940 $   & $ -0.0619 $   &  $ -0.0108$ \\
%$a_4$ & $  -0.0005 $ & $   0.0050 $  &  $-0.0038 $   &  $ -0.0010 $   & $  0.0431 $   &  $  0.0108$ \\
%$a_5$ & $   0.0000 $ & $  -0.0004 $  &  $ 0.0005 $   &  $  0.0329 $   & $ -0.0005 $   &  $ -0.0008$ \\
%$a_6$ & $  -1.5011 $ & $  -1.4839 $  &  $-1.4508 $   &  $ -1.5474 $   & $ -1.4564 $   &  $ -1.4128$ \\
%
%&&&&&&\\
$i$ & $a_0$                   & $a_1$                 & $a_2$                     & $a_3$                   & $a_4$                   & $a_5$                   & $a_6$      \\
%&&&&&&\\
\hline
%&&&&&&\\
$1$ & $-2.70569\times10^{+1}$ & $-2.54150\times10^{+0}$ & $-4.59592\times10^{-4}$ & $ 4.46992\times10^{-3}$ & $-5.27022\times10^{-4}$ & $ 2.68353\times10^{-5}$ & $-1.50113$ \\
$2$ & $-1.09645\times10^{+1}$ & $-3.81827\times10^{+0}$ & $ 1.79997\times10^{-2}$ & $-4.00723\times10^{-2}$ & $ 5.02873\times10^{-3}$ & $-4.19260\times10^{-4}$ & $-1.48386$ \\
$3$ & $ 1.45160\times10^{+0}$ & $-7.25300\times10^{+0}$ & $-1.15250\times10^{-2}$ & $ 4.54761\times10^{-2}$ & $-3.78671\times10^{-3}$ & $ 4.62772\times10^{-4}$ & $-1.45082$ \\
$4$ & $ 6.17102\times10^{+0}$ & $-1.15201\times10^{+1}$ & $-1.08160\times10^{-2}$ & $ 1.93951\times10^{-1}$ & $-1.02466\times10^{-3}$ & $ 3.29115\times10^{-2}$ & $-1.54743$ \\
$5$ & $ 7.93110\times10^{+0}$ & $-1.62517\times10^{+1}$ & $ 1.03977\times10^{-1}$ & $-6.19113\times10^{-2}$ & $ 4.30871\times10^{-2}$ & $-5.16600\times10^{-4}$ & $-1.45643$ \\
$6$ & $ 7.11511\times10^{+0}$ & $-1.54875\times10^{+1}$ & $ 5.54672\times10^{-2}$ & $-1.08431\times10^{-2}$ & $ 1.08004\times10^{-2}$ & $-7.96518\times10^{-4}$ & $-1.41275$ \\
\hline
% \bottomrule[1.0pt]
\end{tabular*}
% \end{tabular}
%\end{ruledtabular}
\end{table*}
\renewcommand{\arraystretch}{1.0}

\begin{figure}[t]
% \begin{center}
%\hspace{14mm}
%\includegraphics[width=\columnwidth, angle=0]{rp_57Cu_58Zn_grace_contri.eps}
% \includegraphics[width=9.0cm, angle=0]{rp_57Cu_58Zn_grace_contri.eps}
\includegraphics[width=8.7cm, angle=0]{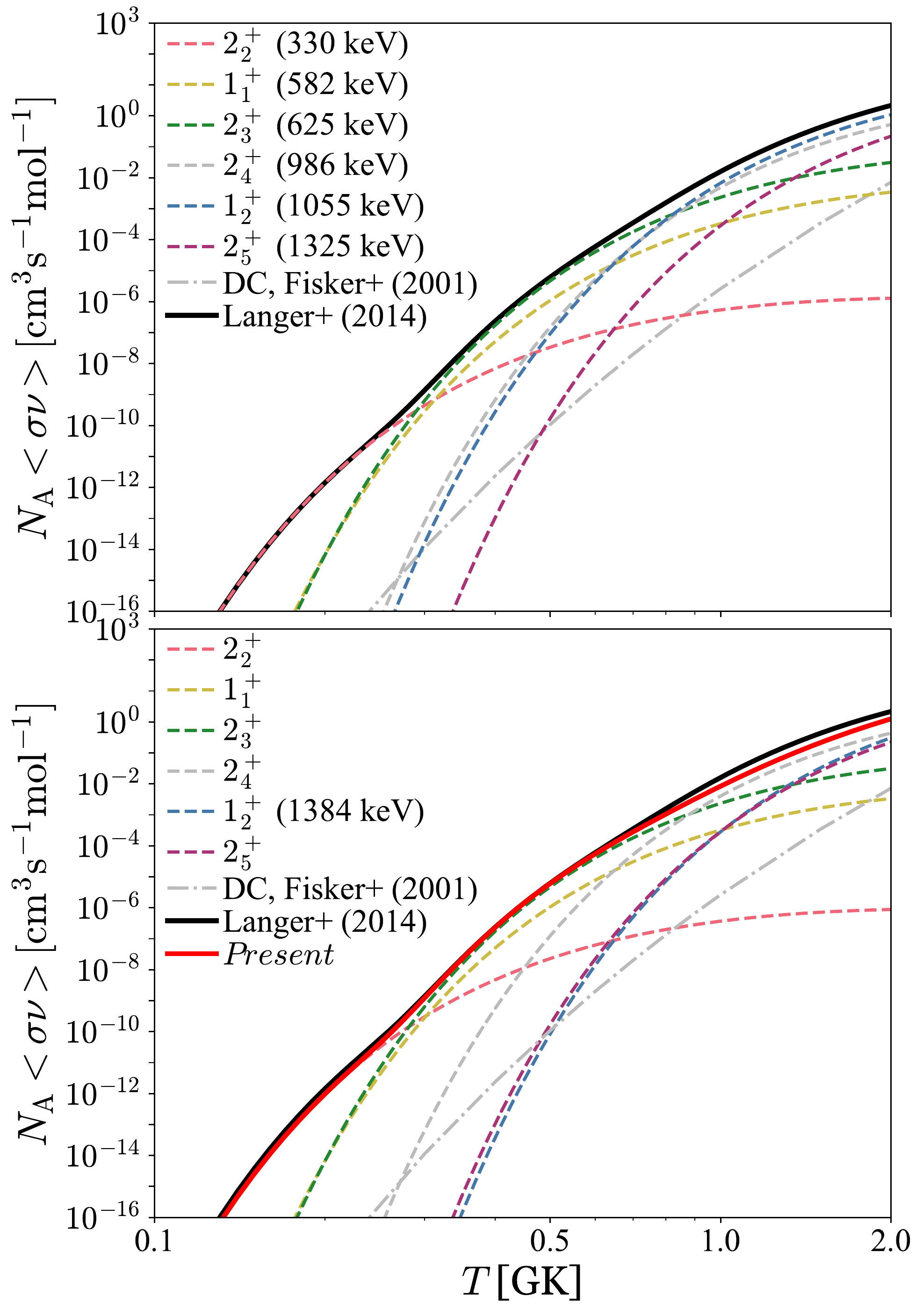}
% \plotone{57Cu_58Zn_contri.pdf}
% \vspace{-12mm}
\singlespace
\caption{\label{fig:rp_57Cu_58Zn_contri}{\footnotesize The $^{57}$Cu(p,$\gamma$)$^{58}$Zn thermonuclear reaction rates. Top Panel: The main contributing resonances of proton captures on the $3/2^-_\mathrm{g.s.}$ state of $^{57}$Cu in the temperature region of XRB interest are indicated as dashed color lines with the respective resonance energies. Bottom Panel: The updated main contributing resonances with full \emph{pf} shell-model space calculation for $\Gamma_\gamma$ widths and spectroscopic factors, and with the resonance energy of the $1^+_2$ state using IMME formalism. See details in the text and Table~\ref{tab:58Zn}.}}
% (A color version of this figure is available in the online journal.)
% \end{center}
\end{figure}

\begin{figure}[t]
% \begin{center}
% \hspace{7mm}
%\includegraphics[width=\columnwidth, angle=0]{rp_57Cu_58Zn_grace.eps}
% \plotone{57Cu_58Zn_python.pdf}
\includegraphics[width=8.7cm, angle=0]{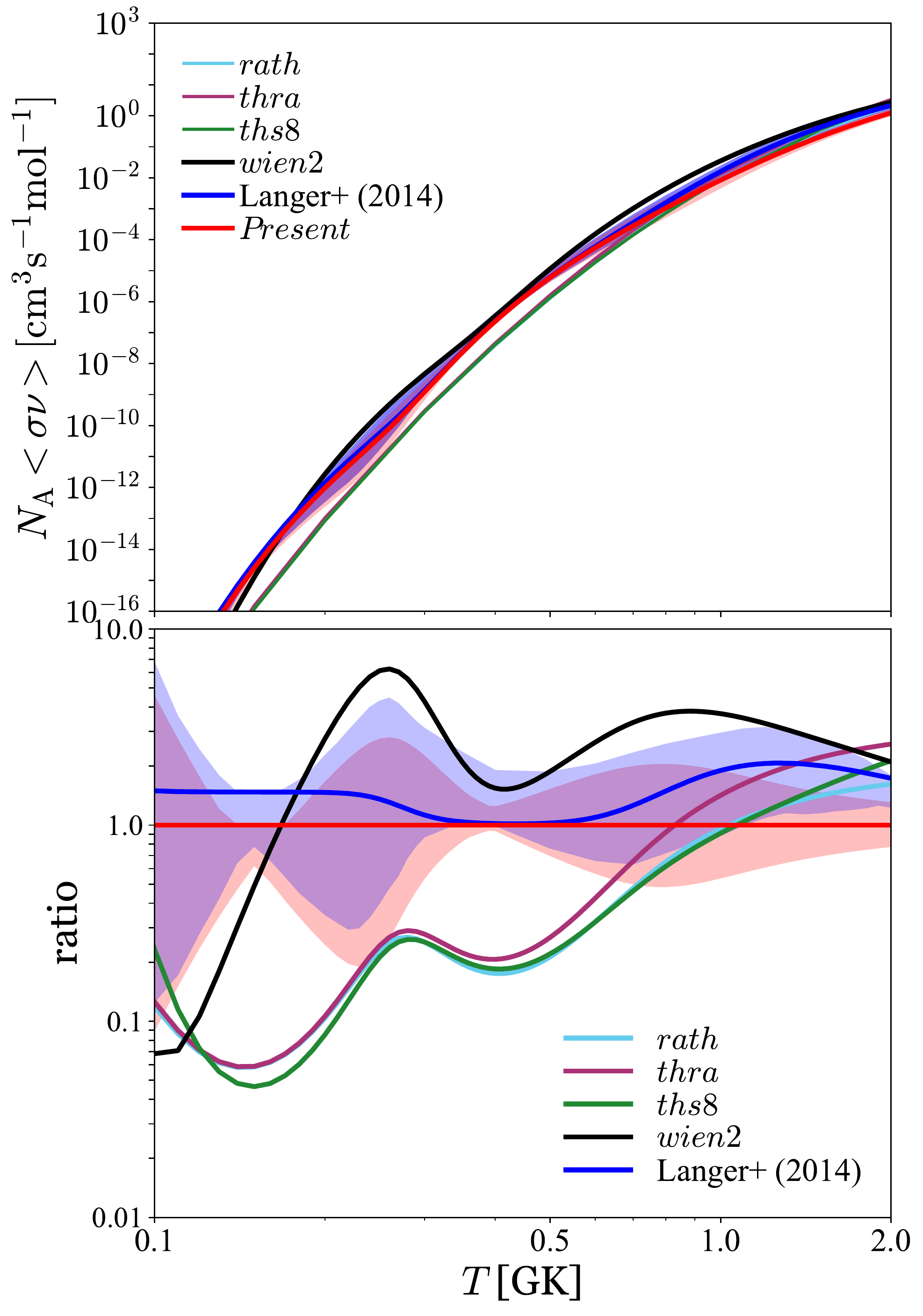}
% \vspace{-12mm}
\caption{\label{fig:rp_57Cu_58Zn}{\footnotesize The comparison of $^{57}$Cu(p,$\gamma$)$^{58}$Zn thermonuclear reaction rates. Top Panel: the \emph{rath}, \emph{thra}, \emph{ths8}, and \emph{wien}2 rates are the available rates compiled by \citet{Cyburt2010} and \emph{wien}2 is the recommended rate published in part of the JINA REACLIB~v2.2 release. All available rates in JINA REACLIB v2.2 define $S_\mathrm{p}$($^{58}$Zn)~$=2.277$~MeV.
%\emph{Bottom panel}: The ratios of Langer et al. (or wien2) rate to the present rate. The uncertainties of Langer et al. and the present rates are indicated as blue and red zones, respectively. 
Bottom Panel: the comparison of the \emph{Present} rate with Langer et al. rate and with the reaction rates compiled in the JINA REACLIB v2.2. The uncertainties of Langer et al. and the present rates are indicated as blue and red zones, respectively.}}
% (A color version of this figure is available in the online journal.)
% \vspace{-3mm}
% \end{center}
\end{figure}

We reproduce Langer et al. rate~\citep{Langer2014} taking into account contributions from the $1^+_2$, $2^+_4$, and $2^+_5$ resonance states, which are dominant at temperature region $0.8\lesssim T$(GK) $\lesssim 2$, see the top panel in Fig.~\ref{fig:rp_57Cu_58Zn_contri}. Other contributing resonances to Langer et al. rate for temperature $T\lesssim 0.8$~GK are also included in Fig.~\ref{fig:rp_57Cu_58Zn_contri}. The \emph{Present} rate and the respective main contributing resonances with updated $\Gamma_\mathrm{p}$ and $\Gamma_\gamma$ widths based on a full \emph{pf}-model space are plotted in the bottom panel of Fig.~\ref{fig:rp_57Cu_58Zn_contri}. We find that with the new energy of the $1^+_2$ state, estimated from the IMME formalism, the contribution of this resonance to the total rate reduces and becomes even less dominant than the contribution of the $2^+_4$ resonance state at temperature regime $0.8\lesssim T$(GK) $\lesssim 2$.
%We find that the new resonance energy estimated from IMME formalism reduces the contribution of the $1^+_2$ resonance state, becoming lower than the contribution of the $2^+_4$ resonance state \textcolor{red}{at temperature regime $0.8\lesssim T$(GK) $\lesssim 2$}.

The comparison of the \emph{Present} rate with Langer et al. rate and with other reaction rates compiled into JINA REACLIB v2.2 by \citet{Cyburt2010} is shown in Fig.~\ref{fig:rp_57Cu_58Zn}. The Hauser-Feshbach statistical model rates, i.e., \emph{rath}\footnote{Produced by \citet{Rauscher2000} using \textsc{Non-Smoker} code with FRDM mass input \citep{Moller1995}.}, \emph{thra}\footnote{\emph{Ibid.}, with ETFSI-Q mass input \citep{Pearson1996}.}, and \emph{ths8}\footnote{Produced by T. Rauscher using \textsc{Non-Smoker} code as part of JINA REACLIB since the v1.0 release \citep{Cyburt2010}.} are very close to one another from $0.1$ to $2.0$~GK, and they are lower than the \emph{Present} rate up to an order of magnitude at temperature $T \lesssim 0.9$~GK. Due to the reduction of the contribution from the $1^+_2$ resonance state, the \emph{Present} rate is up to a factor of two lower than Langer et al. rate from $0.8$ to $2$~GK covering the typical maximum temperature of GS~1826$-$24 burster, and up to a factor of four lower than the \emph{wien}2 rate \citep{Forstner2001} recommended by JINA REACLIB v2.2, see the comparison in the respective ratio in the bottom panel of Fig.~\ref{fig:rp_57Cu_58Zn}. 

% rath = RAuscher & THielemann
% thra = RAuscher & THielemann
% ths  = Thomas Rauscher
%\paragraph{\label{sec:uncertainty}Uncertainty of the $^{61}$Ga(p,$\gamma$)$^{62}$Ge reaction rates} 
By taking into account the uncertainty of $S_\mathrm{p}$($^{58}$Zn), we estimate and list the uncertainty of \emph{Present} $^{57}$Cu(p,$\gamma$)$^{58}$Zn reaction rate as upper and lower limits in Table~\ref{tab:rates}. Both upper and lower limits are shown as red zone in Fig.~\ref{fig:rp_57Cu_58Zn}, whereas the uncertainty of Langer et al. rate is indicated as blue zone. Even if the uncertainty due to the order of $1^+_1$ and $2^+_3$ states would have been removed, the uncertainty of $S_\mathrm{p}$($^{58}$Zn) propagated from the measured $^{58}$Zn mass \citep{Seth1986} is still dominant and persistent. Note that this is the first $^{57}$Cu(p,$\gamma$)$^{58}$Zn reaction rate constructed from important experimental information supplemented with the full $pf$-shell space shell-model calculation that yields converged resonance energies, $\Gamma_\gamma$, and spectroscopic factors; and the uncertainty is clearly identified, whereas the Hauser-Feshbach statistical model rates may include unknown systematic errors because of their limited capability in estimating level densities of nuclei near to the proton drip line.

\section{Implication on multi-zone X-ray burst models}
\label{sec:Astro}

We explore the influence of the \emph{Present} $^{57}$Cu(p,$\gamma$)$^{58}$Zn reaction rate on characterizing the XRB light curves of the GS~1826$-$24 X-ray source \citep{Makino1988,Tanaka1989} and burst ash composition after an~episode~of XRBs 
%, and the fluences and recurrence time of SAX~J1808.4$-$3658 photospheric radius expansion (PRE) burster 
based on one-dimensional multi-zone hydrodynamic XRB models. 
%The GS~1826$-$24 X-ray burster was discovered by \citet{Makino1988,Tanaka1989} with the \emph{Ginga} satellite in 1988 September, was termed as ``clocked burster'' by \citet{Ubertini1999} due to its almost consistent accretion rate, recurrence time, and light curve profile, and was called ``textbook burster'' by \citet{Bildsten2000} due to the agreement with theory.
The theoretical XRB models matched with the GS~1826$-$24 clocked burster \citep{Ubertini1999} are instantiated by the \textsc{Kepler} code~\citep{Weaver1978,Woosley2004,Heger2007} and were used by \citet{Heger2007} to perform the first quantitative comparison with the observed GS 1826$-$24 light curve. Later, the GS~1826$-$24 XRB models were used by \citet{Cyburt2016} and by \citet{Jacobs2018} to study the sensitivity of $(\alpha,\gamma)$, ($\alpha$,p), (p,$\gamma$), and (p,$\alpha$) nuclear reactions. The GS~1826$-$24 XRB models are continuously updated and were recently used by \citet{Goodwin2019b} and by \citet{Johnston2020} to study the high density properties of accreted envelopes of GS~1826$-$24 clocked burster.
%and SAX~J1808.4$-$3658 bursters, Goodwin2019a, Johnston2018
The XRB models are fully self-consistent, which take into account of the correspondence between the evolution in astrophysical conditions and the feedback of nuclear energy generation in substrates of accreted envelope. Throughout an episode of outbursts, which may consist of a series of bursts with either an almost consistent or progressively increasing recurrence time, the models are capable to keep updating the evolution of chemical inertia and thermal configurations that drive the nucleosynthesis in the accreted envelope of an accreting neutron star. 

The XRB models simulate a grid of Lagrangian zones \citep{Weaver1978,Woosley2004,Heger2007}, and each zone independently contains its own isotopic composition and thermal properties. We implement the time-dependent mixing length theory~\citep{Heger2000} to describe the convection transferring heat and nuclei between these Lagrangian zones. \textsc{Kepler} uses an adaptive thermonuclear reaction network that automatically includes or discards the respective reactions out of the more than 6000 isotopes provided by JINA REACLIB v2.2~\citep{Cyburt2010}. % as long as the abundance of a specific isotope reaches the predefined threshold. 

% \subsection{GS~1826$-$24 clocked burster model}

We adopt the XRB model from~\citet{Jacobs2018} to compare with the observed burst light curves of the GS~1826$-$24 clocked burster. The model had been used by \citet{Jacobs2018} in a recent sensitivity study of nuclear reactions. To match the modeled light curve with the observed light curve and recurrence time, $\Delta t_\mathrm{rec}=5.14\pm0.7$~h, of \emph{Epoch Jun 1998} of GS~1826$-$24 burster, we adjust the accreted $^1$H, $^4$He, and CNO metallicity fractions to $0.71$, $0.2825$, and $0.0075$, respectively. The accretion rate is tuned to a factor of $0.122$ of the Eddington-limited accretion rate, $\dot{M}_\mathrm{Edd}$. This adjusted XRB model with the associated nuclear reaction library (JINA REACLIB v2.2) characterizes the \emph{baseline} model in this work. Note that the \emph{wien}2 rate is the recommended $^{57}$Cu(p,$\gamma$)$^{58}$Zn reaction rate in JINA REACLIB v2.2. Other XRB models that adopt the same astrophysical configurations but implement 
the \emph{Present} $^{57}$Cu(p,$\gamma$)$^{58}$Zn; 
or the corrected $^{55}$Ni(p,$\gamma$)$^{56}$Cu \citep{Valverde2019}; 
or the \emph{Present} $^{57}$Cu(p,$\gamma$)$^{58}$Zn and Valverde et al. corrected $^{55}$Ni(p,$\gamma$)$^{56}$Cu; 
or Langer et al. $^{57}$Cu(p,$\gamma$)$^{58}$Zn, Valverde et al. corrected $^{55}$Ni(p,$\gamma$)$^{56}$Cu, and \citep{Kahl2019} $^{56}$Ni(p,$\gamma$)$^{57}$Cu;
or the \emph{Present} $^{57}$Cu(p,$\gamma$)$^{58}$Zn, Valverde et al. corrected $^{55}$Ni(p,$\gamma$)$^{56}$Cu, and Kahl et al. $^{56}$Ni(p,$\gamma$)$^{57}$Cu
reaction rates are denoted as \emph{Present}$^\dag$, \emph{Present}$^\ddag$, \emph{Present}$^\spadesuit$, \emph{Present}$^\heartsuit$, and \emph{Present}$^\S$ models, respectively. The \emph{Present}$^\heartsuit$ and \emph{Present}$^\S$ models implement a factor of $0.120$ of $\dot{M}_\mathrm{Edd}$ for the accretion rate in order to obtain a modeled recurrence time close to the observation, proposing that either \emph{Present} or Langer et al. $^{57}$Cu(p,$\gamma$)$^{58}$Zn reaction rate, which is lower than the \emph{wien}2 rate, shortens the recurrence time by up to $5$\%.
%model is best fit to the light curve of GS~1826$-$24 burster. The accretion rate is $3.325\times10^{-9}~M_{\odot}$yr$^{-1}$. 
%, see Table~\ref{tab:rates} and Fig.~\ref{fig:rp_57Cu_58Zn}.

We then simulate a series of 40 consecutive XRBs for \emph{baseline}, \emph{Present}$^\dag$, \emph{Present}$^\ddag$, \emph{Present}$^\spadesuit$, \emph{Present}$^\heartsuit$, and \emph{Present}$^\S$ models; and only the last 30 bursts are summed up with respect to the time resolution and then averaged to yield a burst light-curve profile. The first 10 bursts simulated from each model are excluded because these bursts undergo a transition from a chemically fresh envelope with unstable burning to an enriched envelope with chemically burned-in burst ashes and stable burning. Throughout the transition, the enriched burst ashes are recycled in the succeeding burst heating which gradually stabilize the following bursts. The averaging procedure applied on the modeled light curves is similar to the method performed by~\citet{Galloway2017} to produce an averaged light-curve profile from the observed data set of \emph{Epoch Jun 1998}.
%which consists of 5 bursts.
%and \emph{Mar 2007}, which consists of 5 and 9 bursts, respectively. 
The epoch was recorded by the \emph{Rossi X-ray Timing Explorer} (RXTE) Proportional Counter Array~\citep{Galloway2004,Galloway2008,MINBAR} and were compiled into the Multi-Instrument Burst Archive\footnote{\href{https://burst.sci.monash.edu/minbar/}{https://burst.sci.monash.edu/minbar/}} by \citet{MINBAR}. 

The yielded burst luminosity, $L_\mathrm{x}$, from each model is transformed and related to the observed flux, $F_\mathrm{x}$, via the relation~\citep{Johnston2020},
\begin{eqnarray}
\label{eq:flux}
F_\mathrm{x} =&& \frac{L_\mathrm{x}}{4\pi d^2\xi_\mathrm{b}(1+z)^2} \, ,
\end{eqnarray}
where $d$ is the distance; $\xi_\mathrm{b}$ takes into account of the possible deviation of the observed flux from an isotropic burster luminosity due to the scattering and blocking of the emitted electromagnetic wave by the accretion disc~\citep{Fujimoto1988,He2016}; and the redshift, $z$, re-scales the light curve when transforming into an observer's frame. The $d$ and $\xi_\mathrm{b}$ are combined to form the modified distance $d\sqrt{\xi_\mathrm{b}}$ by assuming that the anisotropy factors of burst and persistent emissions are degenerate with distance. We include the entire burst timespan of an averaged observational data to fit our modeled burst light curves of each model to the observed light curve. The best-fit $d\sqrt{\xi_\mathrm{b}}$ and $(1+z)$ factors of the \emph{baseline}, \emph{Present}$^\dag$, \emph{Present}$^\ddag$, \emph{Present}$^\spadesuit$, \emph{Present}$^\heartsuit$, and \emph{Present}$^\S$ modeled light curves to the averaged-observed light curve and recurrence time of \emph{Epoch Jun 1998} are 
$7.28$~kpc and $1.29$, %\emph{baseline}, 
$7.32$~kpc and $1.29$, %\emph{Present}$^\dag$, 
$7.32$~kpc and $1.29$, %\emph{Present}$^\ddag$, 
$7.32$~kpc and $1.28$, %\emph{Present}$^\spadesuit$, 
$7.30$~kpc and $1.29$, %\emph{Present}$^\heartsuit$,
$7.62$~kpc and $1.29$, %\emph{Present}$^\S$
respectively. 
%The redshift-distance factors that yield less than 30\% of deviation between modeled and observed burst light curves of the \emph{baseline} and \emph{Present}$^\S$ are plotted in Fig.~\ref{fig:redshift_distance}. The redshift-distance contour reveals that the best-fit redshift-distance factors for matching the modeled light curve and recurrence time to observation are within the best-match range (dark red regions in Fig.~\ref{fig:redshift_distance}). 
Using these redshift factors, we obtain a set of modeled recurrence times which are close to the observation. The recurrence times of \emph{baseline}, \emph{Present}$^\dag$, \emph{Present}$^\ddag$, \emph{Present}$^\spadesuit$, \emph{Present}$^\heartsuit$, and \emph{Present}$^\S$, are 
$4.85$~h, 
$4.91$~h, 
$4.91$~h, 
$4.88$~h, 
$4.96$~h, and 
$4.95$~h, 
respectively. 
Though further reducing the accretion rate for each model improves the matching between modeled and observed recurrence time, all modeled burst light curves remain similar. For instance, the recurrence time of the \emph{Present}$^\S$ model $\Delta t_\mathrm{rec}=4.95$~h is produced with defining accretion rate as $0.120$ $\dot{M}_\mathrm{Edd}$ and the produced burst light curve is similar to other modeled light curves in the present work.
\begin{figure}[t]
\begin{center}
%\hspace{-20mm}
%\includegraphics[width=8.7cm,angle=0]{GS1826_Flux.pdf}
\includegraphics[width=\columnwidth,angle=0]{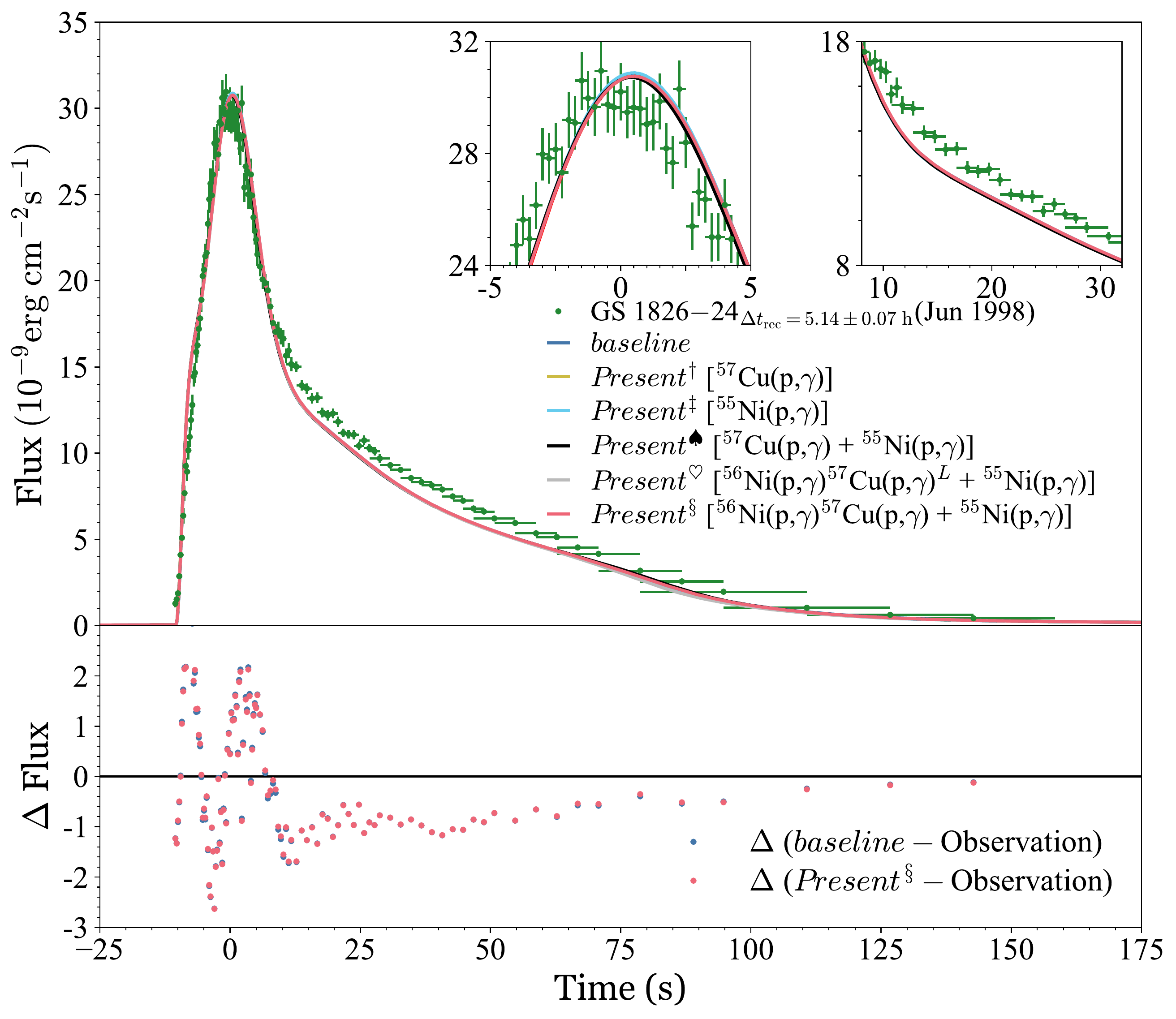}
%\vspace{-15mm}
\singlespace
\caption{\label{fig:Flux_GS1826}{\footnotesize The light curves of GS~1826$-$24 clocked burster as a function of time. Top Panel: the best-fit \emph{baseline}, \emph{Present}$^\dag$, \emph{Present}$^\ddag$, \emph{Present}$^\spadesuit$, \emph{Present}$^\heartsuit$, and \emph{Present}$^\S$ modeled light curves to the observed light curve and recurrence time of \emph{Epoch Jun 1998}. Both insets in the Top Panel magnify the light curve portions at $t=-5$ to 5~s (left inset) and at $t=8$ to 32~s (right inset). Bottom Panel: the deviation between the best-fit \emph{baseline} (or \emph{Present}$^\S$) modeled light curves and the observed light curve.}}
\vspace{-3mm}
\end{center}
\end{figure}

%%%%%%%%%%%%%%%%%%%%%%%%%%%%%%%%%%%%%%%%%%%%%%%%%%%%%%%%%%%%%%%%%
%%% t = -10.15 s onset
%%%%%%%%%%%%%%%%%%%%%%%%%%%%%%%%%%%%%%%%%%%%%%%%%%%%%%%%%%%%%%%%%

\begin{figure*}
% \gridline{\fig{NuclearChart_baseline_A.pdf}{12cm}{}}
\gridline{\fig{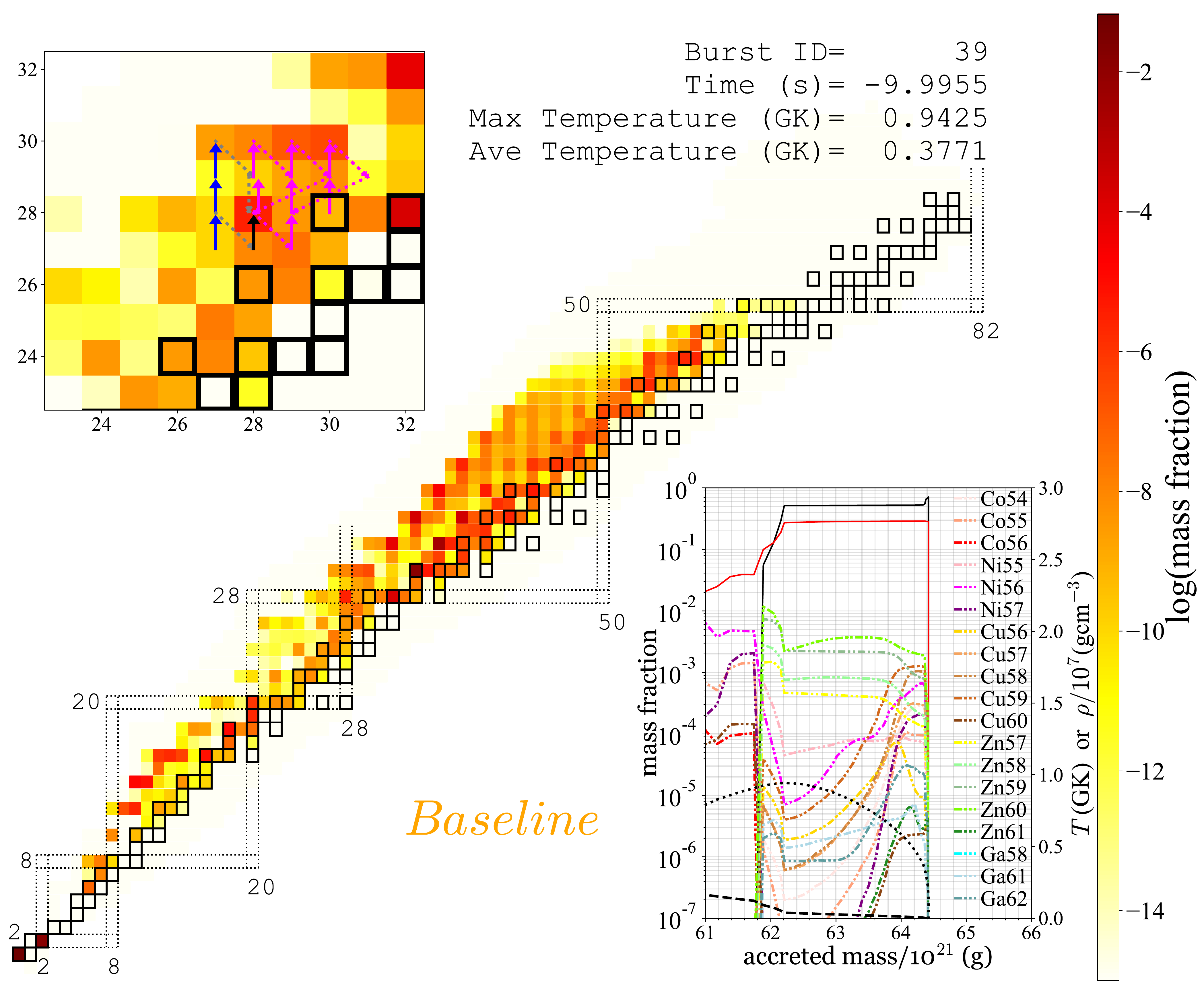}{9cm}{} \fig{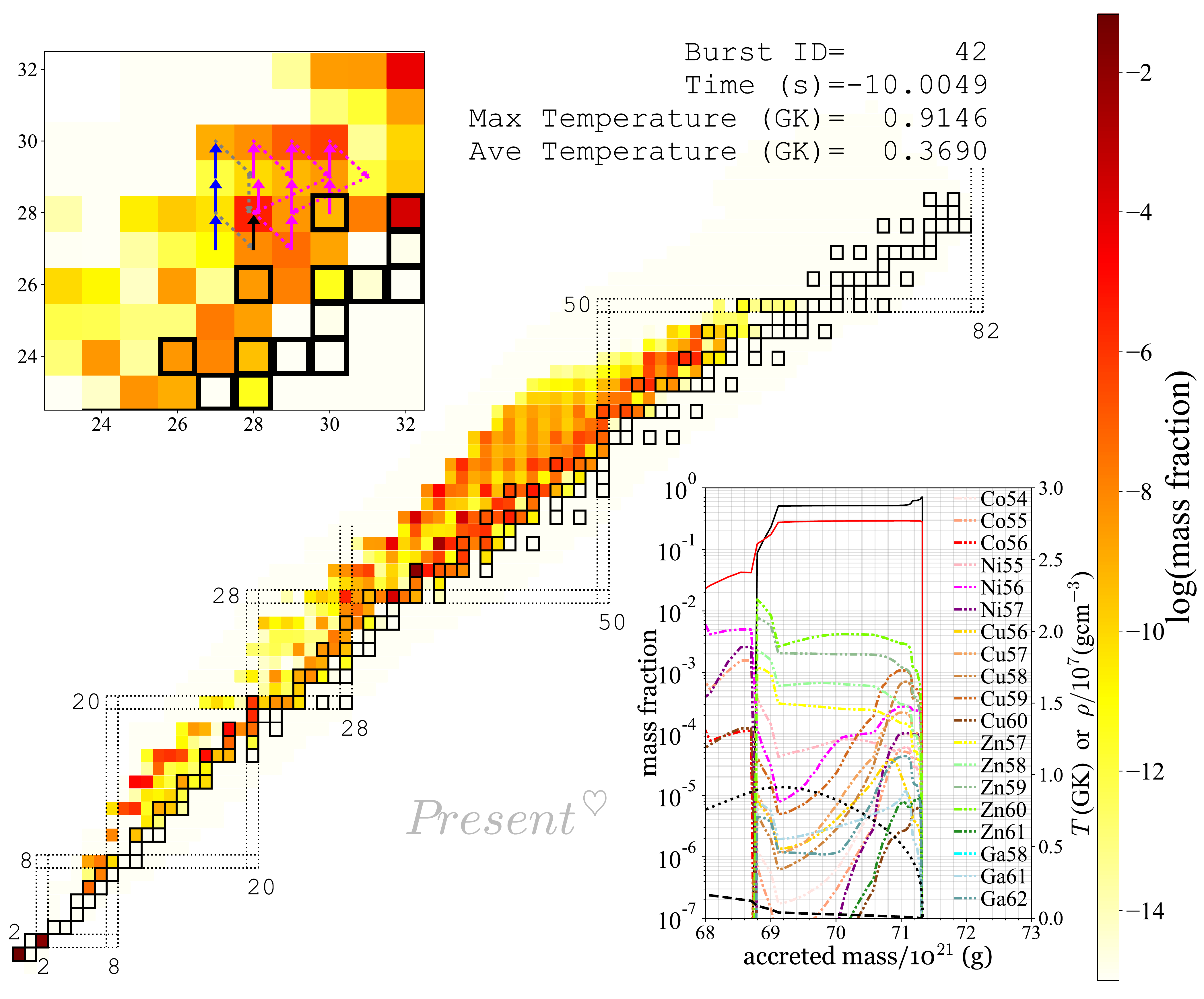}{9cm}{}}
\vspace{-10mm}
% \gridline{\fig{NuclearChart_Ni56Cu57Zn58_Ni55pg_A.pdf}{12cm}{}}
\gridline{\fig{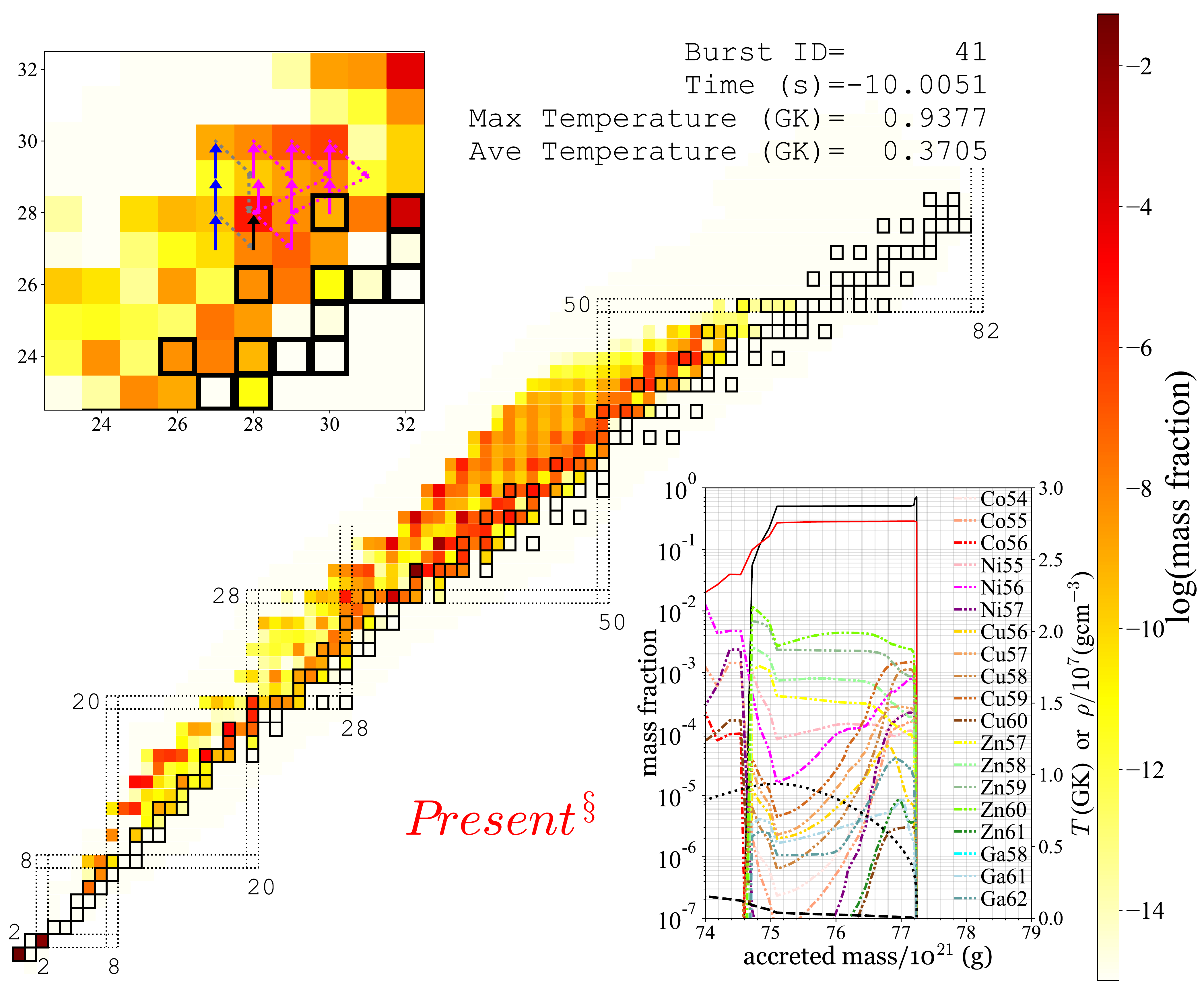}{9cm}{} \fig{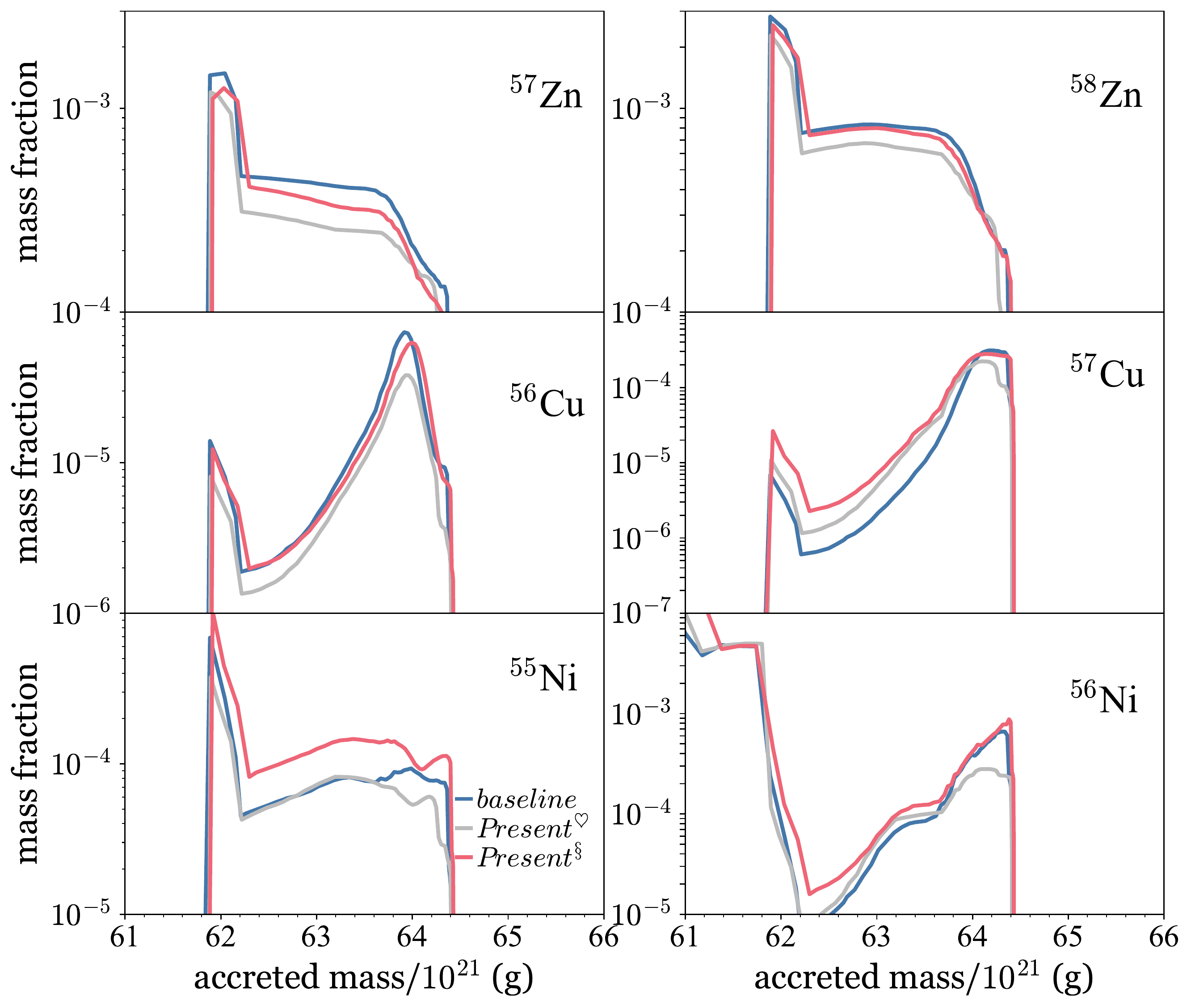}{9cm}{}}
\vspace{-5mm}
\singlespace
\caption{\label{fig:Onset}{\footnotesize The nucleosynthesis and evolution of envelope corresponds to the moment just before the onset of the 39$^\mathrm{th}$ burst for \emph{baseline} (\textsl{Top Left Panel}), of the 42$^\mathrm{nd}$ for \emph{Present}$^\heartsuit$ (\textsl{Top Right Panel}), and of the 41$^\mathrm{st}$ burst for \emph{Present}$^\S$ (\textsl{Bottom Left Panel}) scenarios. The averaged abundances of synthesized nuclei are represented by color tones referring to the right color scale in the nuclear chart of each panel. The black squares are stable nuclei. The top left insets in each panel magnify the regions related to the NiCu and ZnGa cycles. Pointing upward arrows indicate the (p,$\gamma$) reactions, whereas pointing downward arrows show the photodisintegration ($\gamma$,p) reactions. Slanting arrows from left to right depict ($\beta^+ \nu$) decays and long slanting arrows from right to left represent (p,$\alpha$) reactions. These arrows are merely used to guide the eyes. The bottom right insets in each panel present the corresponding temperature (black dotted line) and density (black dashed line) of each mass zone, referring to right the $y$-axis, and the abundances of synthesized nuclei, referring to the left $y$-axis, in the accreted envelope regime where nuclei heavier than CNO isotopes are densely synthesized. The abundances of H and He are represented by black and red solid lines, respectively. \textsl{Bottom Right Panel}: The comparisons of abundances of $^{55}$Ni, $^{56}$Cu, $^{57}$Zn, $^{56}$Ni, $^{57}$Cu, and $^{58}$Zn of \emph{baseline}, \emph{Present}$^\heartsuit$, and \emph{Present}$^\S$ at the respective time snapshot. These abundances are plotted with respect to the mass coordinate of the \emph{baseline} accreted envelope regime.}}
\end{figure*}

%%%%%%%%%%%%%%%%%%%%%%%%%%%%%%%%%%%%%%%%%%%%%%%%%%%%%%%%%%%%%%%%%
%%% t = 0 s burst peak
%%%%%%%%%%%%%%%%%%%%%%%%%%%%%%%%%%%%%%%%%%%%%%%%%%%%%%%%%%%%%%%%%

\begin{figure*}
% \gridline{\fig{NuclearChart_baseline_B.pdf}{12cm}{}}
\gridline{\fig{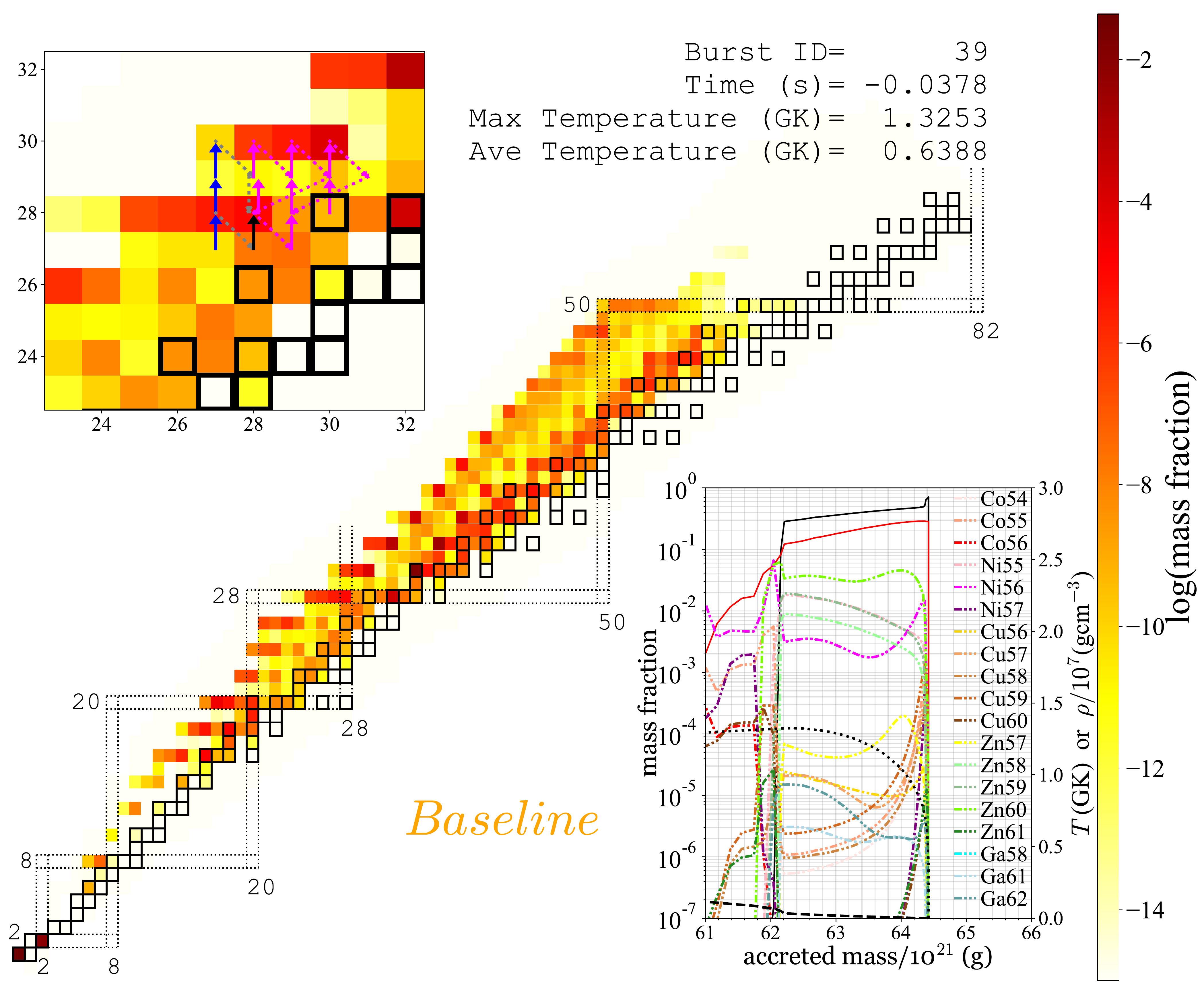}{9cm}{} \fig{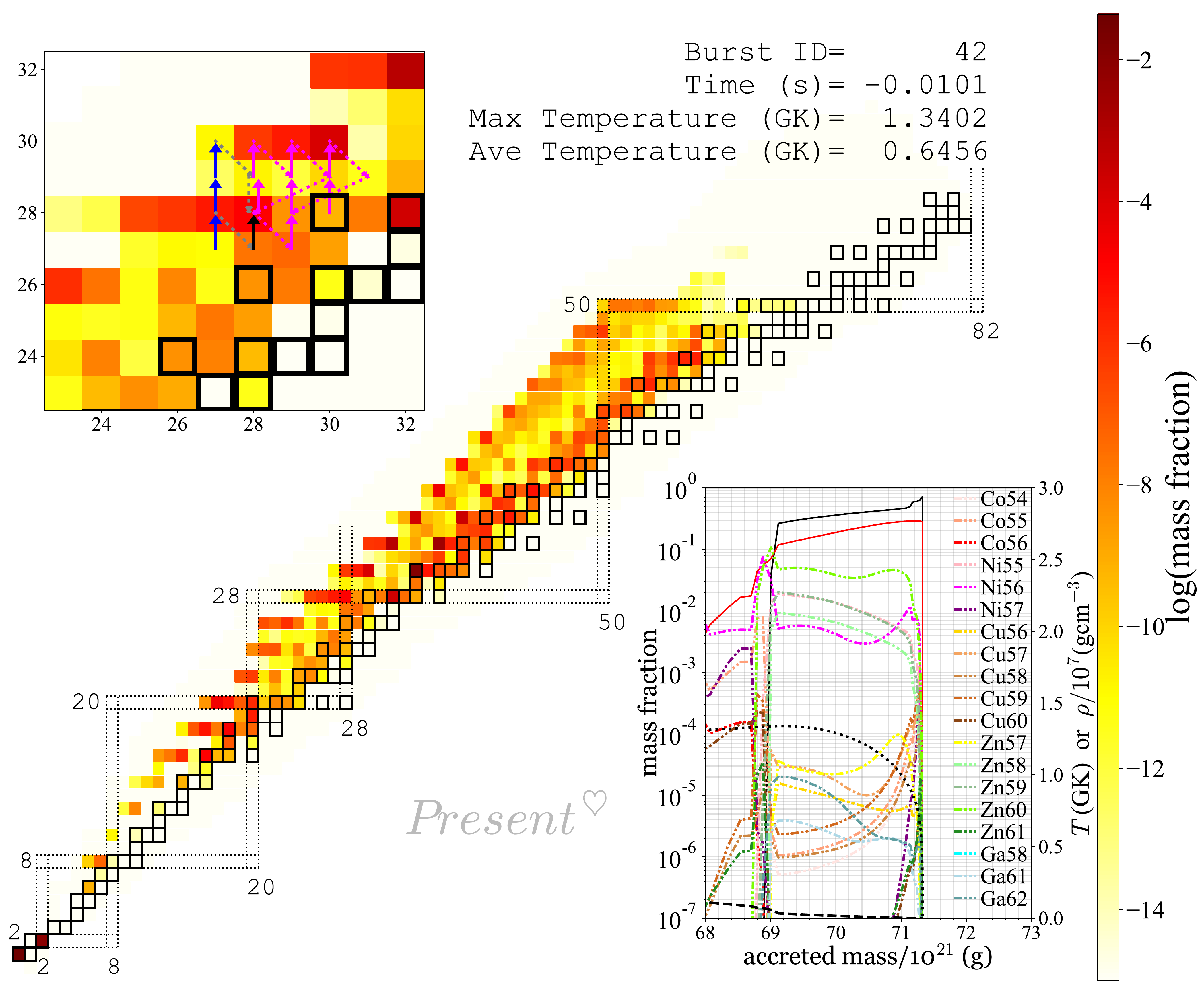}{9cm}{}}
\vspace{-10mm}
% \gridline{\fig{NuclearChart_Ni56Cu57Zn58_Ni55pg_B.pdf}{12cm}{}}
\gridline{\fig{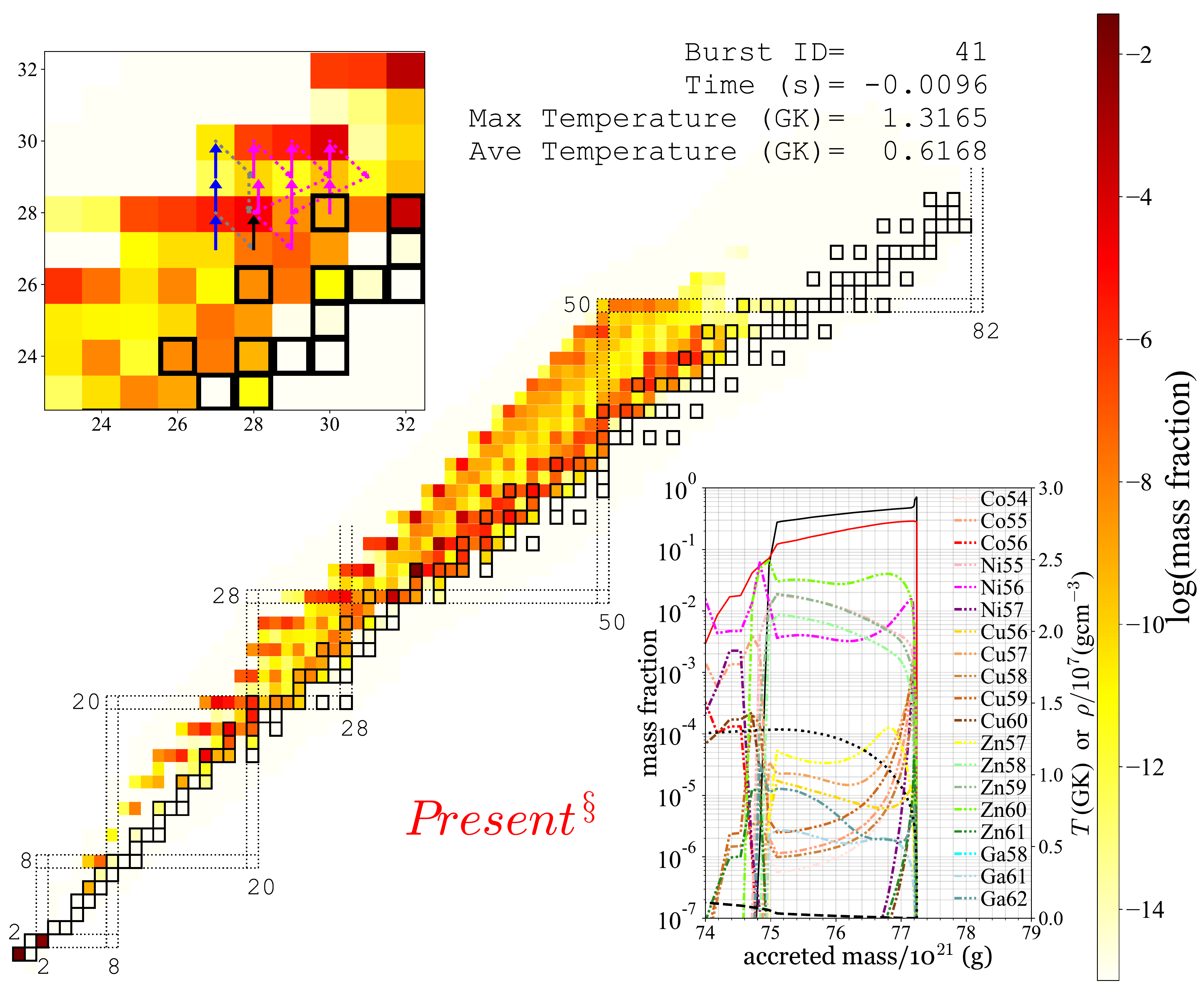}{9cm}{} \fig{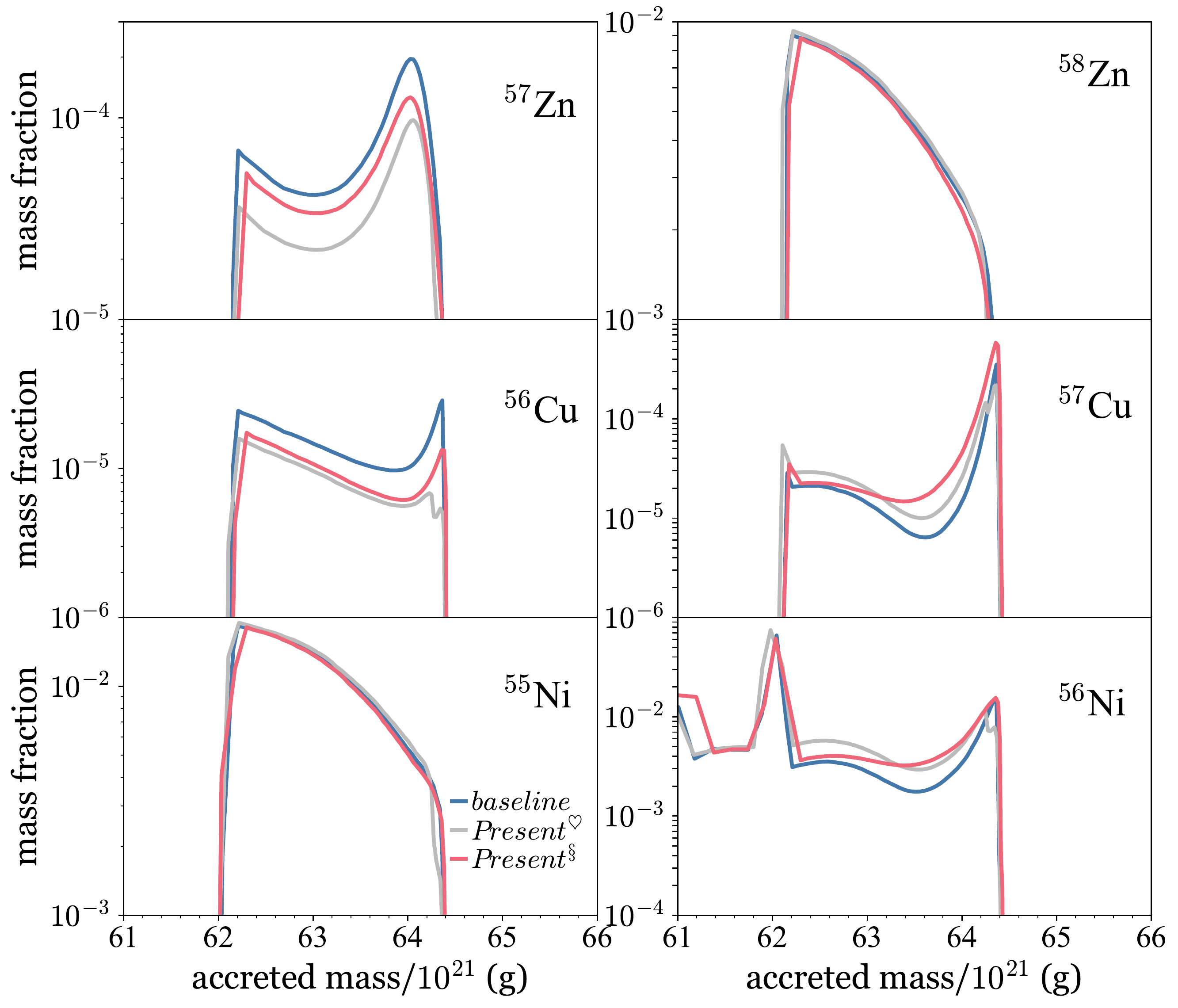}{9cm}{}}
\vspace{-10mm}
\caption{\label{fig:Peak}{\footnotesize The nucleosynthesis and evolution of envelope corresponds to the moment at the immediate vicinity of the burst peak for \emph{baseline} (\textsl{Top Left Panel}), \emph{Present}$^\heartsuit$ (\textsl{Top Right Panel}), and \emph{Present}$^\S$ (\textsl{Bottom Left Panel}) scenarios. See Fig.~\ref{fig:Onset} for further description.}}
\end{figure*}
%%%%%%%%%%%%%%%%%%%%%%%%%%%%%%%%%%%%%%%%%%%%%%%%%%%%%%%%%%%%%%%%%
%%% t = 14 s after peak
%%%%%%%%%%%%%%%%%%%%%%%%%%%%%%%%%%%%%%%%%%%%%%%%%%%%%%%%%%%%%%%%%

\begin{figure*}
% \gridline{\fig{NuclearChart_baseline_C.pdf}{12cm}{}}
\gridline{\fig{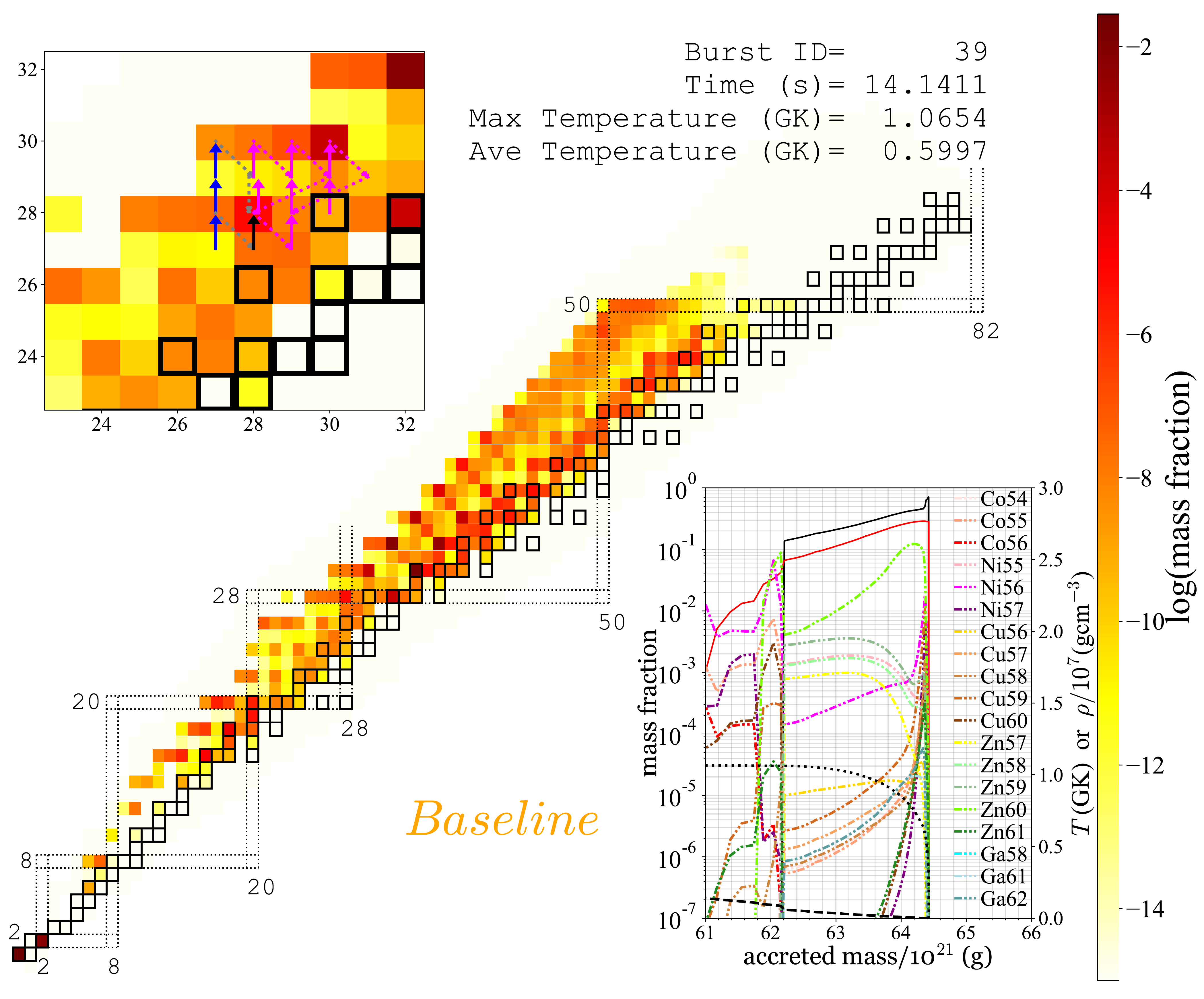}{9cm}{} \fig{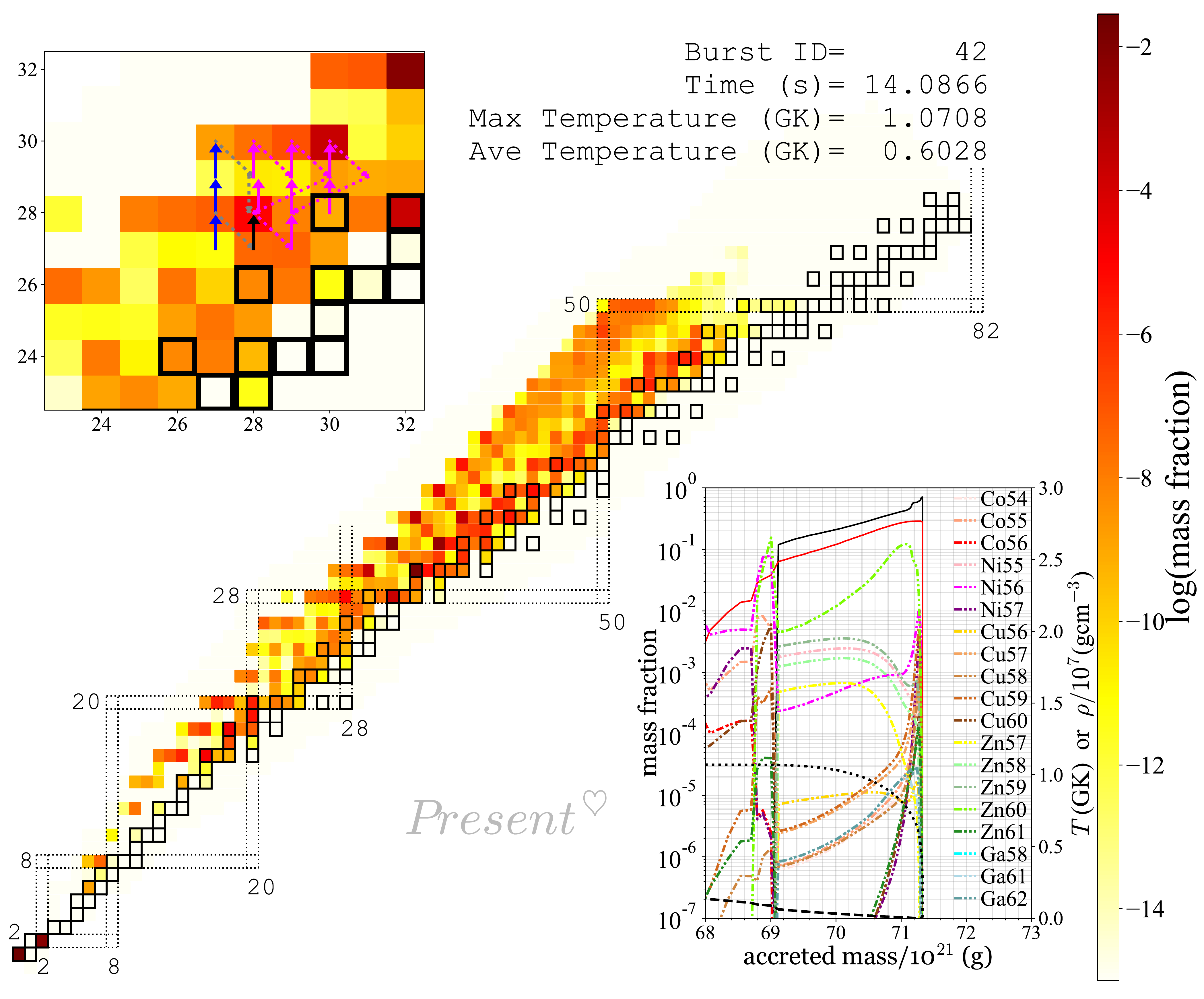}{9cm}{}}
\vspace{-10mm}
% \gridline{\fig{NuclearChart_Ni56Cu57Zn58_Ni55pg_C.pdf}{12cm}{}}
\gridline{\fig{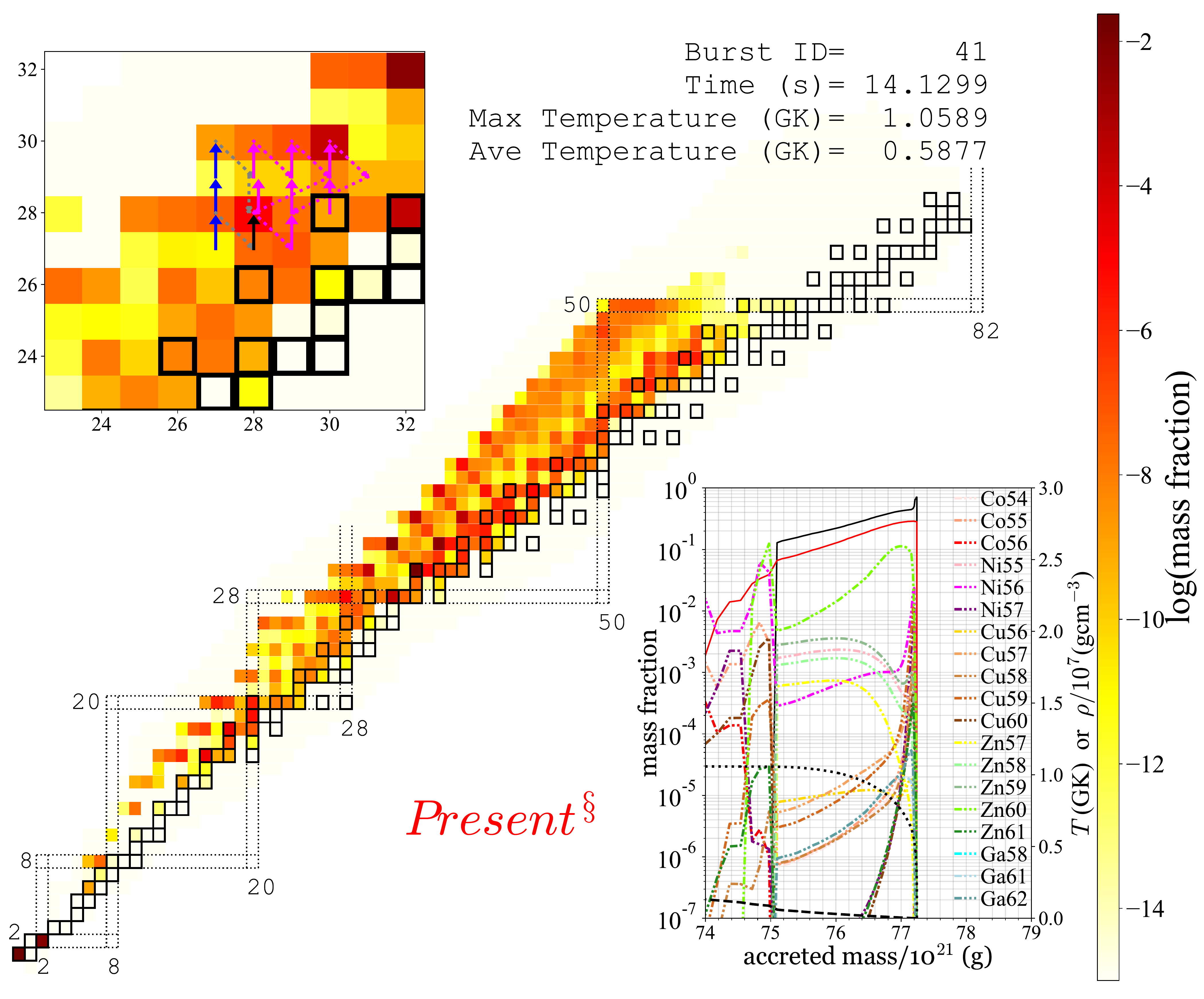}{9cm}{} \fig{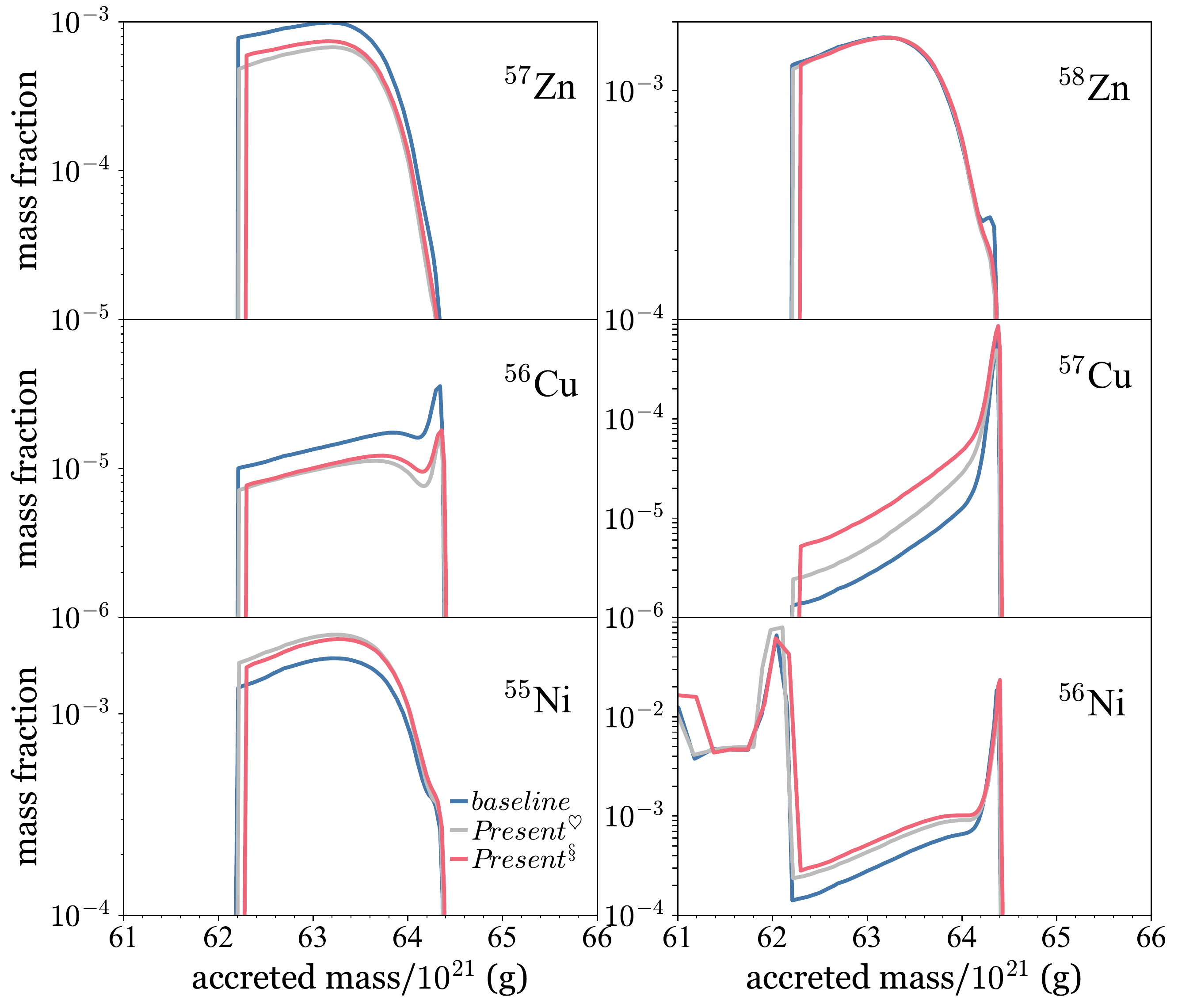}{9cm}{}}
\vspace{-10mm}
\caption{\label{fig:AfterPeak}{\footnotesize The nucleosynthesis and evolution of envelope corresponds to the moment of around 14~s after the burst peak for \emph{baseline} (\textsl{Top Left Panel}), \emph{Present}$^\heartsuit$ (\textsl{Top Right Panel}), and \emph{Present}$^\S$ (\textsl{Bottom Left Panel}) scenarios. See Fig.~\ref{fig:Onset} for further description.}}
\end{figure*}

%%%%%%%%%%%%%%%%%%%%%%%%%%%%%%%%%%%%%%%%%%%%%%%%%%%%%%%%%%%%%%%%%
%%% t = 180 s after peak
%%%%%%%%%%%%%%%%%%%%%%%%%%%%%%%%%%%%%%%%%%%%%%%%%%%%%%%%%%%%%%%%%

\begin{figure*}
% \gridline{\fig{NuclearChart_baseline_C.pdf}{12cm}{}}
\gridline{\fig{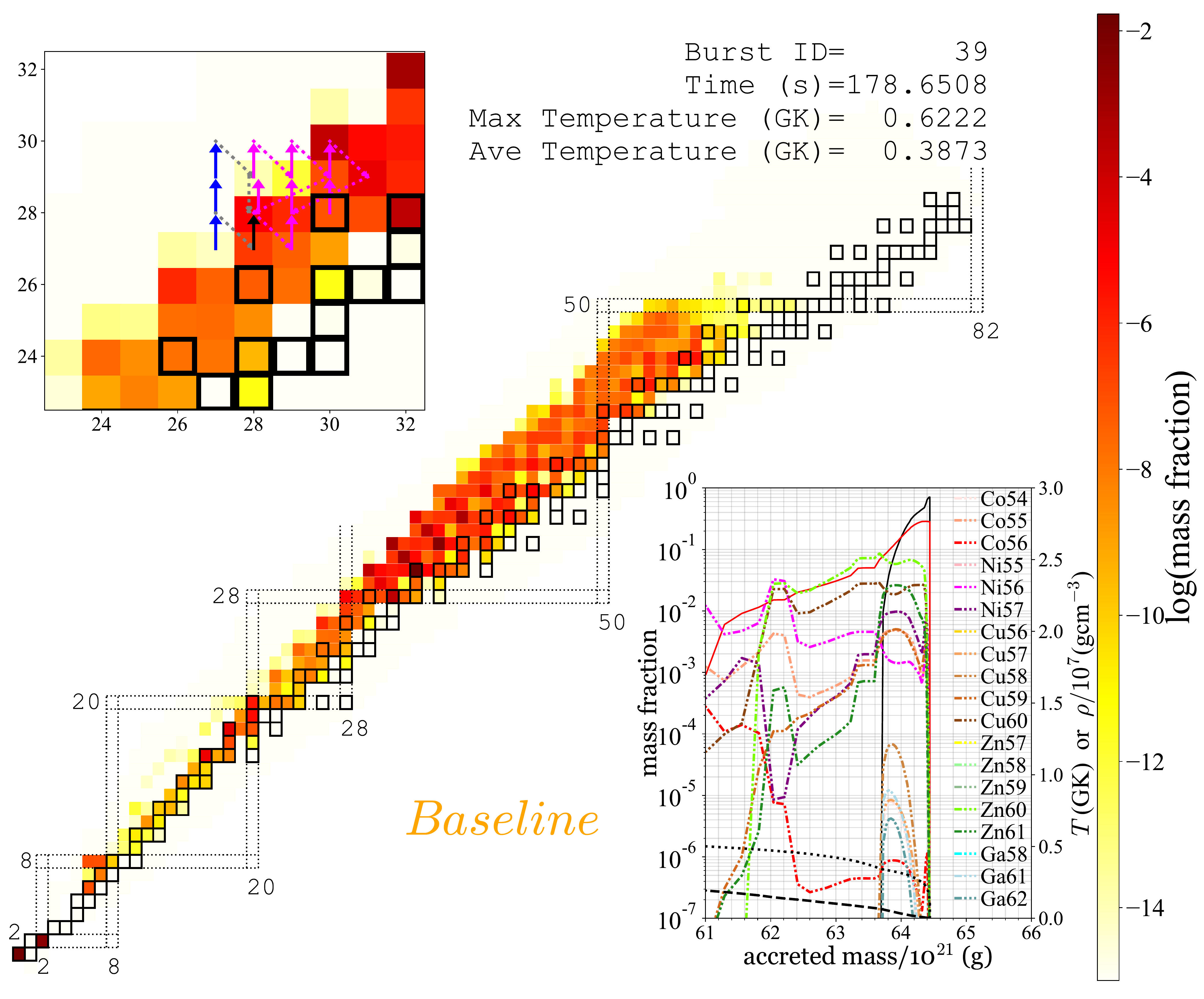}{9cm}{} \fig{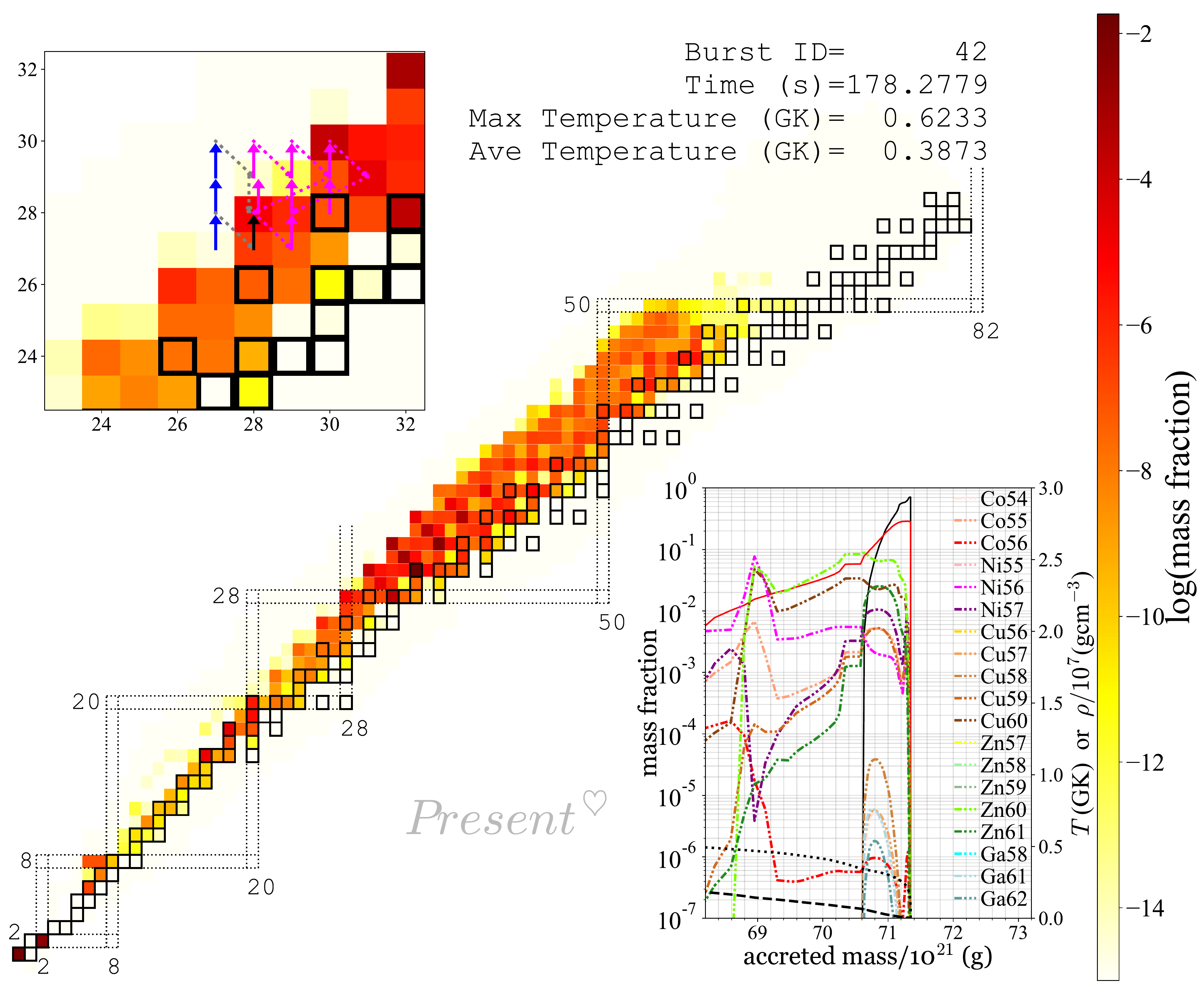}{9cm}{}}
\vspace{-10mm}
% \gridline{\fig{NuclearChart_Ni56Cu57Zn58_Ni55pg_C.pdf}{12cm}{}}
\gridline{\fig{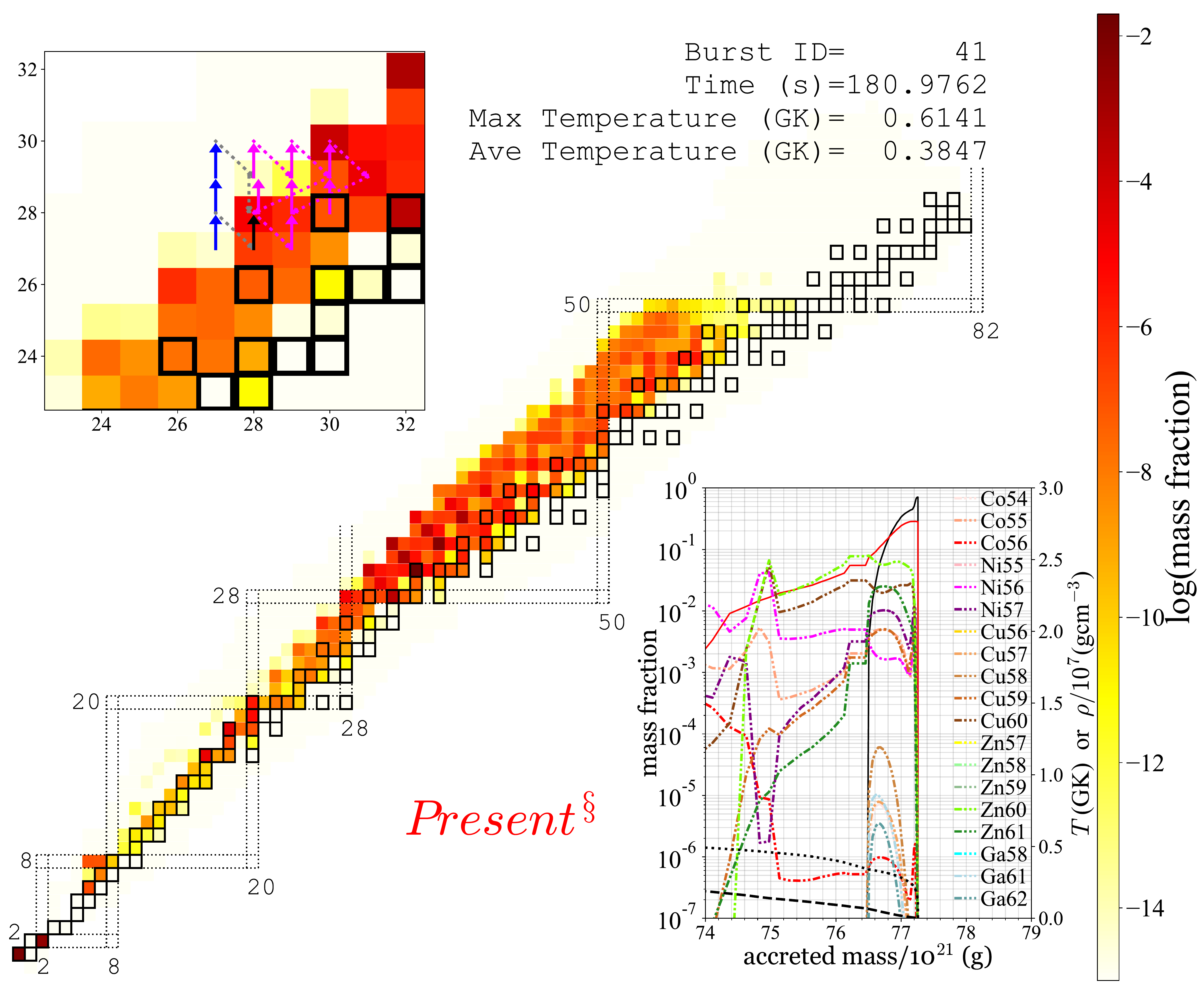}{9cm}{} \fig{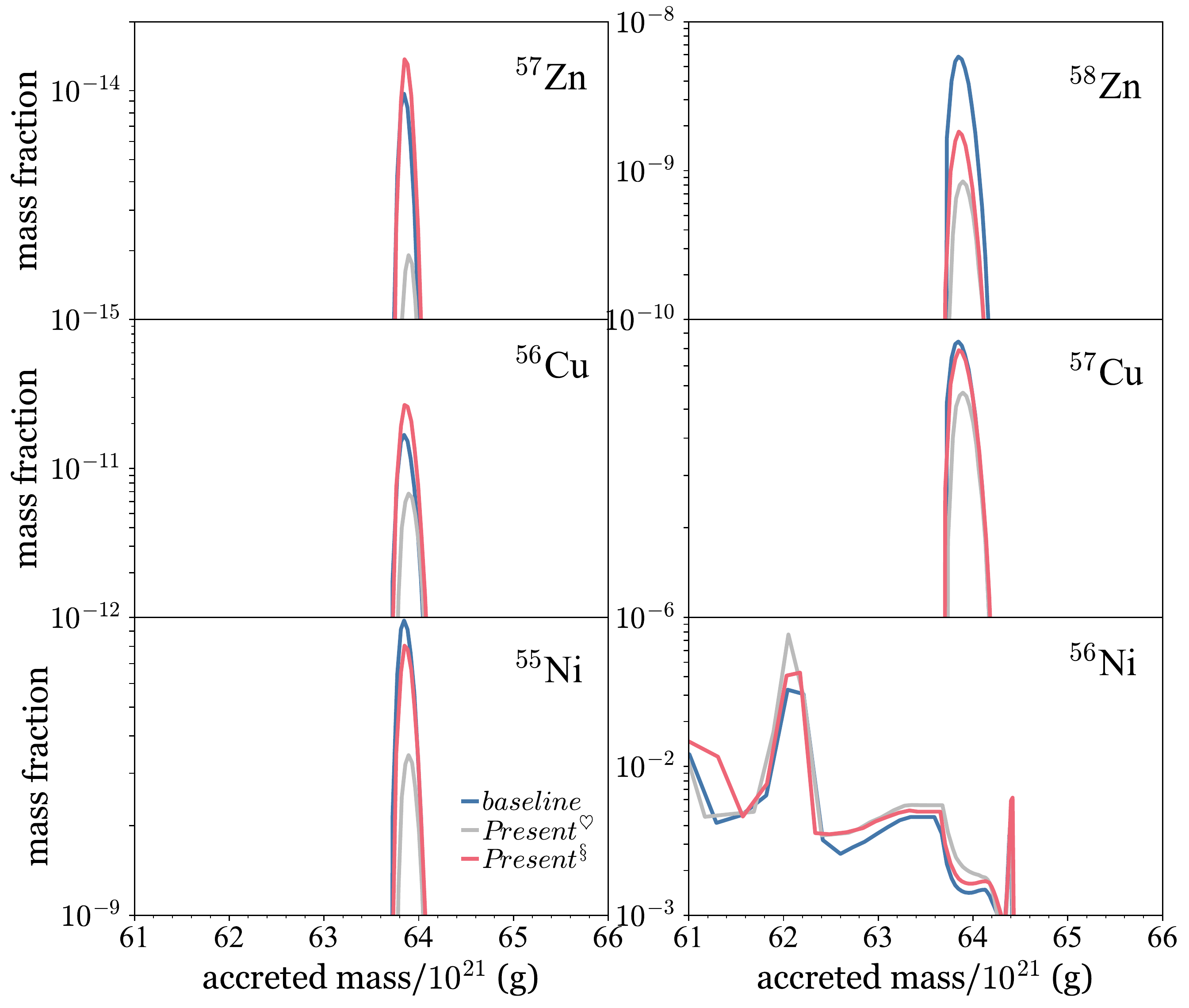}{9cm}{}}
\vspace{-10mm}
\caption{\label{fig:BurstTail}{\footnotesize The nucleosynthesis and evolution of envelope corresponds to the moment of around 180~s after the burst peak for \emph{baseline} (\textsl{Top Left Panel}), \emph{Present}$^\heartsuit$ (\textsl{Top Right Panel}), and \emph{Present}$^\S$ (\textsl{Bottom Left Panel}) scenarios. See Fig.~\ref{fig:Onset} for further description.}}
\end{figure*}

%%%%%%%%%%%%%%%%%%%%%%%%%%%%%%%%%%%%%%%%%%%%%%%%%%%%%%%%%%%%%%%%%

The top panel of Fig.~\ref{fig:Flux_GS1826} illustrates the comparison between the best-fit modeled and observed XRB light curves. The evolution time of light curve is relative to the burst-peak time, $t=0$~s. The overall averaged flux deviations between the observed epoch and each of these theoretical models, \emph{baseline}, \emph{Present}$^\dag$, \emph{Present}$^\ddag$, \emph{Present}$^\spadesuit$, \emph{Present}$^\heartsuit$, and \emph{Present}$^\S$, in units of $10^{-9}\mathrm{erg~cm}^{-2}\mathrm{s}^{-1}$ are 
$1.154$, 
$1.170$, 
$1.172$, 
$1.133$, 
$1.181$, and 
$1.147$,
respectively. The deviations between the \emph{Present}$^\S$ (and \emph{baseline}) and observed light curve throughout the whole timespan of the observed light curve are displayed in the bottom panel of Fig.~\ref{fig:Flux_GS1826}.
%Instead of specifically selecting data close to the burst peak at $t$=$-10\mathrm{~to~} 40$~s~\citep{Meisel2018a}, 
% and compared with normalized to 

The observed burst peak is thought to be located in the time regime $t=-2.5$~s~--~2.5~s (top left inset in Fig.~\ref{fig:Flux_GS1826}), and at the vicinity of the modeled light-curve peaks of \emph{baseline}, \emph{Present}$^\dag$, \emph{Present}$^\ddag$, \emph{Present}$^\spadesuit$, \emph{Present}$^\heartsuit$, and \emph{Present}$^\S$. The modeled light curves of \emph{baseline}, \emph{Present}$^\dag$, \emph{Present}$^\ddag$, \emph{Present}$^\spadesuit$, \emph{Present}$^\heartsuit$, and \emph{Present}$^\S$ at the near-burst-peak region $t=-4.5$~s~--~$5.5$~s are almost indiscernible.
%, whereas the peak of \emph{Present}$^\S$ is 0.25$\times10^{-9}\mathrm{erg~cm}^{-2}\mathrm{s}^{-1}$ lower than the peaks of \emph{baseline}, \emph{Present}$^\dag$, \emph{Present}$^\spadesuit$, and \emph{Present}$^\ddag$.
%the averaged deviation between the observed data and \emph{baseline}, and \emph{Present}$^0$ light curves, is 0.904 and 0.701, respectively, in units of~$10^{-9}\mathrm{erg~cm}^{-2}\mathrm{s}^{-1}$. 

All modeled light curves are less enhanced than the observed light curve at $t=8$~s~--~$80$~s, and the decrement is even augmented at around $t=13$~s and $40$~s, increasing the deviation between the modeled and observed light curves (bottom panel in Fig.~\ref{fig:Flux_GS1826}).
%and improving the reproducibility of the observational data with an averaged deviation of \textcolor{red}{0.301$\times10^{-9}\mathrm{erg~cm}^{-2}\mathrm{s}^{-1}$}. 
From the time regime at $t=78$~s onward until the burst tail end, all modeled burst light curves are enhanced. Overall, all modeled light-curve profiles are similar and note that the observed burst tail is reproduced from $t=78$~s onward until the burst tail end.
%due to the new $^{57}$Cu(p,$\gamma$)$^{58}$Zn reaction rate slightly deviates the modeled light curve from the observation. 

To investigate the microphysics behind the difference between both modeled burst light curves of the \emph{baseline} and \emph{Present}$^\S$ models, we consider the 39$^\mathrm{th}$, the 42$^\mathrm{nd}$, and the 41$^\mathrm{st}$ burst for the \emph{baseline}, \emph{Present}$^\heartsuit$, and \emph{Present}$^\S$ models, respectively. These bursts resemble the respective averaged light curve profile presented in Fig.~\ref{fig:Flux_GS1826}. The reference time of accreted envelope and nucleosynthesis in the following discussion is also relative to the burst-peak time, $t=0$~s.

% rp-process path and 
\paragraph{The moment before and during the onset.} 
After the preceding burst, the synthesized proton-rich nuclei in the accreted envelope go through $\beta^+$ decays and enrich the region around stable nuclei with long half-lives, e.g., $^{60}$Ni, $^{64}$Zn, $^{68}$Ge, and $^{78}$Se, which are the remnants of waiting points. When the accreted envelope evolves to the moment just before the onset of the succeeding XRB, due to the continuing nuclear reactions that occur in unburned hydrogen above the base of the accreted envelope, the temperature of the envelope increases up to a maximum value of about $0.93$~GK at the moment $t=-10$~s for the \emph{baseline}, \emph{Present}$^\heartsuit$, and \emph{Present}$^\S$ scenarios, see Fig.~\ref{fig:Onset}. At the moment just before the onset, some nuclei have already been synthesized and stored in the NiCu cycles, i.e., the NiCu I and II cycles \citep{Wormer1994}, and the sub-NiCu II cycle (Fig.~\ref{fig:Cycles_NiCu}), see the top left and bottom right insets of the top left, bottom left, and top right panels of Fig.~\ref{fig:Onset}.  %$^{56}$Ni(p,$\gamma$)$^{57}$Cu(p,$\gamma$)$^{58}$Zn($\beta^+\nu$)$^{58}$Cu(p,$\gamma$)$^{59}$Zn($\beta^+\nu$)$^{59}$Cu(p,$\alpha$)$^{56}$Ni, and the NiCu II cycle, $^{57}$Ni(p,$\gamma$)$^{58}$Cu(p,$\gamma$)$^{59}$Zn ($\beta^+\nu$)$^{59}$Cu(p,$\gamma$)$^{60}$Zn($\beta^+\nu$)$^{60}$Cu(p,$\alpha$)$^{57}$Ni \citep{Wormer1994}.
Among the isotopes in the NiCu cycles, the highly synthesized nuclei having mass fractions of more than 2$\times10^{-4}$ are $^{59}$Zn, $^{58}$Zn, $^{57}$Zn, $^{58}$Ni, $^{59}$Cu, $^{58}$Cu, $^{57}$Cu, $^{60}$Cu isotopes, and the $^{56}$Ni and $^{60}$Zn waiting points, whereas the $^{56}$Co and $^{60}$Cu isotopes having analogous mass-fraction distributions in the envelope are converted to $^{57}$Ni and $^{61}$Zn, respectively (the lower right insets in the top left, bottom left, and top right panels of Fig.~\ref{fig:Onset}).

% and not regulate the material flow
We find that although the reaction flow induced by the $^{55}$Ni(p,$\gamma$)$^{56}$Cu(p,$\gamma$)$^{57}$Zn branch noticeably bypasses the $^{56}$Ni waiting point and enriches $^{57}$Zn for the \emph{baseline}, \emph{Present}$^\heartsuit$, and \emph{Present}$^\S$ scenarios, it eventually has to go through the $^{57}$Zn($\beta^+\nu$)$^{57}$Cu(p,$\gamma$)$^{58}$Zn branch and combines with the NiCu cycles and then breaks out from the NiCu cycles to the ZnGa cycles, see the upper left insets in the top left, bottom left, and top right panels of Fig.~\ref{fig:Onset}. Due to the rather weak $^{57}$Zn(p,$\gamma$)$^{58}$Ga reaction, the $^{55}$Ni(p,$\gamma$)$^{56}$Cu(p,$\gamma$)$^{57}$Zn reactions and the subsequent $^{57}$Zn($\beta^+\nu$)$^{57}$Cu(p,$\gamma$)$^{58}$Zn branch redirect an appreciable amount material away from the $^{56}$Ni waiting point, but the redirecting branch does not store material. Moreover, the newly corrected $^{55}$Ni(p,$\gamma$)$^{56}$Cu reaction rate is lower than the one recommended in JINA REACLIB v2.2 (Fig.~\ref{fig:rp_55Ni_56Cu}), causing less enrichment of $^{57}$Zn in the \emph{Present}$^\heartsuit$ and \emph{Present}$^\S$ scenarios (bottom right panel of Fig.~\ref{fig:Onset}). This explains why neither the newly corrected nor the recommended $^{55}$Ni(p,$\gamma$)$^{56}$Cu reaction rate exhibiting significant influence on the light curve of GS~1826$-$24 burster and abundances of synthesized heavier nuclei. Also, the corrected $^{55}$Ni(p,$\gamma$)$^{56}$Cu reaction rate is not as influential as claimed by \citet{Valverde2018, Valverde2019}. Note that the one-zone models used by \citet{Valverde2018, Valverde2019} do not reproduce any burst light curves that are matched with observations. We remark that the \emph{baseline} model that uses the recommended $^{55}$Ni(p,$\gamma$)$^{56}$Cu reaction rate in JINA REACLIB v2.2 has already manifested the possibility of the bypassing reaction flow of the $^{56}$Ni waiting point without replacing the recommended rate by \citet{Valverde2019} corrected rate because the recommended $^{55}$Ni(p,$\gamma$)$^{56}$Cu reaction rate is stronger than the Valverde et al. corrected reaction rate, see Fig.~\ref{fig:rp_55Ni_56Cu}. 

At this moment, more than $60$~\% of mass zones in the accreted envelope, where nuclei heavier than CNO isotopes are densely synthesized, is with temperature above $0.8$~GK. The \emph{Present} $^{57}$Cu(p,$\gamma$)$^{58}$Zn reaction rate is up to a factor of $2$ lower than Langer et al. rate from $0.8$~GK to $2$~GK due to the reduction of the domination of $1^+_2$ resonance state (bottom panel of Fig.~\ref{fig:rp_57Cu_58Zn}), reducing the transmutation rate of $^{57}$Cu to $^{58}$Zn. This situation impedes the $^{56}$Ni(p,$\gamma$)$^{57}$Cu(p,$\gamma$)$^{58}$Zn reaction flow while enhances the reaction flow by-passing the important $^{56}$Ni waiting point, causing a higher production of $^{55}$Ni in the \emph{Present}$^\S$ scenario (bottom right panel of Fig.~\ref{fig:Onset}). Meanwhile, Valverde et al. corrected $^{55}$Ni(p,$\gamma$)$^{56}$Cu reaction rate implemented in the \emph{Present}$^\S$ scenario reduces the production of $^{57}$Zn, and induces the reaction flow to $^{57}$Cu. These reaction flows are regulated with new reaction rates and then produce a rather similar $^{58}$Zn abundance in the \emph{baseline} and \emph{Present}$^\S$ scenarios that are about a factor of $1.2$ higher than the $^{58}$Zn abundance in the \emph{Present}$^\heartsuit$.

Note that the productions of $^{55}$Ni, $^{56}$Cu, $^{57}$Zn, $^{56}$Ni, $^{57}$Cu, and $^{58}$Zn based on the \emph{Present}$^\heartsuit$ and \emph{Present}$^\S$ are discernible due to the \emph{correlated influence} among the \emph{Present} (or \citet{Langer2014}) $^{57}$Cu(p,$\gamma$)$^{58}$Zn, \citet{Valverde2019} corrected $^{55}$Ni(p,$\gamma$)$^{56}$Cu, and \citet{Kahl2019} $^{56}$Ni(p,$\gamma$)$^{57}$Cu reaction rates. The continuous impact from the correlated influence among these reactions and $^{59}$Cu(p,$\alpha$)$^{56}$Ni that cycles the reaction flow back to the reaction series in the NiCu cycles since the onset later influences the burst ash composition at the burst tail end.
The mass fraction of $^{57}$Cu in the \emph{baseline} is lower than the one in the \emph{Present}$^\heartsuit$ and \emph{Present}$^\S$ scenarios because the newly updated $^{56}$Ni(p,$\gamma$)$^{57}$Cu by \citet{Kahl2019} implemented in \emph{Present}$^\heartsuit$ and \emph{Present}$^\S$ is about up to a factor of 9 higher than the recommended $^{56}$Ni(p,$\gamma$)$^{57}$Cu rate from JINA REACLIB v2.2 used in \emph{baseline} at temperature region around 1~GK. Nevertheless, the mass fraction of $^{58}$Zn in the \emph{baseline} is about a factor of 1.2 higher than the one in the \emph{Present}$^\S$ scenario. This reflects a stronger flow of $^{57}$Cu(p,$\gamma$)$^{58}$Zn in the \emph{baseline} than in the \emph{Present}$^\S$ scenario. Such stronger flow is because the recommended \emph{wien}2 $^{57}$Cu(p,$\gamma$)$^{58}$Zn reaction rate from JINA REACLIB v2.2 used in \emph{baseline} is about up to a factor of 4 higher than the \emph{Present} $^{57}$Cu(p,$\gamma$)$^{58}$Zn reaction rates at temperature region around 1~GK. Meanwhile, the induced $^{57}$Zn($\beta^+\nu$)$^{57}$Cu flow from the reaction flow by-passing the important $^{56}$Ni waiting point stacks up the abundance of $^{57}$Cu in the \emph{Present}$^\S$ scenario. Hence, a strong flow of the $^{56}$Ni(p,$\gamma$)$^{57}$Cu coupled with a weak flow of the $^{57}$Cu(p,$\gamma$)$^{58}$Zn in the \emph{Present}$^\S$ scenario and the stacked up $^{57}$Cu eventually yield a set of almost similar mass fractions of $^{58}$Zn along the mass zones in the accreted envelope during the onset for both \emph{baseline} and \emph{Present}$^\S$ scenarios. On the other hand, the synthesized nuclei heavier than $^{68}$Se for the \emph{baseline} is almost as extensive as the \emph{Present}$^\S$ scenario, see the nuclear chart in Fig.~\ref{fig:Onset}. This indicates the reaction flow is regulated at the $^{60}$Zn waiting point by the $^{59}$Cu(p,$\alpha$)$^{56}$Ni reaction that competes with the $^{59}$Cu(p,$\gamma$)$^{60}$Zn reaction. Furthermore, after the reaction flow breaks out from the NiCu cycles through the $^{59}$Cu(p,$\gamma$)$^{60}$Zn(p,$\gamma$)$^{61}$Ga branch to the ZnGa cycles \citep{Wormer1994}, it is stored in the ZnGa cycles before surging through the nuclei heavier than $^{68}$Se. We find that the GeAs cycle that involves two-proton sequential capture of $^{64}$Ge consisting of $^{64}$Ge(p,$\gamma$)$^{65}$As(p,$\gamma$)$^{66}$Se reactions could weakly exist in the mid of onset until the moment after burst peak \citep{Lam2022a}, see the nucleosynthesis charts in Figs.~\ref{fig:Onset}, \ref{fig:Peak}, and \ref{fig:AfterPeak}. A new $^{65}$As(p,$\gamma$)$^{66}$Se reaction rate based on a more precise $^{66}$Se mass is desired to constrain the transient period, nonetheless, the fact that the transient existence of the weak GeAs cycle is not ruled out for the GS~1826$-$24 burster.

% \textcolor{red}{and transiently stored at the weak GeAs cycles} 
The ZnGa cycles have been recently investigated by \citet{Lam2022} using the same GS~1826$-$24 clocked burster model as is used in this work and the full $pf$-model space shell model calculation. They found that the GeAs cycle that follows the ZnGa cycles only weakly exists for a brief period, which could last until $t = 21.4$~s~--~$58.6$~s after the burst peak \citep{Lam2022a}. This causes some reactions relevant to the ZnGa cycles becomes decisive in controlling the reaction flow reaching nuclei heavier than Ge and Se isotopes where the extensive H-burning via (p,$\gamma$) reactions occur. These influential reactions are $^{59}$Cu(p,$\gamma$) and $^{61}$Ga(p,$\gamma$), which were identified and marked by \citet{Cyburt2016} as the top four most sensitive reactions on clocked burst light curve. \citet{Lam2022} found that the $^{59}$Cu(p,$\gamma$) and $^{61}$Ga(p,$\gamma$) reactions characterize the burst light curve of the GS~1826$-$24 clocked burster at $t \approx 8$~s~--~$30$~s after the burst peak and the burst tail end. Preliminary results of the investigation of the ZnGa cycles were presented in the Supplemental Material of \citet{Hu2021} prior to \citet{Lam2022} publication.

We notice that the balance between the $^{56}$Ni(p,$\gamma$)$^{57}$Cu and $^{57}$Cu(p,$\gamma$)$^{58}$Zn reactions also redistributes the reaction flow to the NiCu II cycle and then the reaction flow eventually joins with the NiCu I cycle and branches out to the ZnGa cycles at the $^{60}$Zn waiting point or follows the $^{60}$Cu(p,$\gamma$)$^{61}$Zn(p,$\gamma$)$^{62}$Ga reactions branches out to the ZnGa II cycle. Then, the joint reaction flow surges through the proton-rich region heavier than $^{64}$Ge where (p,$\gamma$) reactions actively burn hydrogen and intensify the rise of burst light curve from $t=-10$~s up to %$t$=$0.00^{+0.06}_{-0.06}$~s. 
$t=0$~s (burst peak). 

\paragraph{The moment at the immediate vicinity of burst peak.} As the redistributing and reassembling of reaction flow from the moment of onset until the burst peak regulate a rather similar feature of abundances in the NiCu cycles (the lower right insets in the top left, bottom left, and top right panels of Fig.~\ref{fig:Peak}), and the maximum envelope temperature of the \emph{baseline}, \emph{Present}$^\heartsuit$, and \emph{Present}$^\S$ scenarios are rather similar. These outcomes cause burst peaks of the \emph{baseline}, \emph{Present}$^\heartsuit$, and \emph{Present}$^\S$ scenarios almost close to each other, see the left inset in the upper panel of Fig.~\ref{fig:Flux_GS1826} and the maximum envelope temperatures in Fig.~\ref{fig:Peak}.
%is the account for

% $t$=$14.14^{+0.01}_{-0.01}$~s
%At $t$=$14.14$~s and $T$=$1.06$~GK, the stronger flow of $^{57}$Cu(p,$\gamma$)$^{58}$Zn in \emph{baseline} yields the lower (higher) abundance of $^{57}$Cu ($^{58}$Zn) than the one in \emph{Present}$^\S$ scenario, see Fig.~\ref{fig:AfterPeak}. 
\paragraph{The moment after the burst peak.} 
At $t\!\approx\!14$~s and $T\approx\!1.06$~GK (maximum envelope temperature), the redistributing of reaction flow since the moment of onset mentioned above slightly keeps the reaction flow in NiCu cycles for somehow longer time and slightly delays the reaction flow from passing through the waiting point $^{60}$Zn in the \emph{Present}$^\S$ scenario. The small delay allows the reaction flow to leak out from the NiCu cycles at later time and to burn hydrogen along the way reaching isotopes heavier than $^{68}$Se via (p,$\gamma$) reactions, and this situation mildly deviates the burst light curve of \emph{Present}$^\S$ from the light curves of \emph{baseline} and \emph{Present}$^\heartsuit$. 

\paragraph{The moment at the burst tail end.} The observed burst tail end of \emph{Epoch Jun 1998} of GS~1826$-$24 burster is closely reproduced by the \emph{baseline}, \emph{Present}$^\heartsuit$, and \emph{Present}$^\S$ models, meaning that the H-burning in these models recesses accordingly to produce a set modeled light curves in good agreement with observation. At $t=85$~s, the light curves of \emph{baseline} and \emph{Present}$^\heartsuit$ deviate from the light curve of \emph{Present}$^\S$ about $0.3\times10^{-9}\mathrm{erg~cm}^{-2}\mathrm{s}^{-1}$. Based on the analysis of the influence of $^{57}$Cu(p,$\gamma$)$^{58}$Zn reaction rate, we anticipate that if the actual energies of $1^+_2$ and $2^+_5$ resonance states are even higher than the presently estimated ones using the IMME formalism, the contributions of these two resonance states to the total rate are exponentially reduced, and the $2^+_4$ resonance state becomes the only dominant resonance at $T=1$~GK~--~$2$~GK for the $^{57}$Cu(p,$\gamma$)$^{58}$Zn reaction rate, and thus the modeled burst light curve is more diminished at the burst peak and at $t=8$~s~--~$32$~s, whereas at $t=65$~s~--~$150$~s, the \emph{Present}$^\S$ light curve is more enhanced compared to the \emph{baseline} scenario. 
% \paragraph{The moment at the burst tail end.}}
\begin{figure}[t]
\begin{center}
%\hspace{-20mm}
% \includegraphics[width=8.5cm,angle=0]{AbundanceMassA.pdf}
%\includegraphics[width=7.7cm,angle=0]{AbundanceMassA.pdf}
\includegraphics[width=\columnwidth,angle=0]{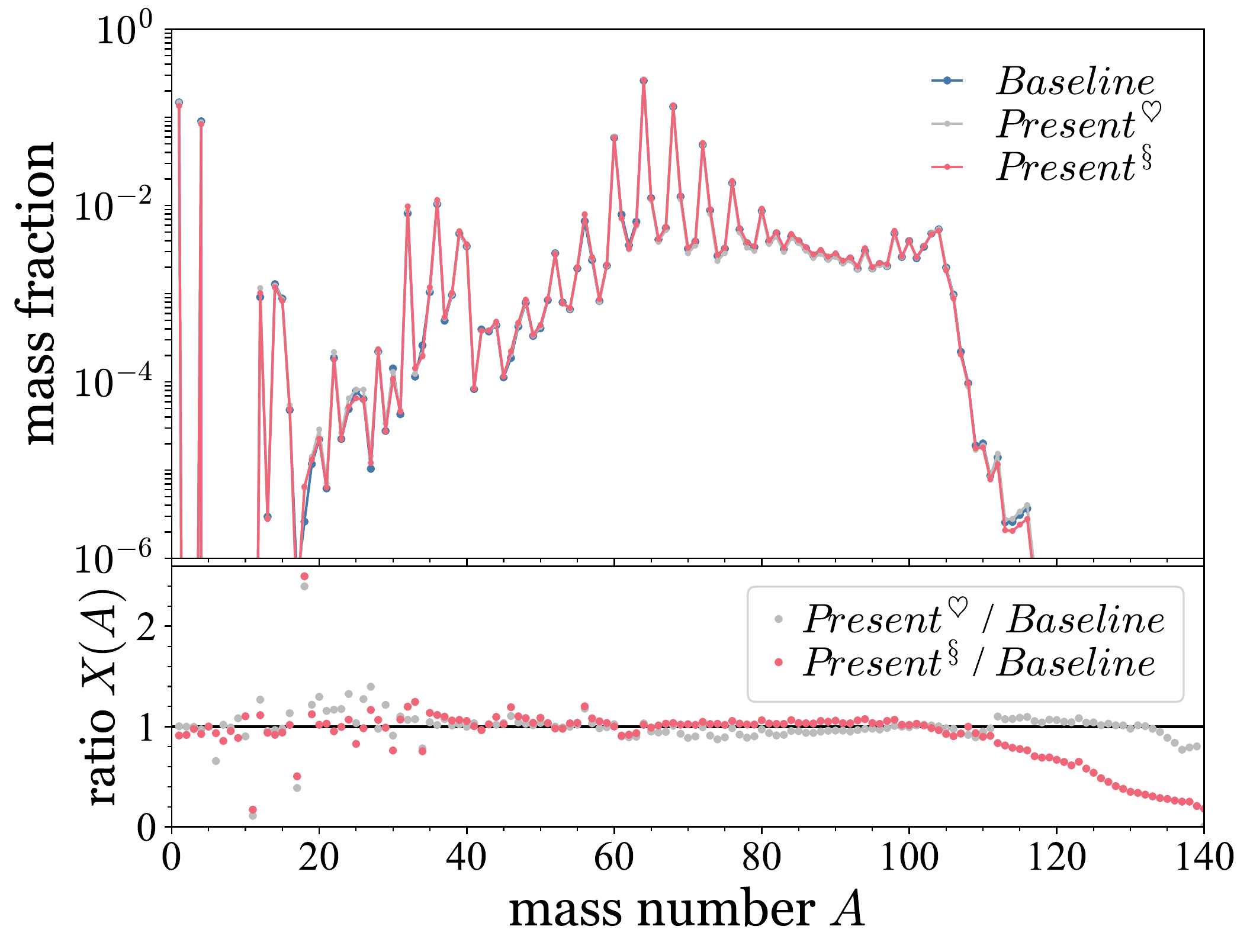}
%\vspace{-15mm}
\singlespace
\caption{{\footnotesize The averaged mass fractions for each mass number at burst tail when $t \approx 180$~s.}}
%(A color version of this figure is available in the online journal.)
\label{fig:abundances}
\end{center}
\end{figure}

%It is worth noting that the modeled burst tail starts to align with the observed burst tail end with merely 0.07$\times10^{-9}\mathrm{erg~cm}^{-2}\mathrm{s}^{-1}$ (\emph{Present}$^\S$) and 0.09$\times10^{-9}\mathrm{erg~cm}^{-2}\mathrm{s}^{-1}$ (\emph{baseline}) deviated from the observed light curve at $t$=78~--~150~s, see the bottom panel of Fig.~\ref{fig:Flux_GS1826}. 

From $t\!\!\approx\!\!14$~s onward until $t\!\approx\!180$~s, the regulation of NiCu cycles gradually deviates for the production of $^{58}$Zn due to the cumulated effect from the correlated influence among the latest $^{56}$Ni(p,$\gamma$)$^{57}$Cu, $^{57}$Cu(p,$\gamma$)$^{58}$Zn, and $^{55}$Ni(p,$\gamma$)$^{56}$Cu reaction rates, despite of the suppression induced by the $^{59}$Cu(p,$\alpha$)$^{56}$Ni reaction, see the bottom right panel in Fig.~\ref{fig:BurstTail}. Although the lower limit of \citet{Langer2014} rate at $0.23$~GK~$\lesssim T \lesssim 0.82$~GK is used for the \emph{Present}$^\heartsuit$ model, the \emph{Present}$^\heartsuit$ model still produces a set of $^{55}$Ni, $^{56}$Cu, and $^{57}$Zn abundances lower than the ones of \emph{baseline} and \emph{Present}$^\S$ models. Also, both \emph{baseline} and \emph{Present}$^\S$ models produce similar $^{55}$Ni, $^{56}$Cu, and $^{57}$Zn abundances. This indicates the cumulated impact that is generated from the difference of a factor of $2$ in temperature regime $T=0.8$~--~$2$~GK between the \emph{Present} and Langer et al. $^{57}$Cu(p,$\gamma$)$^{58}$Zn reaction rates. Meanwhile, the correlated influence on the syntheses of nuclei in the NiCu cycles is also manifested due to the \emph{Present} $^{57}$Cu(p,$\gamma$)$^{58}$Zn, $^{59}$Cu(p,$\alpha$)$^{56}$Ni, Kahl et al. $^{56}$Ni(p,$\gamma$)$^{57}$Cu, and Valverde et al. $^{55}$Ni(p,$\gamma$)$^{56}$Cu reaction rates since the onset at $t = -10$~s.

The compositions of burst ashes generated by these three models are presented in Fig.~\ref{fig:abundances}. The cumulated impact from the regulated NiCu I, II, and sub-II cycles based on the \emph{Present} and Langer et al. $^{57}$Cu(p,$\gamma$)$^{58}$Zn reaction rates manifests on the abundances of burst ashes. Using the \emph{Present} $^{57}$Cu(p,$\gamma$)$^{58}$Zn reaction rate, the production of $^{12}$C is reduced to a factor of $0.2$, and thus the remnants from the hot CNO cycle, e.g., nuclei $A=17$ and $18$ are affected up to about a factor of $0.5$ and $2.5$, respectively. The abundances of the daughters of SiP, SCl, and ArK cycles are reduced (increased) up to a factor of $0.7$ ($1.2$). The total abundance of $^{56}$Ni and its remnant is increased up to a factor of $1.2$ due to the correlated influence between the new $^{56}$Ni(p,$\gamma$)$^{57}$Cu, $^{57}$Cu(p,$\gamma$)$^{58}$Zn, and $^{55}$Ni(p,$\gamma$)$^{56}$Cu reaction rates. Meanwhile, the abundances of nuclei $A=64$~--~$104$ produced by \emph{Present}$^\S$ are closer to \emph{baseline} than the ones produced by \emph{Present}$^\heartsuit$. Furthermore, the abundances of nuclei $A=105$~--~$140$ are decreased up to a factor of 0.2 (red dots in the bottom panel of Fig.~\ref{fig:abundances}). Note that, the \emph{Present} $^{57}$Cu(p,$\gamma$)$^{58}$Zn reaction rate produces a different set of burst ash composition deviating from the one generated by Langer et al. $^{57}$Cu(p,$\gamma$)$^{58}$Zn reaction rate, especially the burst ash composition of $sd$-shell nuclei from $A=20$~--~$34$. Due to Langer et al. $^{57}$Cu(p,$\gamma$)$^{58}$Zn reaction rate, the abundances of nuclei $A=65$~--~$84$ are reduced up to a factor of $0.9$ and the abundances of nuclei $A=100$~--~$134$ are somehow closer to \emph{baseline} than the ones generated from the \emph{Present} $^{57}$Cu(p,$\gamma$)$^{58}$Zn reaction rate.

The noticeable difference in the burst ash compositions from the \emph{Present} and from Langer et al. $^{57}$Cu(p,$\gamma$)$^{58}$Zn reaction rates exhibits the sensitivity of the $^{57}$Cu(p,$\gamma$)$^{58}$Zn reaction in influencing the burst ash composition that eventually affects the composition of the neutron-star crust. Therefore, the presently more constrained $^{57}$Cu(p,$\gamma$)$^{58}$Zn coupled with the latest $^{56}$Ni(p,$\gamma$)$^{57}$Cu and $^{55}$Ni(p,$\gamma$)$^{56}$Cu reaction rates constricts the burst ash composition which is the initial input for studying superburst \citep{Gupta2007}.

\section{Summary and conclusion}
\label{sec:summary}
A theoretical study of $^{57}$Cu(p,$\gamma$)$^{58}$Zn reaction rate is performed based on the large-scale shell-model calculations in the full \emph{pf}-model space using GXPF1a and its charge-dependent version, cdGX1A, interactions. We present a detailed analysis of the energy spectrum of $^{58}$Zn on the basis of the IMME concept with the aim to determine the order of $1^+_1$ and $2^+_3$ states of $^{58}$Zn that are dominant in the $^{57}$Cu(p,$\gamma$)$^{58}$Zn reaction rate at $T=0.3$~--~$0.8$~GK. As no firm assignment can be done due to the lack of experimental information on $^{58}$Cu spectrum, we test an alternative assignment to the previously adopted one. We have also estimated the energy of $1^+_2$ state of $^{58}$Zn based on the presently available candidate for isobaric analogue states of $^{58}$Cu and $^{58}$Ni, which were experimentally determined, and the theoretical IMME $c$ coefficient. We estimate the $1^+_2$ state of $^{58}$Zn to be higher than the one predicted by the isospin conserving interaction \emph{pf}-shell interaction, GXPF1a. The dominance of the $1^+_2$ state in the $^{57}$Cu(p,$\gamma$)$^{58}$Zn reaction rate at $T=0.8$~--~$2$~GK is exponentially reduced. Throughout the course of a clocked burst, more than $60$~\% of the mass zones in the accreted envelope is heated to $T=0.8$~--~$1.6$~GK. The clocked XRBs of the GS~1826$-$24 burster is more sensitive to the $^{57}$Cu(p,$\gamma$)$^{58}$Zn reaction rate at the temperature range $0.8\lesssim T$~(GK)~$\lesssim 1.6$~GK. Thus, the resonance energy of the dominant $1^+_2$ state determining the $^{57}$Cu(p,$\gamma$)$^{58}$Zn reaction rate at $T=0.8$~--~$1.6$~GK is important in influencing the extent of synthesized nuclei during clocked bursts of GS~1826$-$24 burster.
%obtained from the isospin non-conserving interaction of \emph{pf}-shell nuclei, cdGX1A

Using the newly deduced $^{57}$Cu(p,$\gamma$)$^{58}$Zn, the newly corrected $^{55}$Ni(p,$\gamma$)$^{56}$Cu, and the updated $^{56}$Ni(p,$\gamma$)$^{57}$Cu reaction rates, we find that five combinations of these three reactions yield a set of light-curve profiles similar to the one generated by the \emph{baseline} model based on the \citet{Forstner2001} and \citet{Fisker2001} reaction rates which are labeled as \emph{wien}2 and \emph{nfis}, respectively, in JINA REACLIB v2.2. 
Nevertheless, the correlated influence on the nucleosyntheses exhibits that the $^{57}$Cu(p,$\gamma$)$^{58}$Zn reaction is critical in characterizing the burst ash composition. Constraining the $^{57}$Cu(p,$\gamma$)$^{58}$Zn reaction rate lower than a factor of $5$ difference in between the \emph{Present} and \emph{wien}2 $^{57}$Cu(p,$\gamma$)$^{58}$Zn reaction rates and than a factor of $2$ difference in between the \emph{Present} and Langer et al. $^{57}$Cu(p,$\gamma$)$^{58}$Zn reaction rates at the temperature regime relevant for XRBs is important for us to have a more constrained initial neutron-star crust composition.
%Nevertheless, the abundances of nuclei $A=60$, $64$, and $68$ between the \emph{baseline} and \emph{Present}$^\S$ scenarios in the burst ashes are quantitatively different.
%These two effects are due to the redistributing and reassembling of reaction flows in the NiCu cycles that reduce the influence of the $^{57}$Cu(p,$\gamma$)$^{58}$Zn reaction on the burst light curve and affect the nucleosyntheses of $^{57}$Cu, $^{58}$Zn, and some nuclei in the ZnGa cycles. 
We remark that the observed burst tail end of \emph{Epoch Jun 1998} of GS~1826$-$24 burster is closely reproduced by all models of the present work with the slightly adjusted astrophysical parameters. 
% which cannot be achieved with the GS~1826$-$24 models instantiated via MESA code.
% \emph{Present}$^\S$ and \emph{baseline} 

Furthermore, we find that the redistributing and reassembling of reaction flows in the NiCu cycles also diminish the impact of $^{55}$Ni(p,$\gamma$)$^{56}$Cu reaction though this by-passing reaction partially diverts material from the $^{56}$Ni waiting point, the reaction flow eventually joins with the NiCu cycles and leaks out to the ZnGa cycles. Indeed, as indicated by the one-dimensional multi-zone hydrodynamic XRB model matching with the GS~1826$-$24 clocked burster, implementing the \emph{nfis} $^{55}$Ni(p,$\gamma$)$^{56}$Cu reaction rate has already manifested the bypassing reaction flow of the $^{56}$Ni waiting point without the implementation of \citet{Valverde2019} $^{55}$Ni(p,$\gamma$)$^{56}$Cu reaction rate.

In addition, we notice that the weak GeAs cycle involving the two-proton sequential capture on $^{64}$Ge, following the $^{64}$Ge(p,$\gamma$)$^{65}$As(p,$\gamma$)$^{66}$Se branch may exist shortly around the mid of onset until after the burst peak. The period of this transient existence may depend on the precise determination of the $S_\mathrm{p}$($^{66}$Se) value. The \emph{Present} $^{57}$Cu(p,$\gamma$)$^{58}$Zn reaction rate, which is more constrained than \citet{Langer2014} reaction rate, was used by \citet{Lam2022a} to study the weak GeAs cycles, and was also recently used by \citet{Hu2021} to study the prevail influence of the newly deduced $^{22}$Mg($\alpha$,p)$^{25}$Al.

% \acknowledgments
We are very grateful to N. Shimizu for suggestions in tuning the \textsc{KShell} code at the PHYS\_T3 (Institute of Physics) and QDR4 clusters (Academia Sinica Grid-computing Centre) of Academia Sinica, Taiwan, to D. Kahl for checking and implementing the newly updated $^{56}$Ni(p,$\gamma$)$^{57}$Cu reaction rate, to B. Blank for implementing the IMME framework, to M. Smith for using the Computational Infrastructure for Nuclear Astrophysics, to B. A. Brown for using the \textsc{NuShellX@MSU} code, and to J. J. He for fruitful discussion.
This work was financially supported by 
the Strategic Priority Research Program of Chinese Academy of Sciences (CAS, Grant Nos. XDB34000000 and XDB34020204) 
%the Major State Basic Research Development Program of China (2016YFA0400503 and 2016YFA0400504) 
and National Natural Science Foundation of China (No. 11775277). 
We are appreciative of the computing resource provided by the Institute of Physics (PHYS\_T3 cluster) and the ASGC (Academia Sinica Grid-computing Center) Distributed Cloud resources (QDR4 cluster) of Academia Sinica, Taiwan. 
Part of the numerical calculations were performed at the Gansu Advanced Computing Center. 
YHL gratefully acknowledges the financial supports from the Chinese Academy of Sciences President's International Fellowship Initiative (No. 2019FYM0002) and appreciates the laptop (Dell M4800) partially sponsored by Pin-Kok Lam and Fong-Har Tang during the pandemic of Covid-19.
A.H. is supported by the Australian Research Council Centre of Excellence for Gravitational Wave Discovery (OzGrav, No. CE170100004) and for All Sky Astrophysics in 3 Dimensions (ASTRO 3D, No. CE170100013).
N.A.S. is supported by the IN2P3/CNRS, France, Master Project -- ``Exotic nuclei, fundamental interactions and astrophysics''.
%Part of the numerical calculations were performed using the cluster in the Institute of Physics, Academia Sinica, Taiwan


% \bibliography{ApJ_v00}{}
% \bibliographystyle{aasjournal}



% \end{CJK*}
\end{document}